\documentclass[%
aip,
amsmath,amssymb,
preprint,
groupedaddress,
]{revtex4-1}

\usepackage{graphicx}
\usepackage{dcolumn}
\usepackage{bm}
\usepackage[mathlines]{lineno}
\usepackage{cancel}

\usepackage[T1]{fontenc}
\usepackage{mathptmx}
\usepackage{etoolbox}
\usepackage{xcolor}
\usepackage[normalem]{ulem}
\usepackage{amsthm}

\makeatletter
\def\@email#1#2{%
 \endgroup
 \patchcmd{\titleblock@produce}
  {\frontmatter@RRAPformat}
  {\frontmatter@RRAPformat{\produce@RRAP{*#1\href{mailto:#2}{#2}}}\frontmatter@RRAPformat}
  {}{}
}%
\makeatother

\begin{document}

\preprint{AIP/123-QED}

\title[Physics of Fluids]{Physics-informed Kolmogorov--Arnold Network \\with Chebyshev Polynomials for Fluid Mechanics}
\author{Chunyu Guo}

\author{Lucheng Sun}%

\author{Shilong Li}

\author{Zelong Yuan\textsuperscript{*}}
 \email{yuanzelong@hrbeu.edu.cn}

\author{Chao Wang}

\affiliation{%
	\textsuperscript{1}College of Shipbuilding Engineering, Harbin Engineering University, Harbin, 150001, P.R. China
}%
\affiliation{\textsuperscript{2}Qingdao Innovation and Development Base, Harbin Engineering University, Qingdao, 266000, P.R. China
}%
\affiliation{\textsuperscript{3}Nanhai Institute of Harbin Engineering University, Harbin Engineering University, Sanya, 572024, P.R. China
}%
\affiliation{\textsuperscript{4}National Key Laboratory of Hydrodynamics, Harbin Engineering University, Harbin, 150001, P.R. China
}%

\date{\today}

\begin{abstract}
Solving partial differential equations (PDEs) is essential in scientific forecasting and fluid dynamics. Traditional approaches often incur expensive computational costs and trade-offs in efficiency and accuracy. Recent deep neural networks have improved the accuracy but require high-quality training data. Physics-informed neural networks (PINNs) effectively integrate physical laws to reduce the data reliance in limited sample scenarios. A novel machine-learning framework, Chebyshev physics-informed Kolmogorov--Arnold network (ChebPIKAN), is proposed to integrate the robust architectures of Kolmogorov--Arnold networks (KAN) with physical constraints to enhance the calculation accuracy of PDEs for fluid mechanics. We study the fundamentals of KAN, take advantage of the orthogonality of Chebyshev polynomial basis functions in spline fitting, and integrate physics-informed loss functions that are tailored to specific PDEs in fluid dynamics, including Allen--Cahn equation, nonlinear Burgers equation, Helmholtz equations, Kovasznay flow, cylinder wake flow, and cavity flow. Extensive experiments demonstrate that the proposed ChebPIKAN model significantly outperforms the standard KAN architecture in solving various PDEs by effectively embedding essential physical information. These results indicate that augmenting KAN with physical constraints can alleviate the overfitting issues of KAN and improve the extrapolation performance. Consequently, this study highlights the potential of ChebPIKAN as a powerful tool in computational fluid dynamics and propose a path toward fast and reliable predictions in fluid mechanics and beyond.
\end{abstract}


\maketitle

\section{\label{sec:level1}introduction}

Solving partial differential equations (PDEs) is crucial to make accurate physical predictions of dynamic systems and obtain high-fidelity, effective physical information. In fluid dynamics, computational fluid dynamics (CFD) is an important type of method to solve dynamic systems that consist of partial differential equations and obtain fluid dynamics data. CFD includes numerical techniques such as finite-difference scheme, finite-volume method, and finite-element method \cite{moin1998direct}. Although classical CFD approaches can be sufficiently precise, they require a significant amount of computational resources to numerically implement. Although many accelerated algorithms and reduced-order models have been proposed to improve the computational efficiency of CFD \cite{PIROZZOLI2021110408,KOMEN2017565,SHOEYBI20105944,BHAGANAGAR2002200,ZHANG2021110153,CLERICI2024113191,XIAO20121848}, they also sacrifice the accuracy of PDE calculations. In practical engineering applications, some physical information of a complicated dynamic system, such as initial conditions, boundary conditions, and governing equations, are unknown \cite{jin2021nsfnets}, and render conventional CFD techniques inadequate for accurate fluid simulation. Even when various methods are used to enhance computational efficiency, numerical calculations are time-consuming. Therefore, alternative methods are required to achieve fast flow field predictions.

In recent years, machine learning (ML) techniques have quickly gained attention and have been rapidly developed in fluid mechanics research due to their minimal requirements for physical information and fast response capabilities. At present, in mathematics and practical engineering, deep neural networks have been applied to solving partial differential equations due to their excellent nonlinear fitting and interpolation capabilities. Srinivasan et al.\cite{srinivasan2019predictions} evaluated the ability of deep neural networks to predict temporal turbulent flows, generated training data using a nine-equation shear flow model, and tested the multi-layer perceptron (MLP) and long short-term memory (LSTM) networks. The results indicate that LSTM is better than MLP in predicting turbulence statistics and dynamic behaviors since LSTM can use the sequential characters of data with relatively small relative errors. This exploratory study established the foundation for the future development of accurate and effective data-driven sub-grid scale models to apply them to more complex large-eddy simulations of wall turbulence. Bhatnagar et al.\cite{bhatnagar2019prediction} proposed a convolutional-neural-network (CNN)-based model to predict the airfoil flow field, where Reynolds-averaged Navier Stokes (RANS) solutions on airfoil profiles were used as training data. This model automatically extracts mapping features with minimal human supervision and can more rapidly predict the velocity and pressure fields than RANS solvers while enhancing prediction capabilities through specific convolution operations and parameter sharing. X Zhang et al.\cite{10.1063/5.0195824} proposed a Residual Graph Convolutional Network for Initial Flow Field Setting (RGCN-IFS) that provides high-quality initial flow fields for CFD simulations, effectively bridging the accuracy gap between intelligent surrogate models and conventional CFD approaches while maintaining computational efficiency. Santos et al.\cite{santos2020poreflow} introduced PoreFlow Net, which is a three-dimensional (3D) convolutional neural network architecture that significantly accelerates the flow field prediction process by extracting the spatial relationship between porous media morphology and fluid velocity field. This approach alleviates the need for costly numerical simulations and demonstrates the successful application of physics-based machine learning models in digital rock physics. Kim et al.\cite{kim2020prediction} used a CNN framework to predict the local heat flux through direct-numerical-simulation data and demonstrated that the accuracy was preserved even at higher Reynolds numbers. These results enhanced the understanding of turbulence heat transfer and provided a novel method for turbulence modeling. W. Chen et al.\cite{CHEN2021110666} proposed a physics-informed machine learning framework for efficient reduced-order modeling of parametric PDEs. Through performance comparisons with traditional methods, the framework demonstrates significant potential for the efficient reduced-order solution of PDEs. Erichson et al.\cite{erichson2020shallow} proposed a data-driven method that used shallow neural networks to reconstruct flow fields from limited measurement data, which relieved the requirement for complex preprocessing. Kashefi et al.\cite{kashefi2021point} proposed a deep learning framework based on the PointNet architecture to predict the flow field in irregular domains, where point cloud inputs were used to embed the spatial positions into CFD quantities. This method effectively preserves geometric information and trains much faster than traditional CFD solvers without decreasing the prediction accuracy.

Although applying deep neural networks to predict fluid mechanics have achieved satisfactory results and reduced computational costs, the accuracy of neural networks continues to rely on the support of a large amount of high-fidelity data. Currently, the scarcity of sample data is increasingly hindering the development of neural networks. The physics-informed neural network (PINN) \cite{raissi2019physics} was proposed and quickly gained widespread attention due to its "benevolent" requirement for data. From the perspective of the loss function, the PINN framework can be considered an extension of an artificial neural network, where the incorporation of a physical-information-based penalty term significantly enhances the adherence of neural network to physical laws \cite{alzubaidi2023survey}. Employing physical laws, PINN can effectively perform in scenarios with insufficient measurement data to reduce the excessive dependence of classical data-driven methods on training data \cite{arzani2021uncovering,chen2021physics,ramabathiran2021spinn,hanrahan2023studying,xu2023practical}. In terms of numerical computations, PINN also exhibits excellent capabilities in forward problem solving and field inversion. Moreover, the meshless nature of PINN is particularly suitable for wave equations, which facilitates its application to various initial and boundary conditions \cite{rasht2022physics}. PINN has seamless encoding and formulation capabilities under various constraint conditions, is meshless, and simple to implement. These features make it highly effective in solving laminar flows with strong pressure gradients and achieving high accuracy in turbulence modeling \cite{eivazi2022physics}. In terms of fluid experiments, H. Wang et al.\cite{wang2022dense} accurately reconstructed high-resolution velocity fields in sparse particle image velocimetry (PIV) experimental data using PINN constrained by Navier--Stokes (NS) equations. The impacts of the activation function, optimization algorithm, and hyperparameters were systematically evaluated through two-dimensional (2D) Taylor decaying vortices and a 2D turbulent channel flow to explore the ability of PINN to reconstruct wall-bounded turbulence. 

Based on the excellent capability of PINN in solving physical equations, numerous improved network architectures based on PINN have been successively proposed. Inspired by the application of neural networks in computer vision, Levi D et al.\cite{MCCLENNY2023111722} designed self-adaptive weights for PINN, enabling it to autonomously identify and focus on difficult fitting regions during the training process. Through comparative validation, their approach demonstrated superior performance over traditional PINN. S. Lin et al.\cite{LIN2022111053} appended a completely new neural network after the original PINN, incorporating measurements of conservative quantities into the mean squared error loss. This approach demonstrated higher accuracy and generalization ability in standard fluid dynamics cases. Yuan L et al.\cite{yuan2022pinn} proposed auxiliary physics-informed neural networks (A-PINNs), where auxiliary output variables are introduced to represent the numerical integrals in the governing equations and can be automatically differentiated instead of integral operators to avoid the limitation of using integral discretization. Using this neural network to solve nonlinear integral differential equations yields better accuracy than PINN. Inspired by the finite element method, Rezaei S et al.\cite{rezaei2022mixed} sought to improve the performance of the existing PINNs by advocating the use of spatial gradients of the principal variables as the output of the separable neural network. By properly designing the network architecture in PINN, deep learning models can solve unknown problems in heterogeneous domains without initial data from other sources. Recent work by S. Zhao et al.\cite{ZHAO2025114125} directly encodes the Large-Eddy Simulation (LES) equations into neural operators for three-dimensional incompressible turbulent flow simulation, with the resulting LESnets model demonstrating accuracy comparable to conventional methods while exhibiting strong generalization capabilities to unseen flow states. The physics-informed neural-network-based topology optimization (PINNTO), combines PINN with topology optimization, and solves several inherent defects due to the addition of PINN \cite{jeong2023physics}. The accuracy of PINN also highly depends on the selection and distribution of training sample points. Y Liu et al.\cite{liu2024adaptive} proposed a novel adaptive sampling algorithm called expected-improvement residual-based adaptive refinement (EI-RAR), which combines an attention mechanism with sample point generation using a new expected-improvement function. This approach addresses the limitation of traditional adaptive sampling algorithms, which typically focus only on sample points in the solution domain. Furthermore, the expected-improvement gradient-adaptive sampling algorithm, which integrates residual gradient information, was proposed to enhance the nonlinear fitting accuracy of PINN in discontinuous solution regions. W Zhou et al.\cite{10.1063/5.0180770} proposed an enhanced PINN framework that synergistically integrates residual-based adaptive sampling, adaptive loss weighting, and differential evolution optimization algorithms, demonstrating its effectiveness through several benchmark test cases. J. Bai et al.\cite{bai2023physics} extended the PINN algorithm to the solid mechanics and proposed the LSWR loss function based on the least square weighted residual (LSWR) method. Comparisons with energy-based and collocation loss functions have demonstrated the potential of the LSWR loss function in accurately predicting stress and displacement fields. H. Gao et al.\cite{GAO2021110079} employed elliptic coordinate mapping to achieve coordinate transformation between irregular physical domains and regular reference domains, to leverage powerful classic CNN backbones for solving steady-state PDEs. The designed physics-informed geometry-adaptive convolutional neural networks (GhyGeoNet) has been demonstrated to have significant research value. Recently, S Wang et al.\cite{wang2025simulatingthreedimensionalturbulencephysicsinformed} successfully employed PINNs to simulate three-dimensional fully turbulent flows at high Reynolds numbers, marking a significant step toward more flexible and continuous computational fluid dynamics. L Yang et al.\cite{LIU2023112291} research on PINN transfer learning tasks indicates that in the future, PINN would be more easily applied to new physical cases, and the development of physical fields would to some extent, depend on its progress.

Although numerous neural network architectures have been developed to enhance the performance of PINN for different problems, their fundamental structure remains rooted in the original PINN framework. PINN encounters the issues of gradient vanishing during the training process and is highly sensitive to the selection of various hyperparameters. However, extensive parameters require manual tuning, and there is currently no established optimal parameter selection guideline for PINN \cite{faroughi2024physics}. Inspired by Kolmogorov--Arnold representation theorem, Kolmogorov--Arnold networks (KANs) have fundamentally transformed the traditional MLP architecture by fitting the weight coefficients of the learning activation function with learnable univariate function spline curves at the edges instead of using fixed activation functions at the nodes as in conventional multilayer perceptron (MLP) \cite{liu2024kan}. Thus, KANs require no additional nonlinear activation functions between network nodes. This design renders KANs more physically consistent and intuitively aligned with human cognition. In traditional neural network development, the nonlinear mapping between inputs and outputs is often an obscure mapping that lacks interpretability, and the use of nonlinear activation functions cannot be easily explained from a physical perspective. Thus, traditional neural networks function as "black boxes" and make their results inherently difficult to interpret and less convincing \cite{de2018greedy}. Since the learning process of the spline fitting of KANs is grounded in mathematical theorems, the network architecture of KANs inherently possesses physical significance.
From a machine learning perspective, incorporating physical information as constraints in neural networks is a significant development trend. This approach offers several advantages such as introducing physical constraints to the loss function to enhance the physical interpretability of neural networks \cite{willard2022integrating}. However, KANs may still require a substantial amount of training data to achieve ideal results since there are no physical information constraints. Inspired by the concepts of PINN, incorporating machine learning techniques with prior knowledge such as physical information is a critical research approach \cite{li2022machine}. Although KAN has inherent physical significance, we believe that it is essential to incorporate physical information constraints into KAN, which is similar to the enhancements in PINN compared with conventional deep neural networks, and a similar approach has also been employed by L Yang et al.\cite{YANG2021109913} in their study of partial differential equations with noisy data, where they combine physical information with Bayesian Neural Networks (BNN) to achieve improved predictive accuracy. In this study, we propose a novel Kolmogorov--Arnold network architecture based on Chebyshev polynomials, i.e., Chebyshev physics-informed Kolmogorov--Arnold network (ChebPIKAN). The proposed ChebPIKAN framework adopts the Kolmogorov--Arnold network as the baseline network architecture and incorporates essential physical information based on PINN. KAN can adaptively learn the nonlinear activation functions in comparison with the classical neural network and automatically prune the basis functions and weight coefficients to determine the optimal network architecture. The proposal of KAN fundamentally solves the problem of poor generalization caused by the global optimization of traditional neural networks, since the spline curves learned by KAN are essentially local, and the fitted splines learned by the network are independent \cite{liu2024kan}. In recent research findings, KAN has also been demonstrated its effectiveness in the field of physics \cite{KOENIG2024117397,WANG2025117518}. Specifically, the Kolmogorov–Arnold-Informed Neural Network (KINN), proposed by Y Wang et al.\cite{WANG2025117518}, provides a novel approach for solving both forward and inverse problems of PDEs. Tests on multiple PDE benchmarks demonstrate that KINN surpasses traditional MLP methods in accuracy and convergence speed, offering a more efficient and accurate machine learning framework for PDE solutions. The proposed ChebPIKAN framework uses KAN as its foundational architecture, inherits the advantages of the KAN structure, and achieves enhanced performance due to the incorporation of physical information.

The remainder of this work is organized as follows. In Section II, we will introduce the mathematical foundation of the KAN architecture and describe the physical information loss incorporated into each physical PDE. Section III compares the results of PDE predictions using KAN with the Chebyshev basis function and ChebPIKAN to highlight the advantages of integrating physical information into neural networks and the benefits of KAN architecture over traditional MLP architectures. ChebPIKAN, enhanced with physical information outperforms both KAN and PINN in solving the accuracy problem. Section IV discusses the relevant work of ChebPIKAN, its impact, and future development. Section V summarizes the conclusions.

\section{\label{sec:level1}Methodology}

In this section, we will describe our implementation of the development of Physics-informed Kolmogorov--Arnold networks with Chebyshev polynomials and their use to solve partial differential equations. First, we provide an overview of the KAN architecture. Then, we describe our use of Chebyshev polynomials as alternative basis functions. Finally, we introduce our implementation of Physics-informed KANs with Chebyshev polynomials by applying different physical loss functions.

\subsection{\label{sec:level2}Kolmogorov--Arnold Networks}

The design of Kolmogorov--Arnold networks (KANs) is inspired by Kolmogorov--Arnold representation theorem. This theorem posits that any multivariate continuous function $\bm{f}$ defined on a bounded domain can be expressed as a finite composition of continuous univariate functions. A smooth $\bm{f}: [0, 1]^n \to \mathbf{R} $ is represented as
\begin{equation}
	\bm{f} ( \bm{x} ) = \bm{f} ( x _ { 1 } , \cdots , x _ { n } ) = \sum _ { q = 1 } ^ { 2 n + 1 } \Phi _ { q } \{ \sum _ { p = 1 } ^ { n } \phi _ { q , p } ( x _ { p } ) \},
\end{equation}
where $\phi_{q, p}: [0, 1] \rightarrow \mathbf{R}$ and $\Phi_{q}: \mathbf{R} \rightarrow \mathbf{R}$. The aim is to identify suitable univariate functions $\phi_{q,p} $ and $ \Phi $. $\phi(x)$ represents the operations of the KAN layer's node, and the input of the KAN layer is $\bm{x}$, which leads to the mathematical formula for the $l-th$ KAN layer
\begin{equation}
	{\bm{x}_{l + 1}} = \underbrace {\left( {\begin{array}{*{20}{c}}
				{{\phi_{11}^{(l)}(\cdot)}}&{{\phi_{12}^{(l)}(\cdot)}}& \cdots &{{\phi_{1{N_l}}^{(l)}(\cdot)}}\\
				{{\phi_{21}^{(l)}(\cdot)}}&{{\phi_{22}^{(l)}(\cdot)}}& \cdots &{{\phi_{2{N_l}}^{(l)}(\cdot)}}\\
				\vdots & \vdots &{}& \vdots \\
				{{\phi_{{N_{l + 1}}1}^{(l)}(\cdot)}}&{{\phi_{{N_{l + 1}}2}^{(l)}(\cdot)}}& \cdots &{{\phi_{{N_{l + 1}}{N_l}}^{(l)}(\cdot)}}
		\end{array}} \right)}_{\Phi_l}{\bm{x}_l},
\end{equation}
where $\Phi_l$ is the operation matrix for the $l-th$ KAN layer, which contains learnable parameters.

\subsection{\label{sec:level2} Chebyshev Polynomial Basis Functions}

In KAN layer calculations, spline basis functions, activation functions for fitting, and their weight coefficients $\omega$ must be selected. We define the function as
\begin{equation}
	\phi ( x ) = \omega [ b ( x ) + { spline } ( x ) ],
\end{equation}
where $\omega$ is the learnable weight coefficient of neural network. $x$ is the input value of function $\phi$ instead of a vector that contains multiple arguments, as shown in Section II.A, $spline(x)$ is modeled as a linear combination of B-$splines$ functions, and the activation function is defined by
\begin{equation}
	b ( x ) = { silu } ( x ) = x / ( 1 + e ^ { - x } ).
\end{equation}

The goal of neural networks is to learn appropriate weight coefficients $\omega$ to approximate the target function. However, the B-$spline$ function is not an orthogonal basis function and has a linear correlation, which may cause multi-collinearity issues during the fitting process. Multi-collinearity can make the parameter estimation unstable, increase the variance of the fitting results, and make it difficult to explain the contributions of individual basis function. In practical applications, high-frequency oscillations (such as the Runge phenomenon) likely occur. To address this issue, we use Chebyshev polynomials, which provide orthogonality and enhance the interpolation performance.

Chebyshev polynomials consist of two polynomial sequences related to cosine and sine functions, denoted by $T_n(x)$ and $U_n(x)$. They can be defined in several equivalent manners. The $T_n(x)$ Chebyshev polynomials are derived by the recursive formula
\begin{equation}
	\begin{array} { l }{ T _ { 0 } ( x ) = 1 },  \ { T _ { 1 } ( x ) = x },  \ { T _ { n + 1 } ( x ) = 2 x T _ { n } ( x ) - T _ { n - 1 } ( x ) } \end{array},
	\label{eq:Chebyshev}
\end{equation}
with $x \in [-1, 1]$ and $T_n(x) \in [-1, 1]$. Since the initial input $x$ or the output after a single KAN layer can easily exceed the value range of the Chebyshev polynomials, we apply the tanh normalization function to the output of the KAN layer and the first input $x_0$ to limit it within the range of $[-1, 1]$. Therefore, the input of each KAN layer undergoes additional $\tanh$ normalization
\begin{equation}
	\begin{split}
		\phi _0( x_0 ) & = \omega \{b[\tanh(x_0)]\} + {cheb} [\tanh(x_0)],
	\end{split}
\end{equation}
\begin{equation}
	\begin{split}
		\phi _n( x_n ) & = \omega \{ b\{\tanh[\phi_{n-1}(x_{n-1})]\}\} \\
		 \,& + {cheb} \{\tanh[\phi_{n-1}(x_{n-1})]\},
	\end{split}
	\label{eq:Chebyshevn}
\end{equation}
where $n\geq1$ and $cheb(\cdot)$ are the Chebyshev polynomials in Eq. (\ref{eq:Chebyshev}).

\subsection{\label{sec:level2} Physics-informed KAN Network }

Inspired by the principle of physics-informed neural networks training by minimizing the loss function that enforces adherence to physical laws, we apply physical losses to ChebPIKAN. The loss function is defined as
\begin{equation}
	Loss = Los{s_{data}} + \lambda Los{s_{phy}},
\end{equation}
where $Los{s_{data}}$ and $Los{s_{phy}}$ are the data and physical information loss terms, respectively. $\lambda$ is the weight coefficient for the PDE residual term.

The physics-informed loss(i.e. $Los{s_{phy}}$) should be carefully designed to the specific equations being solved. To examine the performance of ChebPIKAN to solve PDEs in fluid mechanics, we take the two-dimensional incompressible Navier--Stokes equations as the solution objective and embed the corresponding physical loss function.

Navier--Stokes equations constitute a set of equations that govern the motion of viscous incompressible fluids and describe the conservation of mass and momentum. These equations occupy a pivotal position in fluid dynamics, since they are used to characterize the flow behavior of fluid substances such as liquids and air.

The general form of the Navier--Stokes equation is
\begin{equation}
	\frac{\partial \bm{u}}{\partial t}+\bm{u} \cdot \nabla \bm{u}=-\frac{1}{\rho} \nabla p+\nu \nabla^{2} \bm{u},
\end{equation}

\begin{equation}
	\nabla \cdot \bm{u}=0.
\end{equation}

Here, $\rho$ is the density. $t$ is time, and $\bm{u}$ is the velocity vector. In the three-dimensional case. $\bm{u}$ can be expressed as $\bm{u}=u \bm{i}+v \bm{j}+w \bm{k}$. $\nabla$ denotes the gradient operator. $p$ is the pressure. For steady-state flows, the temporal derivative term vanishes.

To comprehensively validate the performance of our proposed PIAKN for PDEs in fluid dynamics. Apart from the classical Navier-Stokes problems of cylinder wake flow and lid-driven cavity flow, we also comprehensively considered the one-dimensional Allen--Cahn equation, Burgers equation, and the two-dimensional Helmholtz equation and Kovasznay flow.

The general form of Allen--Cahn equation is
\begin{equation}
	\frac{\partial u}{\partial t}=d \frac{\partial^{2} u}{\partial x^{2}}+f(u),
	\label{eq:AC}
\end{equation}
where function $u$ represents the material properties. $x$ is the spatial coordinate. $f(u)$ is a function of $u$, and $d$ is the diffusion coefficient.

The Burgers equation can be expressed in its general form as follows,
\begin{equation}
	\frac{\partial u}{\partial t}+u \frac{\partial u}{\partial x}=\nu \frac{\partial^{2} u}{\partial x^{2}},
	\label{eq:BG}
\end{equation}
where $u$ is the fluid velocity, $\nu$ is the viscosity coefficient, and $x$ is the spatial coordinate.

Two-dimensional Helmholtz equation is expressed as
\begin{equation}
	\nabla^{2} \varphi+k^{2} \varphi=0,
	\label{eq:HM}
\end{equation}
where $ \varphi$ is an arbitrary scalar field, and $k$ is the wavenumber. 

Kovasznay flow governing equations can be expressed as follows,
\begin{equation}
	u \frac{\partial u}{\partial x}+v \frac{\partial u}{\partial y}=-\frac{\partial p}{\partial x}+\frac{1}{Re}\left(\frac{\partial^{2} u}{\partial x^{2}}+\frac{\partial^{2} u}{\partial y^{2}}\right),
	\label{HMequation1}
\end{equation}
\begin{equation}
	u \frac{\partial v}{\partial x}+v \frac{\partial v}{\partial y}=-\frac{\partial p}{\partial y}+\frac{1}{Re}\left(\frac{\partial^{2} v}{\partial x^{2}}+\frac{\partial^{2} v}{\partial y^{2}}\right),
	\label{HMequation2}
\end{equation}

These equations are a simplified form of the Navier--Stokes equation in fluid dynamics and describe the conservation of momentum in fluid motion. Here, $Re$ is the Reynolds number, which characterizes the relative contribution of the inertial and viscous forces in fluid dynamics.

We introduce the respective physical loss functions based on each physical differential equations and the KAN using Chebyshev polynomials as basis functions. Then, the proposed the ChebPIKAN framework is established.

For each physical equation, we design a varying number of hidden layers from shallow to deep to compare and select the optimal number of hidden layers. The Adam optimizer is used for the parameter adjustment. Specific parameters are presented in Table~\ref{tab:parameters}, where ILR is the initial learning rate, and $Data_{train}$ and $Data_{test}$ are the number of samples that have been used for training and testing, respectively. It should be noted that, to ensure fairness in neural network performance comparisons and to eliminate potential biases introduced by stochastic factors during training, we deliberately disabled the automatic pruning function. This guarantees that each neural network maintains a predictable number of trainable parameters throughout the experiments.
\begin{table}[!ht]
	\centering
	\caption{training parameters.}
	\label{tab:parameters}
	\resizebox{\linewidth}{!}{
	\begin{tabular}{cccccccc}
		\hline
		Types of PDEs & $Data_{train}$ & $Data_{test}$ & Number of Layers & Optimizer & ILR & Epoch & $\lambda$ \\ \hline
		Allen--Cahn equation & 225 & 1000 & [2] + [5] $\times$ (1$\sim$4) + [1] & Adam & $10^{-3}$ & $10^4$ & 0.1 \\
		Burgers equation & 225 & 1000 & [2] + [5] $\times$ (1$\sim$4) + [1] & Adam & $10^{-3}$ & $10^4$ & 0.1 \\
		Helmholtz equation & 225 & 1000 & [2] + [5] $\times$ (1$\sim$4) + [1] & Adam & $10^{-3}$ & $10^4$ & 0.1 \\
		Kovasznay flow & 225 & 1500 & [2] + [5] $\times$ (1$\sim$4) + [3] & Adam & $10^{-3}$ & $10^4$ & 0.1 \\
		cylinder wake flow & 2000 & 5000 & [3] + [5] $\times$ (1$\sim$4) + [3] & Adam & $10^{-3}$ & $10^4$ & 0.1 \\
		cavity flow & 625 & 1200 & [2] + [20/30/50] $\times$ (1$\sim$4) + [3] & Adam & $10^{-3}$ & $2 \times 10^5$ & 0.1 \\ \hline
	\end{tabular}
}
\end{table}

\begin{figure}[!htbp] 
	\centering
	\includegraphics[height=10cm]{1.pdf}
	
	\caption{A diagram of ChebPIKAN.}
	\label{architecture}
\end{figure}

The two-dimensional input flow conditions are randomly sampled from a database of ground truth at a resolution of 15$\times$15 to form the training dataset, for more complex 2D cases like cavity flow, 25$\times$25 ground truth data is utilized as the training dataset. While such low-resolution data theoretically cannot fully describe the characteristics of the flow field, it more closely aligns with the sparsely discrete data obtained from real experiments through sensors. This presents a significant challenge to the inferential capabilities of the neural network, it forces the neural network to develop robust feature extraction capabilities under information-deficient conditions, better preparing it for practical deployment scenarios. For three-dimensional flow conditions, the amount of real data used for network training is appropriately increased.

The experiments were conducted using computing resources of clusters, specifically using an NVIDIA GeForce RTX 4090 GPU with 24 GB of VRAM.

\section{\label{sec:level1} Physical Models and Training Process of ChebPIKAN Models}

In this section, we present six physical models in our study: the one-dimensional Allen--Cahn and Burgers equations and the two-dimensional Helmholtz equation, Kovasznay flow, unsteady (time-dependent) Navier--Stokes equations, and steady (time-invariant) Navier--Stokes equations. We used Kolmogorov--Arnold Networks (KANs) with Chebyshev basis functions, i.e., our ChebPIKAN approach, to solve these equations and demonstrate the benefits of incorporating physical information into the neural network.

For each equation, we explored six different network architectures for both KAN and ChebPIKAN to assess how the number of hidden layers affects the performance. Throughout this study, we maintained consistent hyperparameters and training strategies for both networks to ensure a fair comparison. Each neural network maintains an identical architecture, with the same number of neurons per hidden layer and polynomials of fixed 10th order.

To evaluate the performance, we used the relative error in the residual contour plot,
\begin{equation}
	\mathbf{u}_{\text{res}} = \left( \frac{\sqrt{(\mathbf{u}_{\text{pred}} - \mathbf{u}_{\text{true}})^2}}{\sqrt{||\mathbf{u}_{\text{true}}||}} \right) \times 100,
	\label{eq:err}
\end{equation}
where $\mathbf{u}_{\text{res}}$, $\mathbf{u}_{\text{true}}$, and $\mathbf{u}_{\text{pred}}$ denote the residual, ground truth, and prediction of solution variables, respectively; $||\cdot||$ is the overall Euclidean-norm value.

\subsection{One-dimensional Allen--Cahn Equation}

Allen--Cahn (AC) equation is a pivotal partial differential equation and is widely applied in materials science, physical chemistry, and geometry. It effectively models the evolution of interfaces during phase transitions with a particular focus on single-phase transitions. This study uses the AC equation to assess the network's capability in handling one-dimensional nonlinear problems with the reference solution derived from CFD simulations. 

Our investigation focuses on a specific scenario of Eq. (\ref{eq:AC}) characterized by
\begin{equation}
	\frac{\partial u}{\partial t}=d \frac{\partial^{2} u}{\partial x^{2}}+5\left(u-u^{3}\right), \quad x \in[-1,1], \quad t \in[0,1],
\end{equation}
where $d=1$, and the initial and boundary conditions are defined as
\begin{equation}
	u(x, 0)=x^{2} \cos (\pi x), \quad u(-1, t)=u(1, t)=-1.
\end{equation}

To align our model with the underlying physical process, we developed corresponding loss terms based on the parameters of physical models. By minimizing the discrepancies in the equation, we can ensure consistency with the physical model. The physical loss term is defined as
\begin{equation}
	Los{s_{PDE}}=d \frac{\partial^{2} u}{\partial x^{2}}+5\left(u-u^{3}\right)-\frac{\partial u}{\partial t}.
\end{equation}

For this study, we consider a computational domain of $[-1, 1]$ and a time interval of $[0, 1]$. In this domain, 225 ground truth data are randomly sampled for the training set. For ChebPIKAN, 800 PDE points are randomly selected to calculate the physical loss. Adam optimizer is used due to its efficiency; $10 ^ 4$ epoch iterations, an initial learning rate of 0.01, and a correction coefficient of 0.95 were used. The learning rate is adjusted downward every 10 steps once the loss curve stabilized.

It is noteworthy that, for the sake of descriptive convenience during the analysis, different numbers of hidden layers are used to represent distinct neural networks. However, the number of hidden layers, as a parameter, does not fundamentally affect the performance of neural networks. In essence, the underlying distinction lies in the variation of the number of trainable parameters. Within the same network architecture, the fitting capacity of the network is positively correlated with the number of trainable parameters. Theoretically, given sufficient training, a greater number of trainable parameters enables the neural network to model more complex problems. In addition, the conventional fully-connected artificial neural networks (ANN) and PINN also serve as comparative neural network models and are trained accordingly. Notably, both PINN and ChebPIKAN implementations adopt identical physical information loss terms and consistently maintain the same weighting coefficients $\lambda$.
\begin{figure}[!htbp]
	\centering
	\includegraphics[height=4cm]{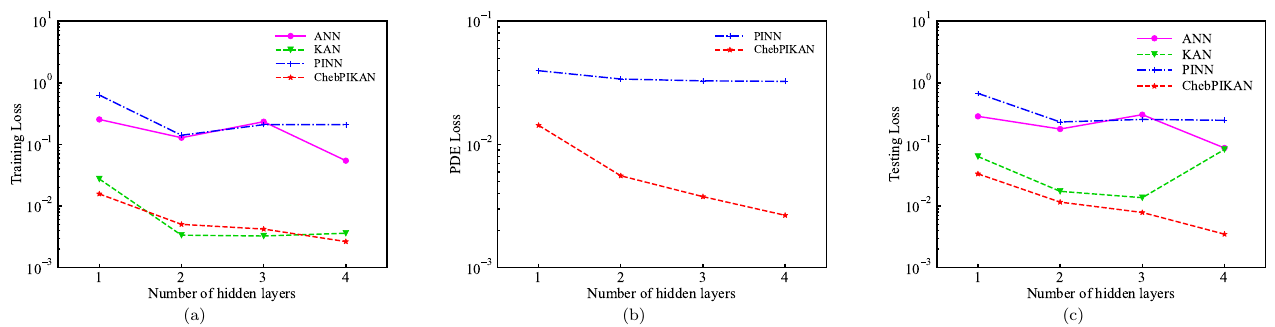}
	\caption{Optimal loss of different neural networks with different hidden layers for the Allen--Cahn equation. (a) Training loss, (b) PDE loss, (c) Testing loss. }
	\label{ACbestloss}
\end{figure}

Figure \ref{ACbestloss} reveals that all KANs exhibit overfitting, which is particularly pronounced in those with three hidden layers. This finding suggests that KANs have suboptimal extrapolation performance. By contrast, ChebPIKANs, which incorporate physical information, effectively mitigate overfitting and notably reduce the testing loss associated with four hidden layers. By integrating physical information, ChebPIKANs enhance the networks' ability to capture the essential characteristics of the model and consequently improve the extrapolation performance. In addition, ChebPIKANs consistently show lower testing loss values than KANs, which aligns with our expectation that the inclusion of physical information strengthens the network performance. We also observe that the training loss and PDE loss remain stable, which confirms that this integration helps prevent overfitting. Ultimately, the most effective ChebPIKAN configuration, which features four hidden layers, achieves loss convergence at approximately $10 ^ {-3} $. Notably, when implementing fully-connected architectures with comparable parameter counts, both conventional ANNs and baseline PINNs exhibit significantly larger training loss values than the two proposed architectures. For PINNs, this training loss metric specifically quantifies only the data-matching component, explicitly excluding any physical constraint terms. The observed performance gap demonstrates these traditional architectures' intrinsic representational limitations when constrained to simple topological designs.
\begin{figure}[!htbp]
	\centering
	\includegraphics[height=10cm]{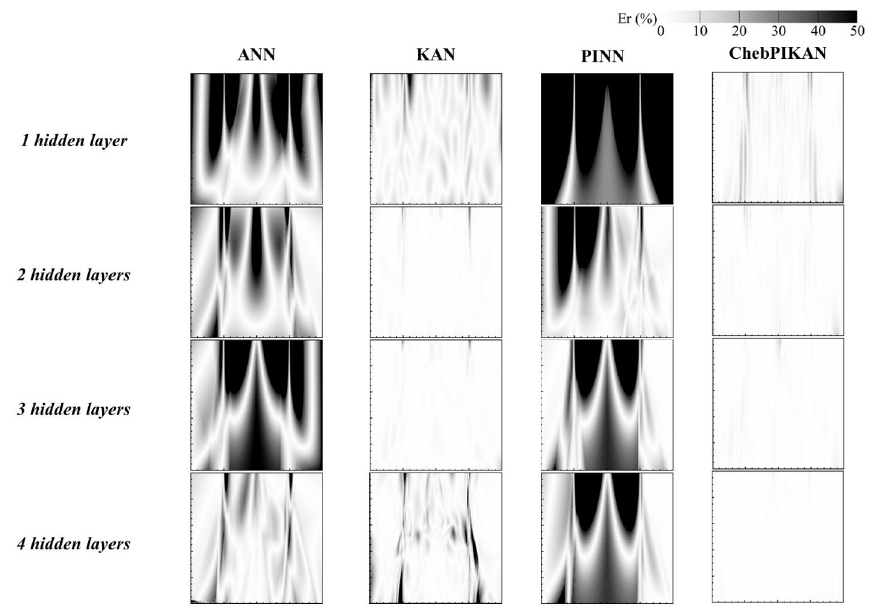}
	\caption{Relative error contours of different neural networks with different hidden layers for the Allen--Cahn equation.}
	\label{ACcontour}
\end{figure}

Figure \ref{ACcontour} presents the error contours demonstrating that ChebPIKANs incorporate physical information to achieve more accurate predictions than KANs, as evidenced by the significantly reduced error distribution and magnitude. While KANs exhibit deteriorating prediction performance with increasing hidden layers due to heightened parameter complexity, ChebPIKANs maintain stability and show steady accuracy improvement, reaching optimal performance with four hidden layers. In contrast, both ANNs and PINNs struggle to extract comprehensive flow field information from real data. Notably, the single-hidden-layer PINN fails to discern latent patterns in sparse data and becomes entirely guided by physics-informed constraints. Although a zero-velocity field technically satisfies the investigated AC equation, this trivial universal solution remains physically meaningless. 
\begin{figure}[!htbp]
	\centering
	\includegraphics[height=5cm]{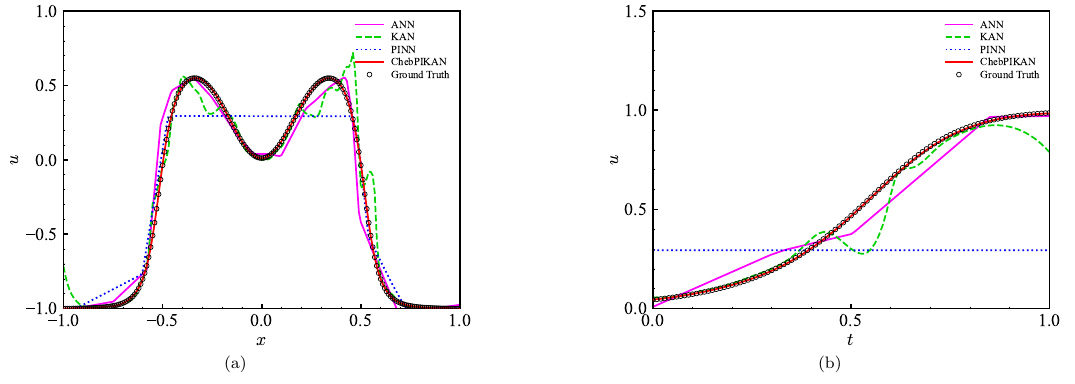}
	\caption{Velocity profiles predicted by different neural networks with four hidden layers for the Allen--Cahn equation. (a) $t$ = 0.5, (b) $x$ = 0.25.}
	\label{AC_line}
\end{figure}

Figure \ref{AC_line} further demonstrates that even with a relatively simple network architecture, the predictions of ChebPIKANs remain closely aligned with the ground truth. Both ANNs and KANs exhibit similar trends, whereas PINNs fail to accurately capture the underlying information from sparse real data and physical equations due to their limited fitting capacity. Consequently, PINNs converge to a particular general solution that merely minimizes the training error.

For clarity, ``$\textup{N}_{\textup{h}}$'' in subsequent figures and tables denotes the number of hidden layers.
\begin{table}[!htbp]
	\centering
	\caption{Relative errors predicted by different neural networks with different hidden layers for the Allen--Cahn equation.}
	\setlength{\tabcolsep}{10pt}
	\begin{tabular}{ccccc}
		\hline
		$\textup{N}_{\textup{h}}$ & ANN & KAN & PINN & ChebPIKAN \\ \hline
		1 & 31.75\% & 6.15\% & 84.74\% & 2.69\% \\
		2 & 17.66\% & 1.01\% & 20.46\% & 0.94\% \\
		3 & 34.31\% & 0.93\% & 23.61\% & 0.73\% \\
		4 & 8.55\% & 6.14\% & 25.41\% & 0.34\% \\ \hline
	\end{tabular}
	\label{tab:ACbestres}
\end{table}
\begin{figure}[!htbp] 
	\centering
	\includegraphics[height=5cm]{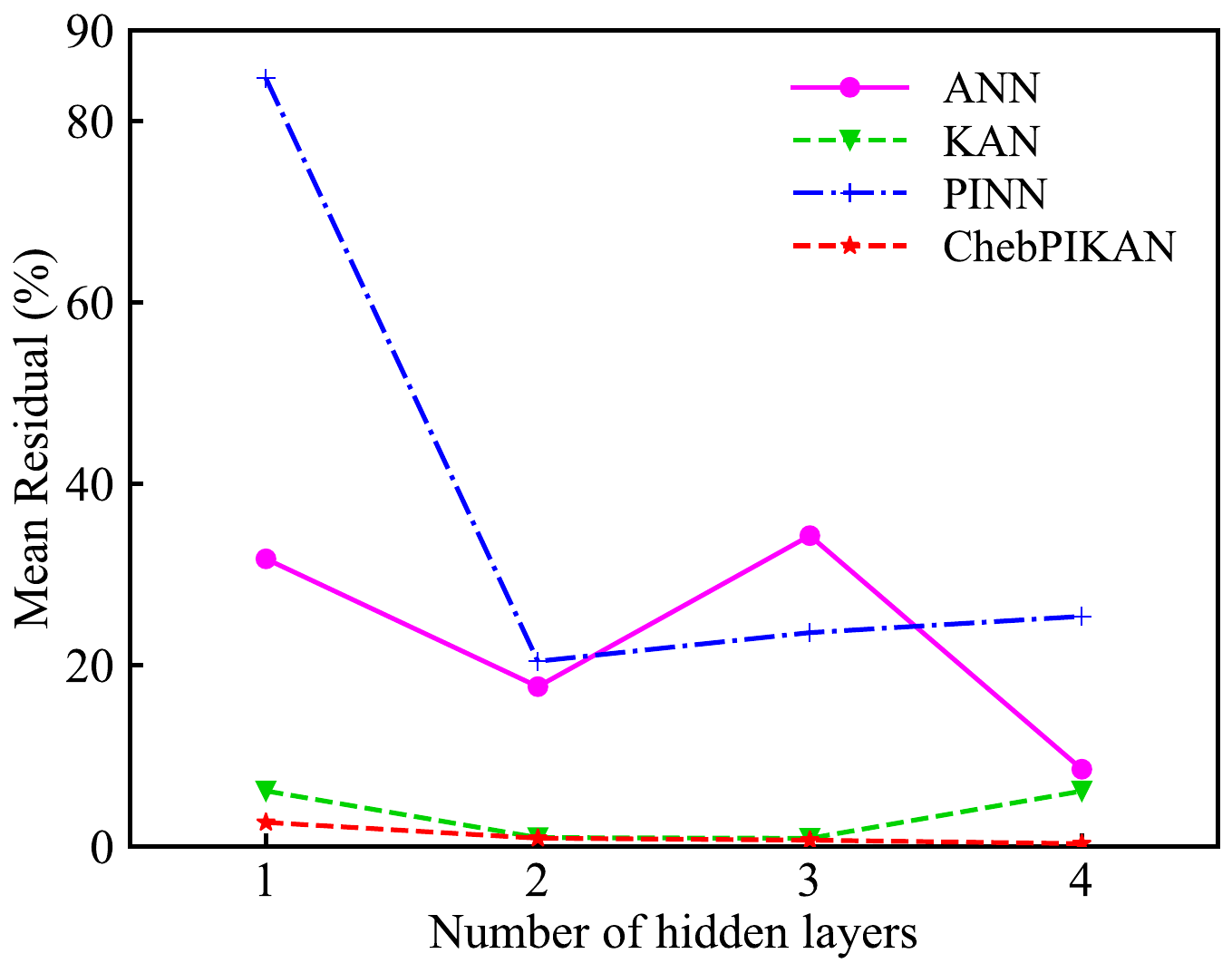}
	\caption{Mean residuals of different neural networks with different hidden layers for the Allen--Cahn equation.}
	\label{ACbestres}
\end{figure}

Table \ref{tab:ACbestres} summarizes the average residuals for KANs, ChebPIKANs, ANNs and PINNs across different hidden layer configurations. ChebPIKANs consistently demonstrate lower mean residuals than KANs, in particular, with four hidden layers, KAN shows increased residuals, but ChebPIKAN does not. As comparative models, both ANNs and PINNs show suboptimal overall performance. However, in terms of the number of trainable parameters, the neural network enhanced by Chebyshev polynomials, as shown in Eq. (\ref{eq:Chebyshevn}), has only one trainable parameter $\omega$ between every two neuron nodes in adjacent network layers. Despite this, it learns significantly more effective information, indicating that the new algorithm effectively improves the information storage capacity per parameter. This is of great significance for large-scale models, as it can enhance information storage efficiency or reduce the number of network parameters, thereby facilitating the transfer of neural networks. In the study of the Allen--Cahn equation, sparse real data are used. Both ANNs and PINNs are entirely unable to effectively predict the flow field with a small-parameter network structure. In particular, the PINNs perform even worse due to the insufficient descriptive capacity of the small-parameter network, failing to effectively learn the physical equations.

We explored the effects of various neural network structures on the network performance by configuring four different network architectures, each of which contained 1-4 hidden layers. Standard ANNs and PINNs with relatively simple architectures exhibit limited applicability to the problem under investigation. Theoretically, deeper networks have stronger nonlinear representative capabilities. However, with four hidden layers, KAN faces challenges due to an increase in parameters, which lead to poor training performance. Conversely, ChebPIKANs exhibit a steady decrease in loss with more hidden layers, which indicate that the integration of physical information enhance the networks' learning capability. The improved overall nonlinear processing ability demonstrates the benefits of incorporating hidden layers into the architecture.

\subsection{One-dimensional Burgers Equation}

Burgers equation is a fundamental equation in soliton theory and represents a diffusion process that includes nonlinear terms. 

In our study, we define the spatial and temporal domains of Eq. (\ref{eq:BG}) as
\begin{equation}
	x \in[-1,1], \quad t \in[0,1],
\end{equation}
$\nu$ is set at $ {0.01}/{\pi} $, and the Dirichlet boundary conditions and initial conditions are specified as
\begin{equation}
	u(-1, t)=u(1, t)=0, \quad u(x, 0)=-\sin (\pi x).
\end{equation}

Due to the shock wave characteristics of the Burgers equation, the solution rapidly changes in specific regions. Neural networks can effectively address these high-gradient areas by adjusting their predictions accordingly. The presence of an accurate analytical solution for this boundary condition validates the performance of the neural network when solving the Burgers equation. In ChebPIKANs, the physical loss associated with Burgers equation is defined as
\begin{equation}
	Los{s_{PDE}}=\frac{0.01}{\pi} \frac{\partial^{2} u}{\partial x^{2}}-\frac{\partial u}{\partial t}-u \frac{\partial u}{\partial x}.
\end{equation}

For the neural network, the inputs of Burgers equation are position $x$ and time $t$, and the output is fluid velocity $u$. The domain and random sampling strategies for the Burgers equation align with those used for the Allen--Cahn equation, and we consistently apply the Adam optimization strategy.
\begin{figure}[!htbp] 
	\centering
	\includegraphics[height=4cm]{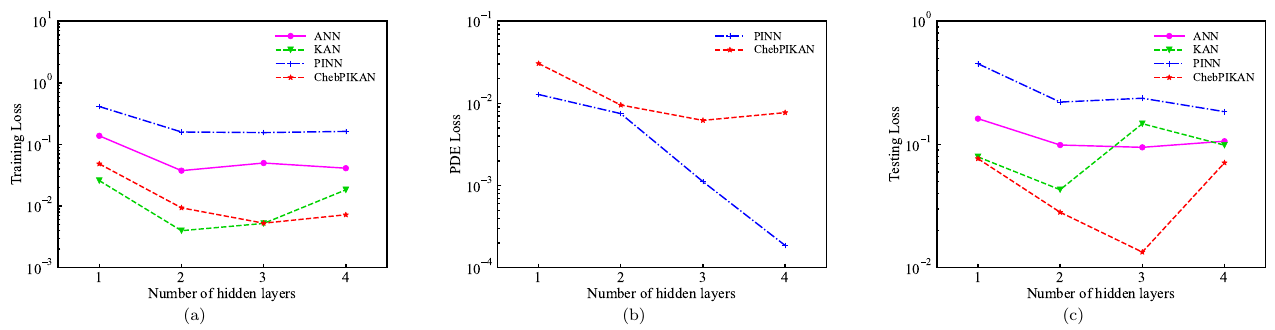}
	\caption{Optimal loss of different neural networks with different hidden layers for the Burgers equation. (a) Training loss, (b) PDE loss, (c) Testing loss.}
	\label{BGbestloss}
\end{figure}

The optimal loss values in Figure \ref{BGbestloss} reveal that the final loss is less than ideal, most network configurations achieve a loss of approximately $10 ^ {-2} $, close to $10 ^ {-1} $. KANs' overfitting is more pronounced in this context than in the Allen--Cahn equation. ChebPIKANs show resilience to overfitting, except with four hidden layers. Burgers equation presents substantial challenges due to its strong nonlinear phenomena and potential for shock wave generation at low viscosity coefficients. Consequently, when the network architecture deepens, the prediction accuracy significantly improves. ChebPIKANs show clear advantages with three hidden layers because this depth enhances its nonlinear processing capabilities; however, the structure with four hidden layers suffers from the complications of having more learning parameters. The training loss and PDE loss curves exhibit a high degree of alignment, which indicates that the inclusion of physical information mitigates overfitting in ChebPIKANs and consequently enhances the generalization. Thus, the network effectively learns the characteristics of the partial differential equations involved. Similar to studies on the Allen-Cahn equation, ANNs and PINNs with simple network architectures remain insufficient for addressing this class of problems.
\begin{figure}[!htbp]
	\centering
	\includegraphics[height=10cm]{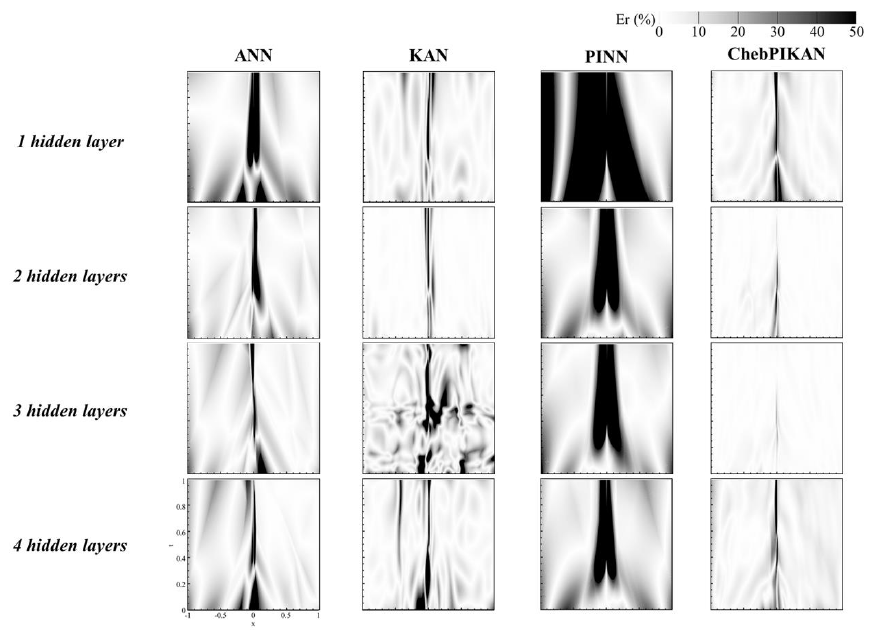}
	\caption{Relative error contours of different neural networks with different hidden layers for the Burgers equation.}
	\label{BGcontour}
\end{figure}

In Figure \ref{BGcontour}, we observe that error concentrations occur at discontinuities in the shear region. Although all networks face challenges in predicting strong shear regions, the error distributions for ChebPIKANs are significantly lower. In the optimal three-layer configuration, all predictions are nearly accurate, except for a small region of error in the strong shear area. Moreover, this study does not artificially enhance prediction accuracy by expanding the neural network architecture to improve fitting capability. Instead, the investigation maintains a comparable number of trainable parameters to ensure a fair comparison.
\begin{figure}[!htbp]
	\centering
	\includegraphics[height=5cm]{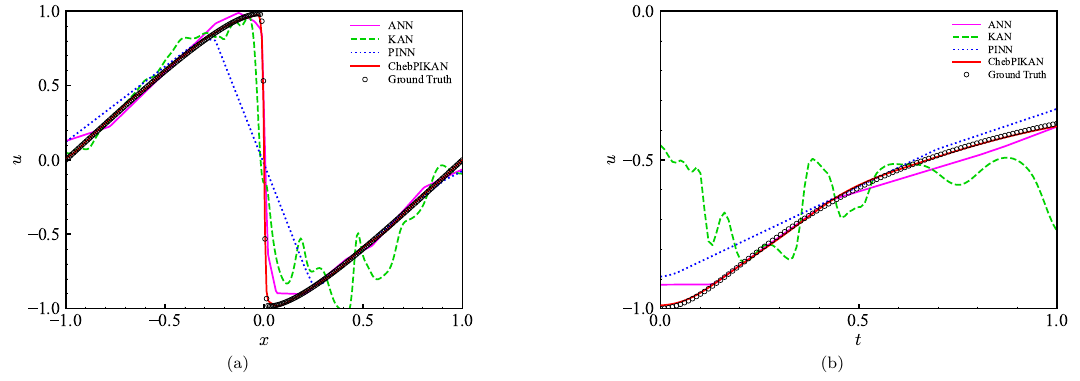}
	\caption{Velocity profiles predicted by different neural networks with three hidden layers under the Burgers equation. (a) $t$ = 0.5, (b) $x$ = 0.5.}
	\label{BG_line}
\end{figure}

The results presented in Figure \ref{BG_line} demonstrate that ChebPIKANs achieve significantly more accurate predictions than other neural networks when handling sharp gradient variations. In contrast, KANs exhibit oscillatory behavior in the learned flow field due to the inherent conflict between their strong fitting capacity and the information deficiency in sparse training data. The loss values further reveal that KANs converge to local optima with respect to the available training data. This observation reinforces that even neural networks with superior approximation capabilities struggle to reconstruct physical fields from limited data, underscoring the necessity of physics-informed guidance in such scenarios.
\begin{table}[!htbp]
	\centering
	\caption{Mean residuals of different neural networks with different hidden layers for the Burgers equation.}
	\setlength{\tabcolsep}{10pt}
	\begin{tabular}{ccccc}
		\hline
		$\textup{N}_{\textup{h}}$ & ANN & KAN & PINN & ChebPIKAN \\ \hline
		1 & 15.23\% & 6.36\% & 58.14\% & 6.50\% \\
		2 & 8.39\% & 2.86\% & 20.87\% & 1.76\% \\
		3 & 8.60\% & 13.67\% & 20.75\% & 1.13\\
		4 & 9.37\% & 7.72\% & 18.91\% & 6.45\% \\ \hline
	\end{tabular}
	\label{tab:BGbestres}
\end{table}
\begin{figure}[!htbp] 
	\centering
	\includegraphics[height=5cm]{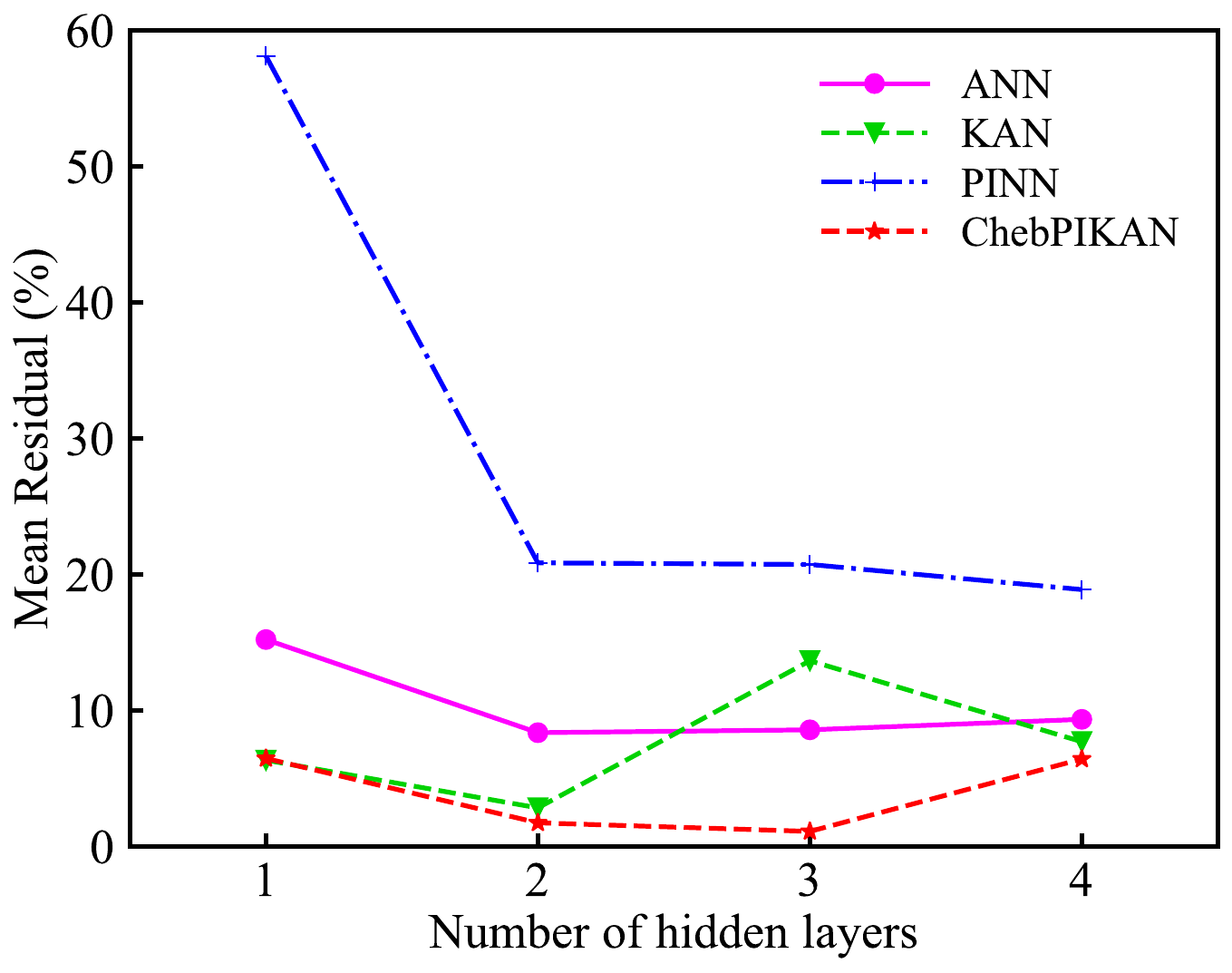}
	\caption{Mean residuals of different neural networks with different hidden layers for the Burgers equation.}
	\label{BGbestres}
\end{figure}

Table \ref{tab:BGbestres} presents the average residuals for four neural network types with various numbers of hidden layers. When there is one hidden layer, KAN and ChebPIKAN have comparable performances, which suggests similar predictive capabilities in solving Burgers equation. However, with more hidden layers, ChebPIKANs exhibit significantly improved performance, which demonstrates their strengths in high-level feature extraction. Regarding the predictive performance demonstrated by the ANNs and PINNs, similar to the findings in the study of the AC equation, the KANs exhibit unstable predictive performance when using three hidden layers. This is a common issue in scenarios with low-density real data, where even a small number of extreme outliers can lead to a completely incorrect fitting of the physical equations. Physics-informed guidance can effectively mitigate the impact of partial erroneous data, and thus, it will play an indispensable role in the future application of artificial intelligence methods in the field of physics.

\subsection{Two-dimensional Helmholtz Equation}

Helmholtz equation is a key partial differential equation in acoustic fields. It describes how a scalar or vector physical quantity changes in a constant external field to enable better understanding and resolution of wave phenomena.

In the study, we focused on the two-dimensional case of Eq. (\ref{eq:HM}) with a wave number defined as $k_{0}=2 \pi n$, where $n=2$. Then, the equation can be simplified to
\begin{equation}
	-u_{x x}-u_{y y}-k_{0}^{2} u=f, \quad \Omega=[0,1]^{2},
\end{equation}
where Dirichlet boundary conditions are specified as
\begin{equation}
	u(x, y)=0, \quad(x, y) \in \partial \Omega,
\end{equation}
and the source term is
\begin{equation}
	f(x, y)=k_{0}^{2} \sin \left(k_{0} x\right) \sin \left(k_{0} y\right).
\end{equation}

In this case, the exact solution of the equation is
\begin{equation}
	u(x, y)=\sin \left(k_{0} x\right) \sin \left(k_{0} y\right).
\end{equation}

In physical applications such as heat conduction and wave equations, $-u_{x x}$ and $u_{y y}$ are the changes in local curvature or temperature in $x$ and $y$ directions, respectively, which reflects the system's diffusion or propagation. Meanwhile, $k_{0}^{2} u$ typically affects the system's equilibrium state and acts as a restoring force or potential energy influence, whereas $f$ is an externally driven source term that represents the external influence or force applied in space and can be understood as the distribution of external stimuli or sources. 

We analyzed the spatial domain defined by $x \in[0, 1]$ and $y \in[0, 1]$ with network inputs at these positions and outputs at velocity $u$. The sampling points and optimization strategies align with the discussed methodology. 

The physical loss for the Helmholtz equation in ChebPIKANs are defined as
\begin{equation}
	Los{s_{PDE}}=k_{0}^{2} \sin \left(k_{0} x\right) \sin \left(k_{0} y\right)+u_{x x}+u_{y y}+k_{0}^{2} u.
\end{equation}
\begin{figure}[!htbp] 
	\centering
	\includegraphics[height=4cm]{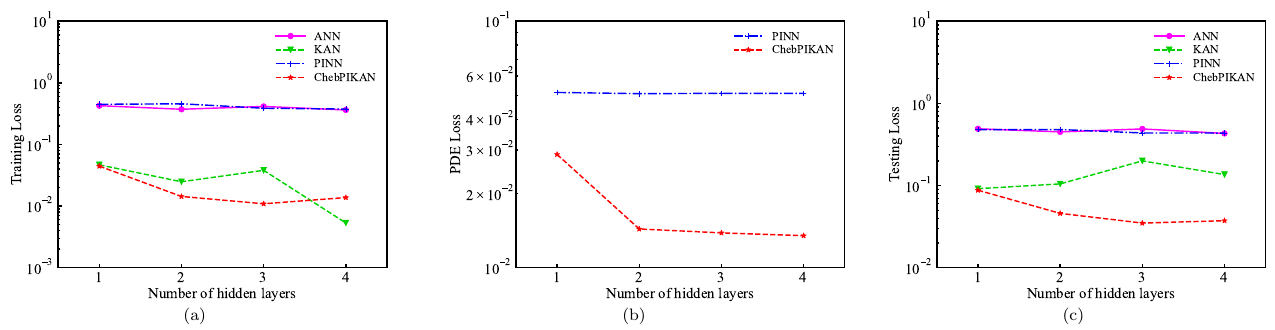}
	\caption{Optimal loss of different neural networks with different hidden layers for the Helmholtz equation. (a) Training loss, (b) PDE loss, (c) Testing loss.}
	\label{HMbestloss}
\end{figure}

Figure \ref{HMbestloss} illustrates the optimal loss values of neural networks, where both conventional ANNs and PINNs still demonstrate suboptimal performance. Notably, KANs exhibit more pronounced overfitting in two-dimensional cases, particularly with four hidden layers. Due to the increased complexity of two-dimensional equations compared to one-dimensional equations, both KAN and ChebPIKAN with four hidden layers improve in performance. The PDE loss and training loss of ChebPIKANs generally show strong consistency, which indicates that the network has effectively learned the equation.
\begin{figure}[!htbp]
	\centering
	\includegraphics[height=10cm]{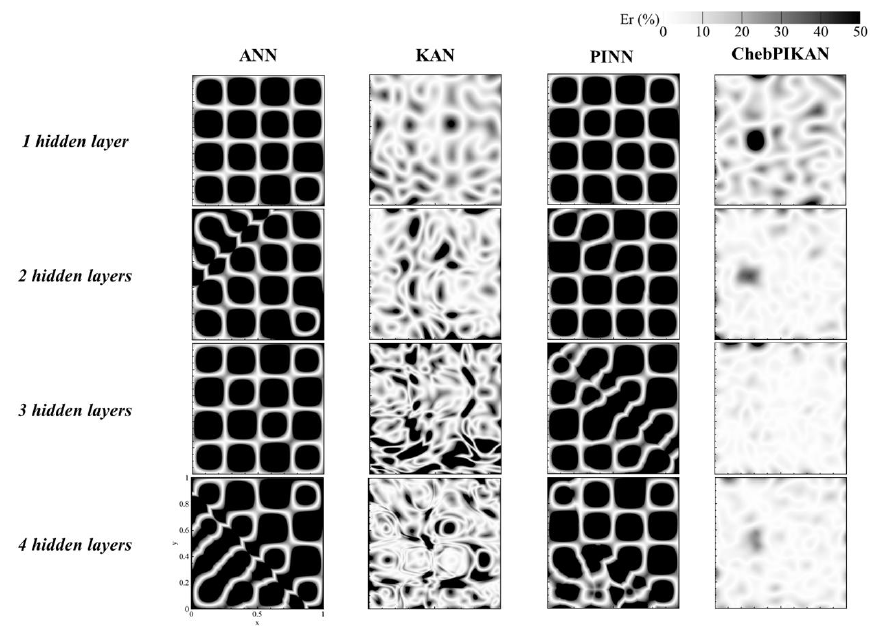}
	\caption{Relative error contours of different neural networks with different hidden layers for the Helmholtz equation.}
	\label{HMcontour}
\end{figure}

As shown in Figure \ref{HMcontour}, when the flow field exhibits complex periodic variations in data gradients, the limited discrete training data prove insufficient to fully characterize the underlying flow physics. In such cases, both ANNs and PINNs fail to produce accurate predictions due to their restricted trainable parameters and consequently weaker approximation capabilities, leading to entirely erroneous flow field reconstructions. By contrast, ChebPIKANs demonstrate superior overall performance compared to KANs. Furthermore, the incorporation of physical constraints in ChebPIKANs substantially improves predictive accuracy. However, persistent errors in certain regions suggest remaining difficulties in resolving non-stationary flow features, which could potentially be mitigated by increasing the sampling density.
\begin{figure}[!htbp]
	\centering
	\includegraphics[height=5cm]{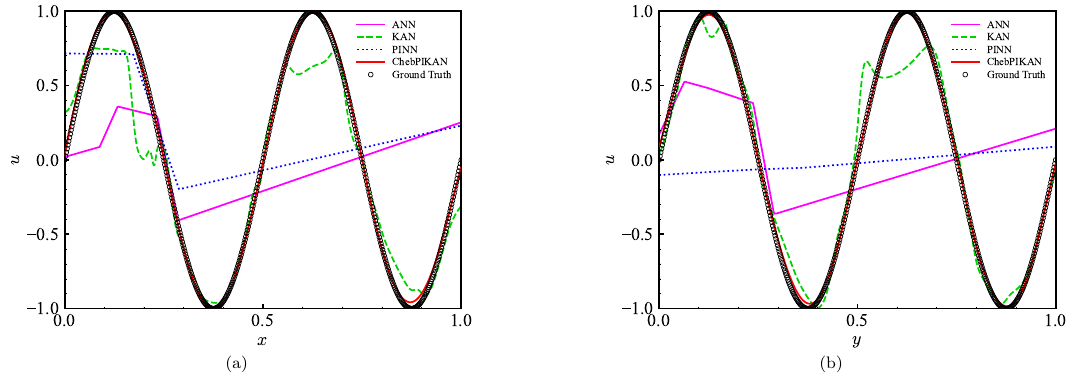}
	\caption{Velocity profiles predicted by different neural networks with four hidden layers for the Helmholtz equation. (a) $y$ = 0.75, (b) $x$ = 0.75.}
	\label{HM_line}
\end{figure}

As shown in Figure \ref{HM_line}, ChebPIKANs demonstrate close agreement with the ground truth. In contrast, KANs lack physics-informed guidance, resulting in limited predictive capability due to sparse training data, and thus fail to autonomously deduce the underlying physical laws governing the entire flow field. Furthermore, both ANNs and PINNs remain unable to achieve sufficient fitting accuracy with their limited trainable parameters.
\begin{table}[!htbp]
	\centering
	\caption{Mean residuals of different neural networks with different hidden layers for the Helmholtz equation.}
	\setlength{\tabcolsep}{10pt}
	\begin{tabular}{ccccc}
		\hline
		$\textup{N}_{\textup{h}}$ & ANN & KAN & PINN & ChebPIKAN \\ \hline
		1 & 80.06\% & 14.39\% & 79.12\% & 12.35\% \\
		2 & 74.80\% & 15.46\% & 79.34\% & 5.55\% \\
		3 & 80.25\% & 29.90\% & 71.81\% & 4.29\\
		4 & 73.45\% & 20.01\% & 69.43\% & 5.01\% \\ \hline
	\end{tabular}
\end{table}
\begin{figure}[!htbp] 
	\centering
	\includegraphics[height=5cm]{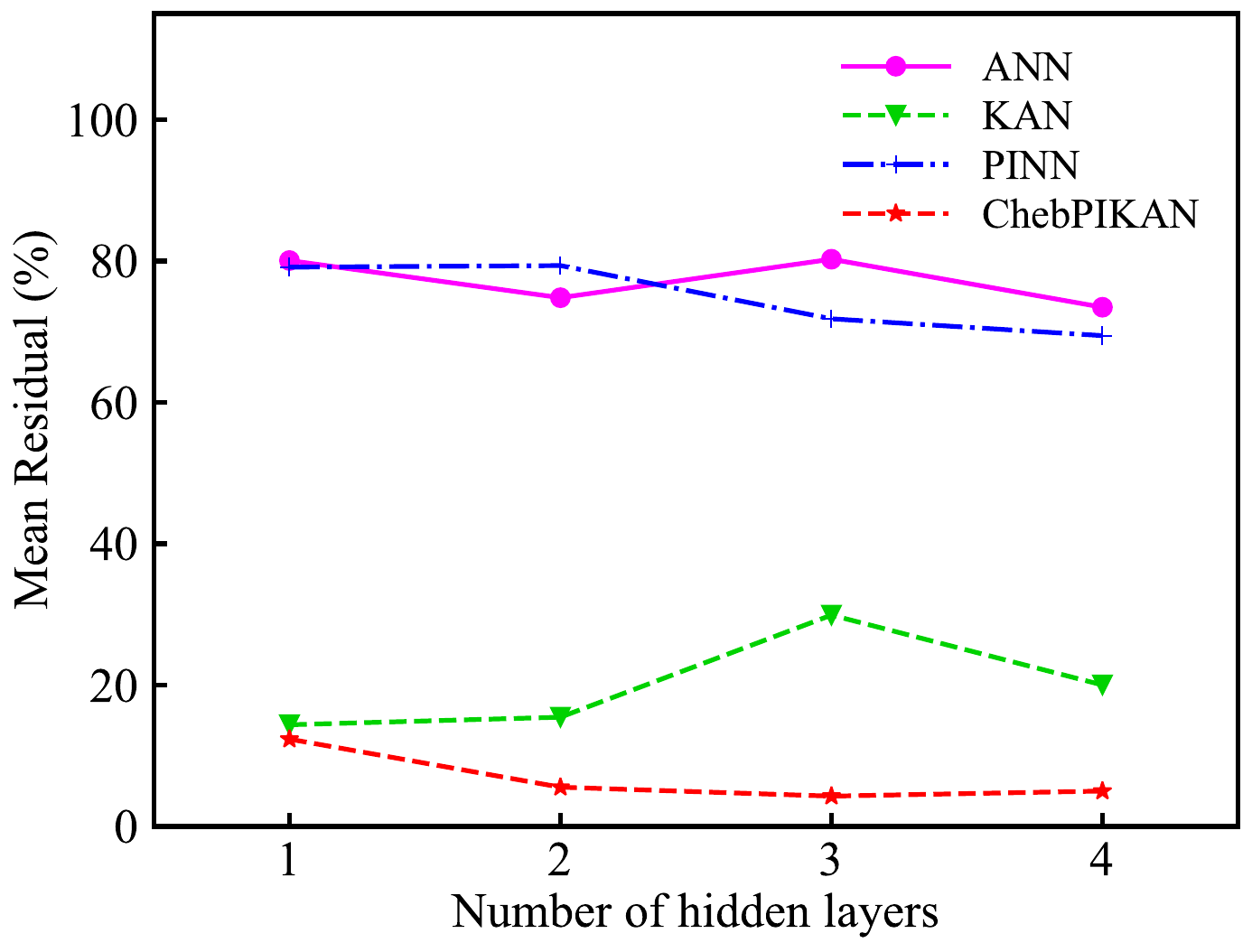}
	\caption{Mean residuals of different neural networks with different hidden layers for the Helmholtz equation.}
	\label{HMbestres}
\end{figure}

For the Helmholtz equation, although its exact solution consists merely of two coordinate-dependent sine functions, the use of low-resolution real data alone is insufficient to accurately capture its characteristics. This presents a certain level of challenge for intelligent prediction algorithms. Regarding the performance across various hidden layers, ChebPIKANs maintain stability and low error rates, and it particularly excels at the second and third layers. By contrast, KANs' performance significantly fluctuates, especially at higher layers, which indicates instability during training and a notable decline with three hidden layers. Both ANNs and PINNs, due to the fact that the functional mappings between network nodes are merely composite functions of linear equations and activation functions, exhibits significantly inferior feature representation capabilities compared to the new algorithm incorporating Chebyshev polynomials, and they are unable to accurately describe its characteristics regardless of whether guided by physical information or not.

\subsection{Two-dimensional Kovasznay Flow}

Kovasznay flow is an idealized model in fluid dynamics and notable for its two-dimensional, steady-state, incompressible, and vortical characteristics. It serves as a vital framework for analyzing the flow stability, transitions to turbulence, and flow control. This model reaches a steady state over time, where the velocity and pressure fields become constant. The governing equations are Eqs. (\ref{HMequation1}) and (\ref{HMequation2}) with Dirichlet boundary conditions,
\begin{equation}
	u(x, y)=0, \quad(x, y) \in [0,1]^{2}.
\end{equation}

The exact solutions for velocity components $u$ and $v$ and pressure $p$ are
\begin{equation}
	u=1-e^{\lambda x} \cos (2 \pi y),
\end{equation}
\begin{equation}
	v=\frac{\lambda}{2 \pi} e^{\lambda x} \sin (2 \pi y),
\end{equation}
\begin{equation}
	p=\frac{1}{2}\left(1-e^{2 \lambda x}\right),
\end{equation}
where $\lambda=\frac{Re}{2}-\sqrt{\frac{Re^{2}}{4}+4 \pi^{2}}$.

The Kovasznay flow is a specific form of the NS equation and thus inherently satisfies the constraints of the NS equation. In our analysis, we define the spatial domain as $x \in[0, 1]$ and $y \in[0, 1]$. The same random sampling and optimization strategies are used as previously mentioned. The neural network model for Kovasznay flow uses two inputs ($x$ and $y$) and three outputs: pressure $p$, velocity $u$, and velocity $v$. In the ChebPIKANs frameworks, the physical losses associated with Kovasznay flow are expressed as
\begin{equation}
	\begin{split}
		Los{s_{PDE}} = \,& u \frac{\partial u}{\partial x}+v \frac{\partial u}{\partial y}+\frac{\partial p}{\partial x}-\frac{1}{R e}\left(\frac{\partial^{2} u}{\partial x^{2}}+\frac{\partial^{2} u}{\partial y^{2}}\right) \\
		+ \,& u \frac{\partial v}{\partial x}+v \frac{\partial v}{\partial y}+\frac{\partial p}{\partial y}-\frac{1}{R e}\left(\frac{\partial^{2} v}{\partial x^{2}}+\frac{\partial^{2} v}{\partial y^{2}}\right) \\
		+ & \frac{\partial u}{\partial x} + \frac{\partial v}{\partial y}.
	\end{split}
\end{equation}

Considering the complexity of Kovasznay flow solutions, prediction errors are often more pronounced at the boundaries. To address this issue, boundary constraints are incorporated into the ChebPIKANs frameworks by selecting random boundary coordinates and applying the governing equations of Kovasznay flow. This integration of boundary constraints enhances the physical information used to compute the PDE loss, which directly improves the accuracy of boundary forecasts and influences the internal predictions.
\begin{figure}[!htbp] 
	\centering
	\includegraphics[height=4cm]{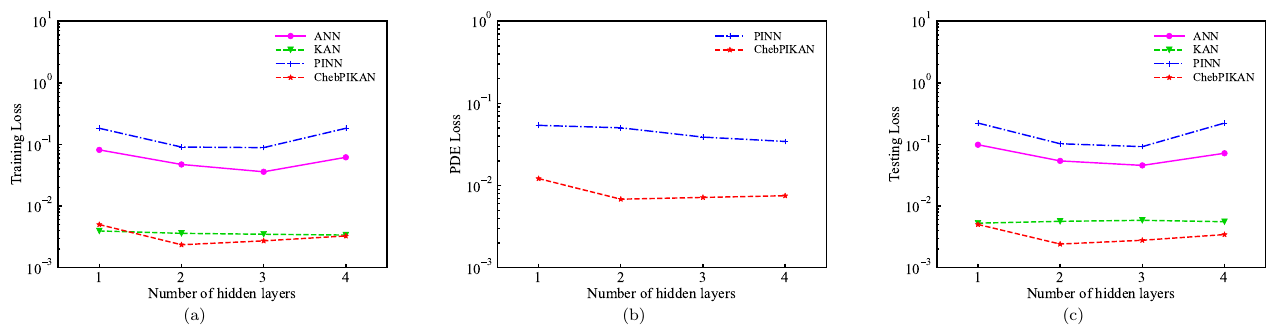}
	\caption{Optimal loss of different neural networks with different hidden layers for the Kovasznay flow. (a) Training loss, (b) PDE loss, (c) Testing loss.}
	\label{KOVbestloss}
\end{figure}

Figure \ref{KOVbestloss} shows the optimal loss values for Kovasznay flow and demonstrates that neither KANs nor ChebPIKANs exhibits severe overfitting. ChebPIKANs demonstrate superior testing loss values compared to KANs, while the performance of ANNs and PINNs is not on the same order of magnitude.
\begin{figure}[!htbp]
	\centering
	\includegraphics[height=10cm]{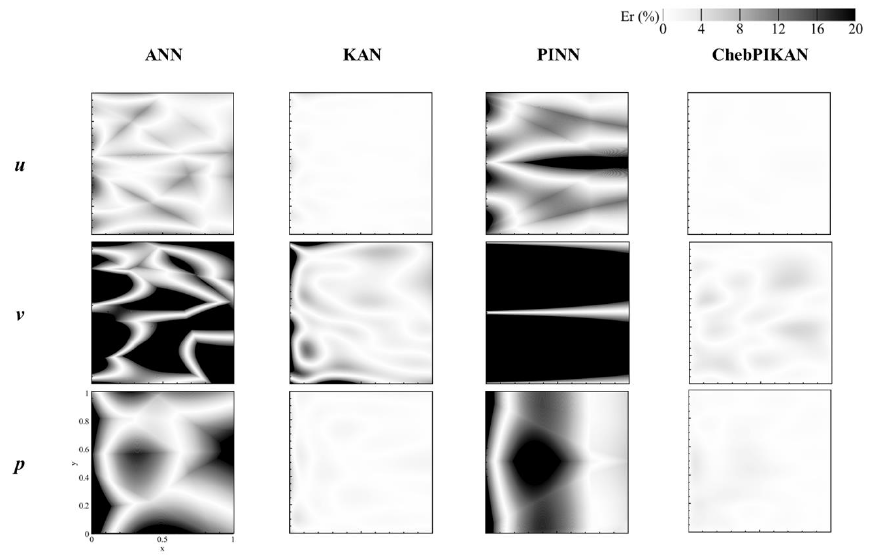}
	\caption{Relative error contours of different neural networks with three hidden layers for the Kovasznay flow.}
	\label{KOVcontour}
\end{figure}

For clarity in illustrating the performance of the four neural networks, Figure \ref{KOVcontour} presents a comparison of error contours exclusively for networks with three hidden layers. The results demonstrate that ANNs and PINNs with smaller architectures remain ineffective in handling physical field problems with sparse real-world data. In contrast, both KANs and ChebPIKANs exhibit strong predictive performance for the u-vector. ChebPIKANs, which are enhanced by additional physical information, exhibit virtually no prediction error. By contrast, KANs show a slight error at the boundary of the strong-shear region on the left. Since the magnitude of $v$ is smaller than that of $u$ and the solution of $v$ exhibits a more complex form, both KANs and ChebPIKANs show lower predictive accuracy for $v$. However, after incorporating boundary constraints, ChebPIKANs show a significant reduction in boundary errors, resulting in markedly lower overall errors than KANs. The predictive performance for $p$ is comparable between the two methods.
\begin{figure}[!htbp]
	\centering
	\includegraphics[height=5cm]{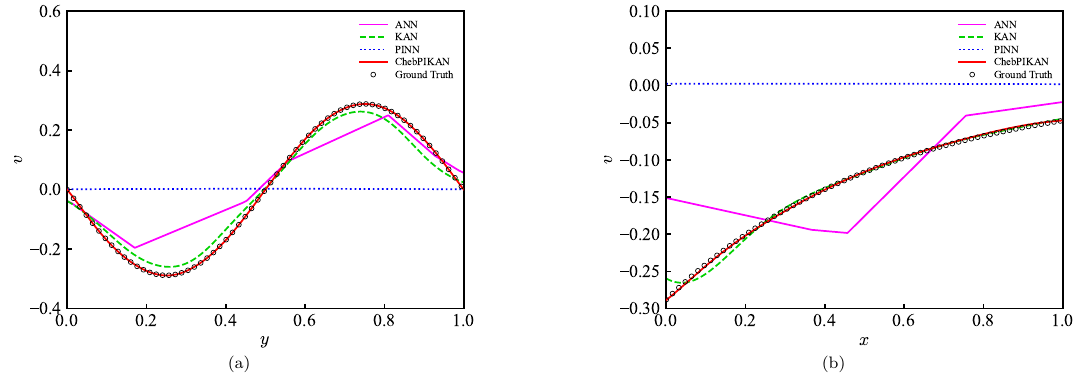}
	\caption{Velocity profiles predicted by different neural networks with three hidden layers for the Kovasznay flow. (a) $x$ = 0, (b) $y$ = 0.25.}
	\label{KOV_line}
\end{figure}

For comparative analysis, we examine the cross-sectional predictions of the $v$-vector component that demonstrates the largest errors. Figure \ref{KOV_line} reveals that both ANNs and PINNs produce results that significantly deviate from the reference values. ChebPIKANs show nearly perfect agreement with the ground truth. KANs exhibit good consistency with the actual solution, particularly at the flow inlet ($x$ = 0) where only minor absolute deviations occur. These KANs results exceed expectations considering the complete absence of physical constraints in the model. Nevertheless, the limited available data prove insufficient for the KANs to fully reconstruct the complete flow field characteristics.
\begin{table}[!htbp]
	\centering
	\caption{Mean residuals of KANs and ChebPIKANs with different hidden layers for the Kovasznay flow.}
	\setlength{\tabcolsep}{8pt}
	\begin{tabular}{ccccccc}
		\hline
		$\textup{N}_{\textup{h}}$ & KAN(u) & ChebPIKAN(u) & KAN(v) & ChebPIKAN(v) & KAN(p) & ChebPIKAN(p) \\ \hline
		1 & 0.33\% & 0.42\% & 4.05\% & 3.37\% & 0.64\% & 1.08\% \\
		2 & 0.40\% & 0.19\% & 3.36\% & 1.77\% & 0.72\% & 0.48\% \\
		3 & 0.35\% & 0.20\% & 3.74\% & 1.41\% & 0.65\% & 0.67\% \\
		4 & 0.32\% & 0.18\% & 3.34\% & 2.71\% & 0.80\% & 0.81\% \\ \hline
	\end{tabular}
\end{table}
\begin{figure}[!htbp] 
	\centering
	\includegraphics[height=5cm]{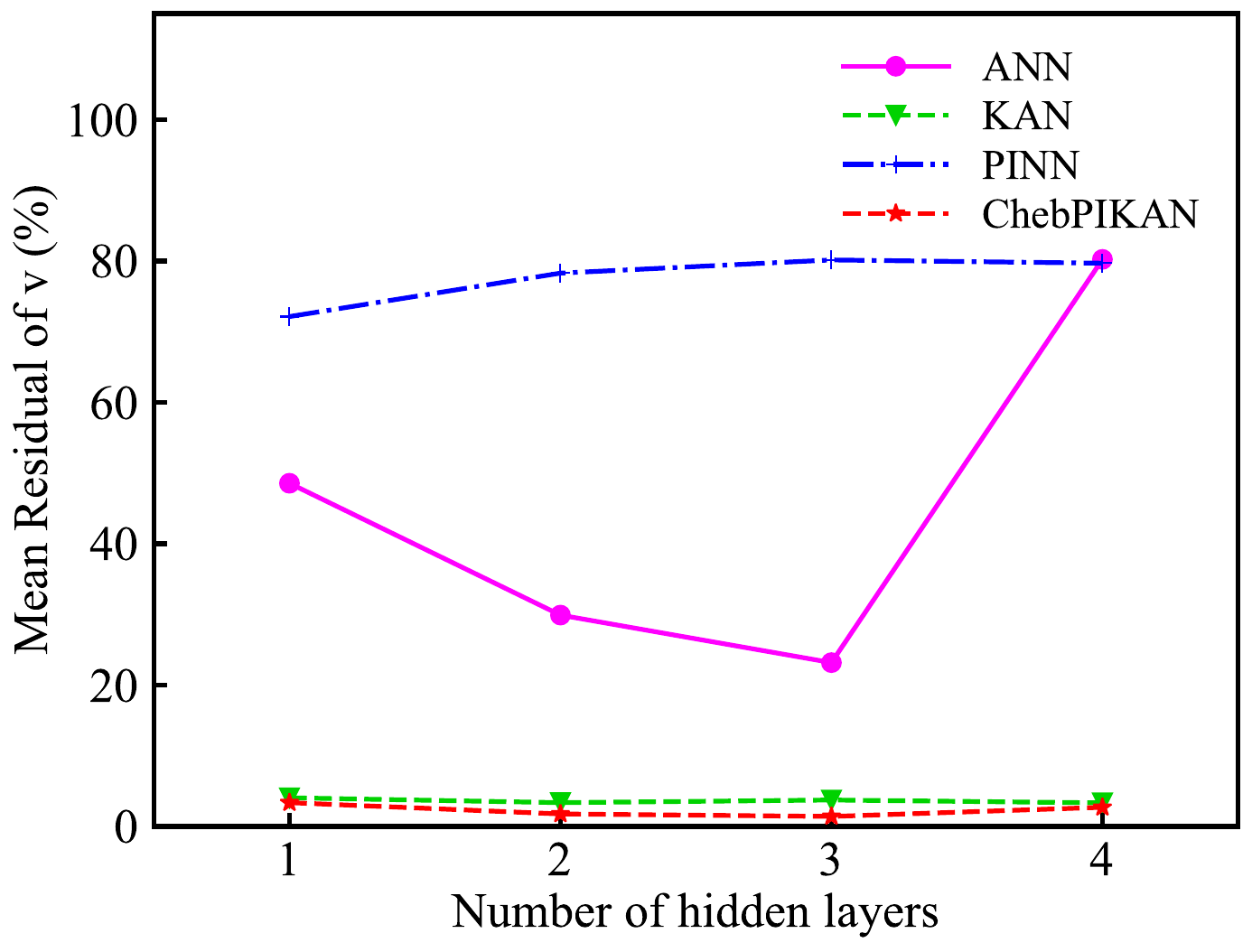}
	\caption{Mean residuals of velocity $v$ of different neural networks with different hidden layers for the Kovasznay flow equation.}
	\label{Kovbestres}
\end{figure}

For the pressure prediction, the simpler form of the pressure solution results in relatively ideal predictions from both networks. Nonetheless, ChebPIKANs demonstrate a significantly smaller overall error than KANs, particularly near the boundary. The accurate solution of Kovasznay flow underscores the strong capability of the KAN architecture, and the addition of physical information sets a solid foundation to tackle other fluid mechanics equations in the future. In the results for $u$ and $v$, the two neural networks exhibit similar performance. For $v$, which exhibits larger average residuals and presents a greater challenge for prediction, ChebPIKANs significantly outperform KANs, particularly with two and three hidden layers. This results highlight the robust model stability of ChebPIKANs, since it better adapts to the complexities introduced by additional hidden layers, especially in predicting $u$ and $v$. By contrast, KANs' performance shows minimal improvement and can become unstable in some cases. For more complex problems, ANNs and PINNs, despite having a comparable number of trainable parameters, are unable to effectively learn the underlying features. The insufficient fitting capacity caused by the limited number of parameters also renders the physics-informed guidance ineffective.

\subsection{Two-dimensional Unsteady Navier--Stokes Equation}

This section focuses on the flow dynamics around a two-dimensional cylinder, which were analyzed in a horizontal plane where external forces are negligible. The governing equations can be simplified as follows,
\begin{equation}
	\frac{\partial u}{\partial t}+u \frac{\partial u}{\partial x}+v \frac{\partial u}{\partial y}=-\frac{\partial p}{\partial x}+\frac{1}{Re}\left(\frac{\partial^{2} u}{\partial x^{2}}+\frac{\partial^{2} u}{\partial y^{2}}\right),
\end{equation}
\begin{equation}
	\frac{\partial v}{\partial t}+u \frac{\partial v}{\partial x}+v \frac{\partial v}{\partial y}=-\frac{\partial p}{\partial y}+\frac{1}{Re}\left(\frac{\partial^{2} v}{\partial x^{2}}+\frac{\partial^{2} v}{\partial y^{2}}\right),
\end{equation}
\begin{equation}
	\frac{\partial u}{\partial x} + \frac{\partial v}{\partial y} = 0.
\end{equation}

The cylindrical wave is examined in the spatial domain defined by $x \in[1, 8]$ and $y \in[-2, 2]$ and the temporal domain $t \in[0, 7]$. The cylinder of interest is centered at (0, 0) with a diameter of $D$ = 1. Considering the complexity of these time-dependent equations, we generated 2,000 training data points and 5,000 test data points. For the ChebPIKAN models, 8000 additional points are sampled to compute the physical loss. The optimizer and optimization strategy are identical to those for other equations. The network has three inputs ($x$, $y$, and $t$) and three outputs (pressure $p$, velocity $u$, and velocity $v$). In the two-dimensional unsteady Navier--Stokes equation, the physical loss is expressed as
\begin{equation}
	\begin{split}
		Los{s_{PDE}} = \,& \frac{\partial u}{\partial t}+u \frac{\partial u}{\partial x}+v \frac{\partial u}{\partial y}+\frac{\partial p}{\partial x}-\frac{1}{Re}\left(\frac{\partial^{2} u}{\partial x^{2}}+\frac{\partial^{2} u}{\partial y^{2}}\right) \\
		+ \,& \frac{\partial v}{\partial t}+u \frac{\partial v}{\partial x}+v \frac{\partial v}{\partial y}+\frac{\partial p}{\partial y}-\frac{1}{Re}\left(\frac{\partial^{2} v}{\partial x^{2}}+\frac{\partial^{2} v}{\partial y^{2}}\right) \\
		+ & \frac{\partial u}{\partial x} + \frac{\partial v}{\partial y}.
	\end{split}
\end{equation}
\begin{figure}[!htbp] 
	\centering
	\includegraphics[height=4cm]{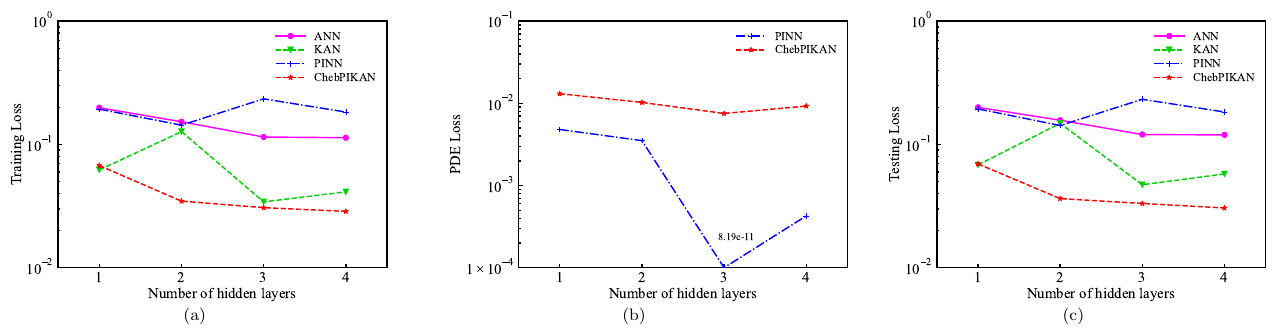}
	\caption{Optimal loss of different neural networks with different hidden layers for the Unsteady Navier--Stokes Equation. (a) Training loss, (b) PDE loss, (c) Testing loss.}
	\label{NSbestloss}
\end{figure}

Due to the challenges in predicting time-dependent two-dimensional equations and the variations in the specific formulation of the loss function, the optimal loss values for KANs and ChebPIKANs are approximately one order of magnitude higher than that of previous models (approximately $10^{-1}$). With the physics-informed weight $\lambda$ = 0.1 selected through preliminary tests, the PDE loss of PINNs remains significantly smaller in magnitude than its test data loss. This indicates that the learning process is predominantly governed by physical information. Within the predetermined network architecture, the fully-connected configuration between PINNs nodes inherently limits their potential fitting capability. Consequently, PINNs fail to effectively learn flow field information through the simultaneous guidance of sparse data and physical constraints, ultimately converging to a specific solution that reduces physics loss while bearing minimal relevance to the target flow field.

ChebPIKANs demonstrate substantially lower losses compared to KANs, revealing their superior predictive capability in handling more complex two-dimensional time-varying flow problems. Among all models, the PDE loss consistently maintains a much smaller magnitude than both training and testing losses, suggesting its relatively minor weighting during the training process. Nevertheless, this physical constraint contributes to enhanced network performance, demonstrating that the incorporation of physical information effectively strengthens the neural network's learning capacity. Notably, the performance degradation observed in KANs stems from excessive learnable parameters generated by their numerous hidden layers. This architectural characteristic adversely affects their modeling capability for the given problem.
\begin{figure}[!htbp]
	\centering
	\includegraphics[height=10cm]{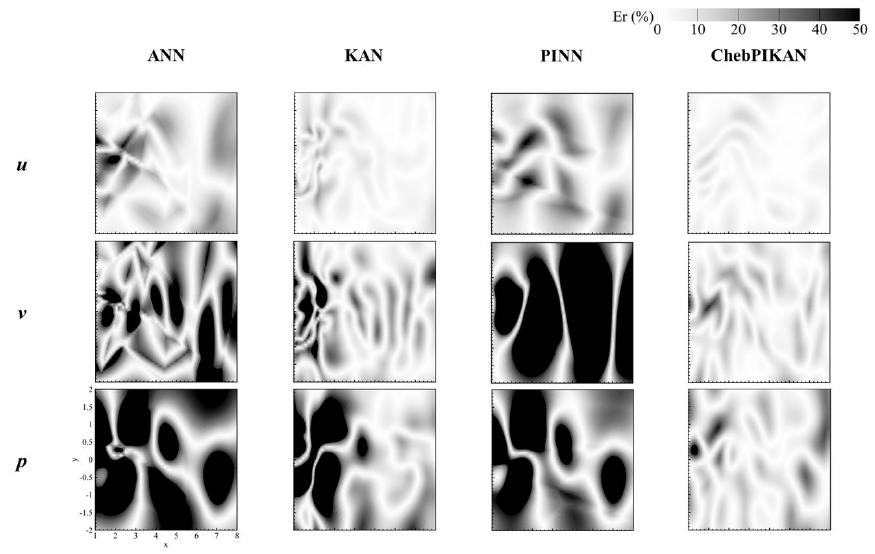}
	\caption{Relative error contours of different neural networks with four hidden layers for the unsteady Navier-Stokes equations.}
	\label{NScontour}
\end{figure}
\begin{figure}[!htbp]
	\centering
	\includegraphics[height=5cm]{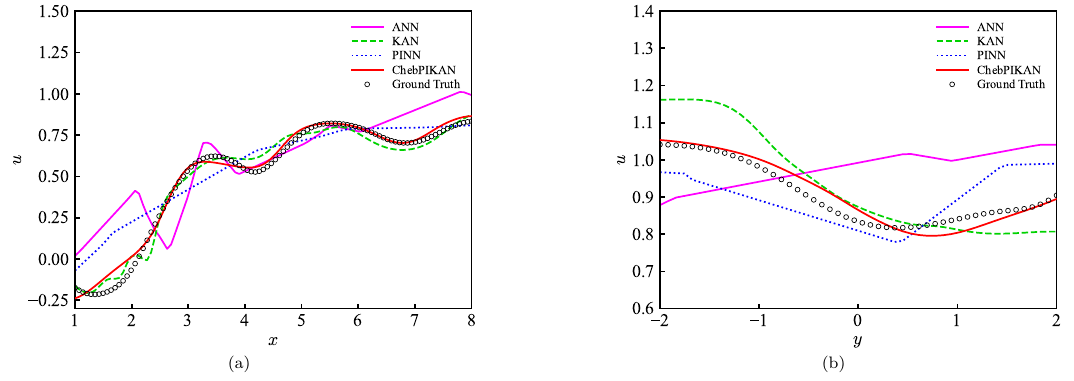}
	\caption{Velocity profiles predicted by different neural networks with four hidden layers for the unsteady Navier-Stokes equations. (a) $y$ = 0, (b) $x$ = 8.}
	\label{NS_line}
\end{figure}

Figure \ref {NScontour} shows prediction error contours at $t=3$. The ANNs' predictions remain ineffective, while the PINNs' results, consistent with their loss values, demonstrate physics-dominated convergence toward a specific solution. The KANs exhibit noticeable oscillatory errors in the high-shear region near the cylindrical boundary, indicating suboptimal learning performance under sparse data conditions that leads to inaccuracies in areas with substantial velocity gradient variations. Although the physics-informed ChebPIKANs show improved performance in this challenging region, the cylinder wake problem presents significantly greater complexity compared to previous test cases. Figure \ref {NS_line} further demonstrates that even the highest-performing ChebPIKANs fail to achieve perfect agreement with the reference solution. To maintain fair performance comparison standards, we deliberately avoid network architecture expansion for error reduction purposes. Instead, we reserve the investigation of larger network architectures for subsequent cavity flow studies, where their enhanced capabilities can be more thoroughly examined. The prediction of normal velocity contour of cylindrical turbulence is relatively small compared with $u$, which makes it challenging to minimize errors. Even a slight discrepancy can significantly increase the relative error due to a small denominator, as shown in Eq. \eqref{eq:err}. Consequently, both KANs and ChebPIKANs exhibit considerable errors in the final prediction of $v$ compared to $u$. The pressure prediction errors primarily concentrate in regions with strong pressure fluctuations, presenting significant challenges for neural networks. These pressure variations largely originate from the negative pressure generated by Bernoulli effects in the flow field. Furthermore, since the pressure magnitude is considerably smaller than that of the velocity component $u$, even minor prediction errors substantially amplify the relative error. Comparative analysis reveals that the physics-informed ChebPIKANs achieve superior performance to KANs, exhibiting both reduced prediction errors and enhanced numerical stability.
\begin{table}[!htbp]
	\centering
	\caption{Mean residuals of KANs and ChebPIKANs with different numbers of hidden layers for the unsteady Navier--Stokes equation.}
	\label{tab:Average residuals_NS}
	\setlength{\tabcolsep}{10pt}
	\begin{tabular}{ccccccc}
		\hline
		$\textup{N}_{\textup{h}}$ & KAN(u) & ChebPIKAN(u) & KAN(v) & ChebPIKAN(v) & KAN(p) & ChebPIKAN(p) \\ \hline
		1 & 7.30\% & 7.38\% & 17.88\% & 17.51\% & 22.09\% & 23.52\% \\
		2 & 15.09\% & 3.34\% & 38.39\% & 8.46\% & 39.45\% & 10.85\% \\
		3 & 3.38\% & 2.76\% & 11.25\% & 9.68\% & 15.33\% & 13.12\% \\
		4 & 4.61\% & 2.96\% & 15.39\% & 7.93\% & 32.55\% & 11.13\% \\ \hline
	\end{tabular}
\end{table}
\begin{figure}[!htbp] 
	\centering
	\includegraphics[height=4cm]{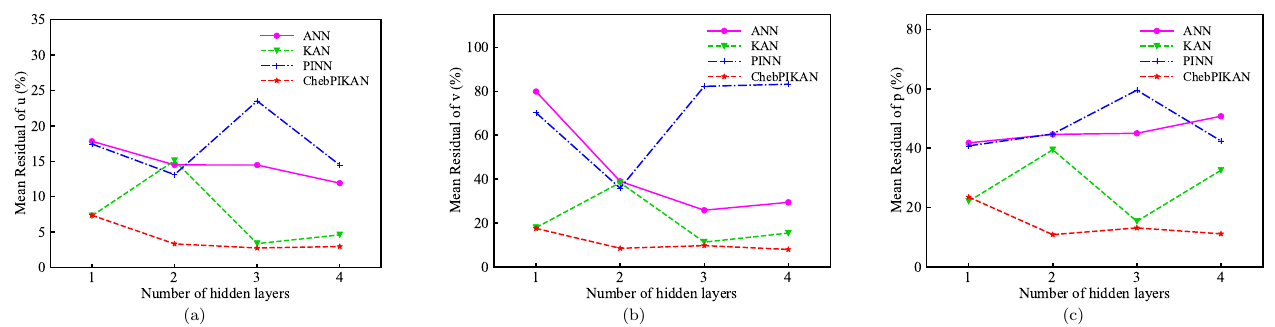}
	\caption{Mean residuals of $u$, $v$, and $p$ of different neural networks with different numbers of hidden layers for the unsteady Navier--Stokes equation.}
	\label{NSbestres}
\end{figure}

In the context of the unsteady Navier--Stokes equations with three-dimensional inputs, ChebPIKANs also demonstrate robust prediction performance, comprehensively outperforming the other three types of neural networks used for comparison. However, for relatively complex problems, although ChebPIKANs exhibit superior fitting capabilities with the same number of trainable parameters, they require a sufficient number of parameters to accurately describe the flow field characteristics. As a result, ChebPIKAN with only one hidden layer performs suboptimally, but its performance shows a positive trend as the network architecture expands.

\subsection{Two-dimensional Steady Navier--Stokes Equation}

This section studies the lid-driven cavity flow at $Re$ = 3200 to assess ChebPIKAN's capability in approximating steady flows. Since the temporal derivative term can be disregarded, the governing equations are simplified to
\begin{equation}
	u \frac{\partial u}{\partial x}+v \frac{\partial u}{\partial y}=-\frac{\partial p}{\partial x}+\frac{1}{Re}\left(\frac{\partial^{2} u}{\partial x^{2}}+\frac{\partial^{2} u}{\partial y^{2}}\right),
\end{equation}
\begin{equation}
	u \frac{\partial v}{\partial x}+v \frac{\partial v}{\partial y}=-\frac{\partial p}{\partial y}+\frac{1}{Re}\left(\frac{\partial^{2} v}{\partial x^{2}}+\frac{\partial^{2} v}{\partial y^{2}}\right),
\end{equation}
\begin{equation}
	\frac{\partial u}{\partial x} + \frac{\partial v}{\partial y} = 0.
\end{equation}

The cavity is a square with a side length of 1. The top boundary is assigned a velocity of $u$ = 1, while all boundaries are impermeable. Considering the complex nature of square cavity flows, which differ significantly from other two-dimensional flow cases, we systematically modified both the neural network structure and training parameters as specified in Table~\ref{tab:parameters}. In the two-dimensional steady Navier--Stokes equation, the physical loss is expressed as
\begin{equation}
	\begin{split}
		Los{s_{PDE}} = \,& u \frac{\partial u}{\partial x}+v \frac{\partial u}{\partial y}+\frac{\partial p}{\partial x}-\frac{1}{Re}\left(\frac{\partial^{2} u}{\partial x^{2}}+\frac{\partial^{2} u}{\partial y^{2}}\right) \\
		+ \,& u \frac{\partial v}{\partial x}+v \frac{\partial v}{\partial y}+\frac{\partial p}{\partial y}-\frac{1}{Re}\left(\frac{\partial^{2} v}{\partial x^{2}}+\frac{\partial^{2} v}{\partial y^{2}}\right) \\
		+ & \frac{\partial u}{\partial x} + \frac{\partial v}{\partial y}.
	\end{split}
\end{equation}

Table~\ref{tab:loss_Cavity30x} presents the optimal training loss values for the four neural network architectures, each configured with 30 nodes per hidden layer.
\begin{table}[!htbp]
	\centering
	\caption{Optimal loss values under different architectures of different neural networks.}
	\label{tab:loss_Cavity30x}
	\setlength{\tabcolsep}{10pt}
	\begin{tabular}{cccccc}
		\hline
		Loss & Name & $\textup{N}_{\textup{h}}$=1 & $\textup{N}_{\textup{h}}$=2 & $\textup{N}_{\textup{h}}$=3 & $\textup{N}_{\textup{h}}$=4  \\ \hline
		Training & ANN & 1.80$\times 10^{-2}$ & 1.40$\times 10^{-3}$ & 2.81$\times 10^{-4}$ & 1.17$\times 10^{-4}$  \\
		~ & KAN & 1.15$\times 10^{-3}$ & 5.66$\times 10^{-5}$ & 9.24$\times 10^{-6}$ & 5.00$\times 10^{-6}$  \\
		~ & PINN & 3.56$\times 10^{-2}$ & 7.33$\times 10^{-3}$ & 1.45$\times 10^{-3}$ & 1.17$\times 10^{-3}$  \\
		~ & ChebPIKAN & 6.02$\times 10^{-3}$ & 4.19$\times 10^{-4}$ & 7.61$\times 10^{-5}$ & 6.89$\times 10^{-5}$  \\
		PDE & PINN & 5.88$\times 10^{-3}$ & 8.02$\times 10^{-3}$ & 1.84$\times 10^{-3}$ & 1.45$\times 10^{-3}$  \\
		~ & ChebPIKAN & 6.22$\times 10^{-3}$ & 7.55$\times 10^{-4}$ & 1.38$\times 10^{-4}$ & 8.32$\times 10^{-5}$  \\
		Testing & ANN & 3.85$\times 10^{-2}$ & 2.41$\times 10^{-2}$ & 2.96$\times 10^{-2}$ & 9.41$\times 10^{-3}$  \\
		~ & KAN & 1.48$\times 10^{-2}$ & 2.49$\times 10^{-2}$ & 2.20$\times 10^{-2}$ & 2.92$\times 10^{-2}$  \\
		~ & PINN & 3.51$\times 10^{-2}$ & 7.36$\times 10^{-3}$ & 4.33$\times 10^{-3}$ & 4.52$\times 10^{-3}$  \\
		~ & ChebPIKAN & 1.15$\times 10^{-2}$ & 2.31$\times 10^{-3}$ & 8.42$\times 10^{-3}$ & 4.45$\times 10^{-3}$  \\ \hline
	\end{tabular}
\end{table}
\begin{figure}[!htbp] 
	\centering
	\includegraphics[height=4cm]{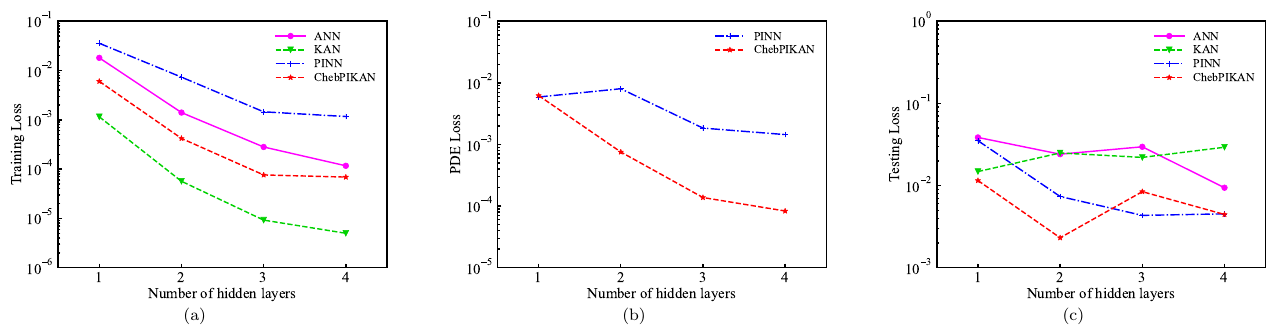}
	\caption{Optimal loss of different neural networks for the steady Navier--Stokes equation. (a) Training loss, (b) PDE loss, (c) Testing loss.}
	\label{Cavity_OlossV30x}
\end{figure}

As clearly illustrated in Figure \ref {Cavity_OlossV30x} (a), the overall optimal training loss exhibits a decreasing trend as the number of hidden layers increases. The results align with the characteristic that neural networks with more trainable parameters exhibit stronger fitting capacity, confirming the fundamental theoretical validity of the network architecture design. During the training processes of both PINNs and ChebPIKANs, the physical information loss and data loss remain within the same order of magnitude. This suggests that data and physical information contribute comparably to guiding the neural network's learning, and the weighting coefficients $\lambda$ for the physical loss terms are appropriately configured. Consequently, neither loss function dominates the learning process, thereby avoiding scenarios where one loss term becomes meaningless or leads to neural network failure. For identical network architectures, KANs achieve the lowest training loss, followed by ChebPIKANs. This demonstrates that, compared to traditional fully connected structures, KANs incorporating Chebyshev polynomials more effectively assimilate known data information, resulting in significantly improved fitting performance.

As evidenced in Figure \ref {Cavity_OlossV30x} (c), PINNs and ChebPIKANs exhibit superior performance on unseen test data compared to standard models. Notably, while KANs achieve the best results on the known training dataset, their performance on the test set is inferior even to PINNs. This suggests that for physical problems where only sparse real-world data is available for training, physics-informed frameworks can leverage minimal ground-truth data as anchor points and reconstruct the entire flow field using governing equations. Such an approach effectively mitigates overfitting and enhances the overall predictive capability over the entire field. Furthermore, the four-hidden-layer KANs, despite delivering optimal training accuracy, perform the worst on the test set. This not only highlights the critical challenge of overfitting in practical applications but also underscores that neural networks with strong fitting capacities—when faced with insufficient data—require carefully designed physics-informed loss functions to improve their real-world performance.
\begin{table}[!htbp]
	\centering
	\caption{Mean residuals of different neural networks for steady Navier--Stokes equation across different architectures.}
	\label{tab:Ave_res_Cavity30x}
	\setlength{\tabcolsep}{12pt}
	\begin{tabular}{ccccc}
		\hline
		Name & Hidden Layers & Er(u) & Er(v) & Er(p)  \\ \hline
		ANN & 1x30 & 7.19\% & 10.41\% & 9.65\%  \\
		~ & 2x30 & 1.96\% & 2.51\% & 2.27\%  \\
		~ & 3x30 & 0.97\% & 1.59\% & 1.27\%  \\
		~ & 4x30 & 1.03\% & 0.89\% & 0.89\%  \\
		KAN & 1x30 & 2.14\% & 2.57\% & 2.74\%  \\
		~ & 2x30 & 3.44\% & 2.27\% & 1.16\%  \\
		~ & 3x30 & 2.78\% & 3.64\% & 2.67\%  \\
		~ & 4x30 & 3.57\% & 3.12\% & 2.80\%  \\
		PINN & 1x30 & 11.39\% & 13.15\% & 8.54\%  \\
		~ & 2x30 & 2.10\% & 2.52\% & 2.34\%  \\
		~ & 3x30 & 0.59\% & 0.71\% & 0.52\%  \\
		~ & 4x30 & 0.50\% & 0.59\% & 0.40\%  \\
		ChebPIKAN & 1x30 & 1.82\% & 2.62\% & 1.26\%  \\
		~ & 2x30 & 0.24\% & 0.35\% & 0.33\%  \\
		~ & 3x30 & 0.25\% & 0.38\% & 0.34\%  \\
		~ & 4x30 & \textbf{0.13\%} & \textbf{0.14\%} & \textbf{0.10\%}  \\ \hline
	\end{tabular}
\end{table}

As shown by the average residuals in Table~\ref{tab:Ave_res_Cavity30x}, ChebPIKANs demonstrate superior performance among four neural networks for cavity flow prediction, achieving errors below 0.15\%. Even with minimal fitting capacity (1×30 hidden layer), ChebPIKANs maintain satisfactory accuracy, while ANNs and PINNs exhibit significantly larger errors. This highlights how Chebyshev polynomial-enhanced KANs substantially improve network fitting capability, enabling physics-uninformed KANs with fewer trainable parameters to outperform ANNs and PINNs. However, as hidden layers increase, the advantages of physics-informed guidance become apparent, with PINNs and ChebPIKANs surpassing ANNs and KANs. Since the training data consists of low-resolution experimental data, the networks must infer complete flow fields from limited measurement points. At the optimal 4×30 hidden layer configuration, ChebPIKANs achieve prediction errors of 0.13\% ($u$), 0.14\% ($v$), and 0.10\% ($p$), representing reductions of 74.79\%, 76.90\%, and 74.22\% respectively compared to PINNs' errors of 0.50\%, 0.59\%, and 0.40\%. These results clearly demonstrate ChebPIKANs' strong advantages for flow prediction using low-resolution experimental data.
\begin{figure}[!htbp] 
	\centering
	\includegraphics[height=10cm]{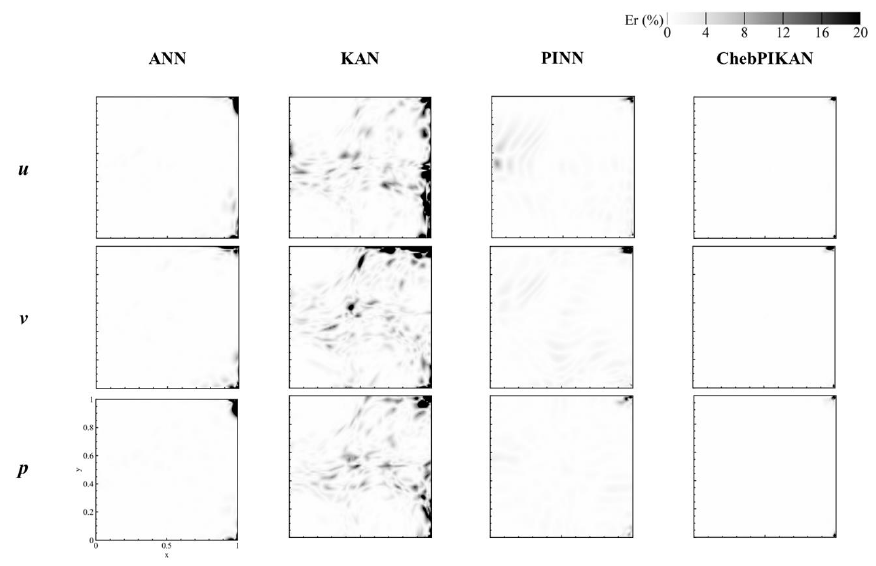}
	\caption{Relative error contours  of u, v and p for different neural networks with 4×30 hidden layers for the steady Navier-Stokes equations.}
	\label{Cavity_Aresfig30x}
\end{figure}

The relative error contours reveal that ChebPIKANs exhibit negligible errors in the interior flow region unaffected by boundary conditions, with only minor discrepancies near fixed boundaries where flow impingement occurs. While KANs demonstrate superior fitting capabilities over traditional fully-connected networks on sparse training datasets, this overfitting to known data points merely constructs an implicit multidimensional surface passing through the sample points, without enhancing overall flow prediction performance. These findings underscore that for practical physics applications, especially in cases where comprehensive physical data are unavailable, incorporating known physical principles to guide neural networks remains an essential component in intelligent physics research.
\begin{figure}[!htbp]
	\centering
	\includegraphics[height=5cm]{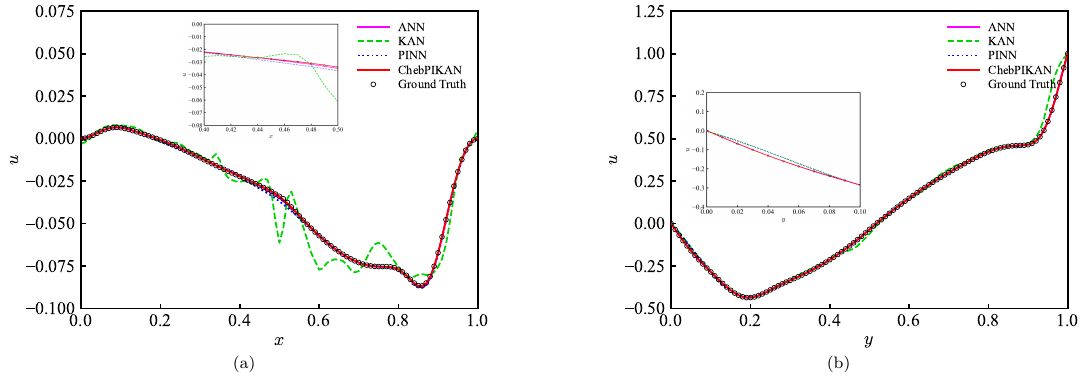}
	\caption{Velocity profiles of $u$ predicted by different neural networks with four hidden layers for the steady Navier-Stokes equations. (a) $y$ = 0.5, (b) $x$ = 0.5.}
	\label{cavity_line}
\end{figure}

Figure \ref {cavity_line} provides further evidence supporting previous conclusions. The KANs exhibit significant overfitting, indicating their unsuitability for the current task while simultaneously demonstrating their remarkable fitting capacity. This strong fitting capability suggests promising potential for purely data-driven applications. In contrast, ANNs display superior generalization performance compared to KANs due to their relatively weaker fitting ability, which prevents them from completely learning all available data information. Both ChebPIKANs and PINNs achieve satisfactory prediction results under this relatively large network framework. These findings indicate that although the fitting capability per trainable parameter in fully-connected architectures may be inferior to newly proposed algorithms, PINNs can still effectively address complex problems when sufficient trainable parameters are employed.

\subsection{The Impact of the Number of Trainable Parameters on Neural Network Performance}

The number of trainable parameters in a neural network significantly impacts its performance. A higher number of trainable parameters generally allows the network to capture more complex patterns and relationships within the data, potentially leading to improved accuracy and generalization. However, this also increases the risk of overfitting, especially when the training dataset is limited, as the network may memorize the training data rather than learning meaningful features. For instance, experimental data with lower resolution, combined with unavoidable experimental errors in real-world testing, may contain some data points that deviate from physical principles. Conversely, a network with too few trainable parameters may lack the capacity to adequately model the underlying data distribution, resulting in underfitting and poor performance. Therefore, optimizing the number of trainable parameters is crucial to achieving a balance between model complexity and generalization capability, ensuring robust performance on both training and unseen data.

A larger number of parameters also leads to another issue, the amount of data that needs to be replicated during network migration becomes substantial. The future trend points toward large-scale models, but an excessive number of parameters may impose higher performance requirements on the devices hosting these models, making widespread adoption less feasible. Therefore, enabling a limited number of parameters to describe more information is a highly meaningful area of focus. The approach introduced by KANs offers significant improvements in this regard. Taking the Navier--Stokes equations as an example, we compare four neural networks under similar parameter scales, examining the relationship between the number of trainable parameters and prediction errors in the velocity $u$.
\begin{figure}[!htbp] 
	\centering
	\includegraphics[height=5cm]{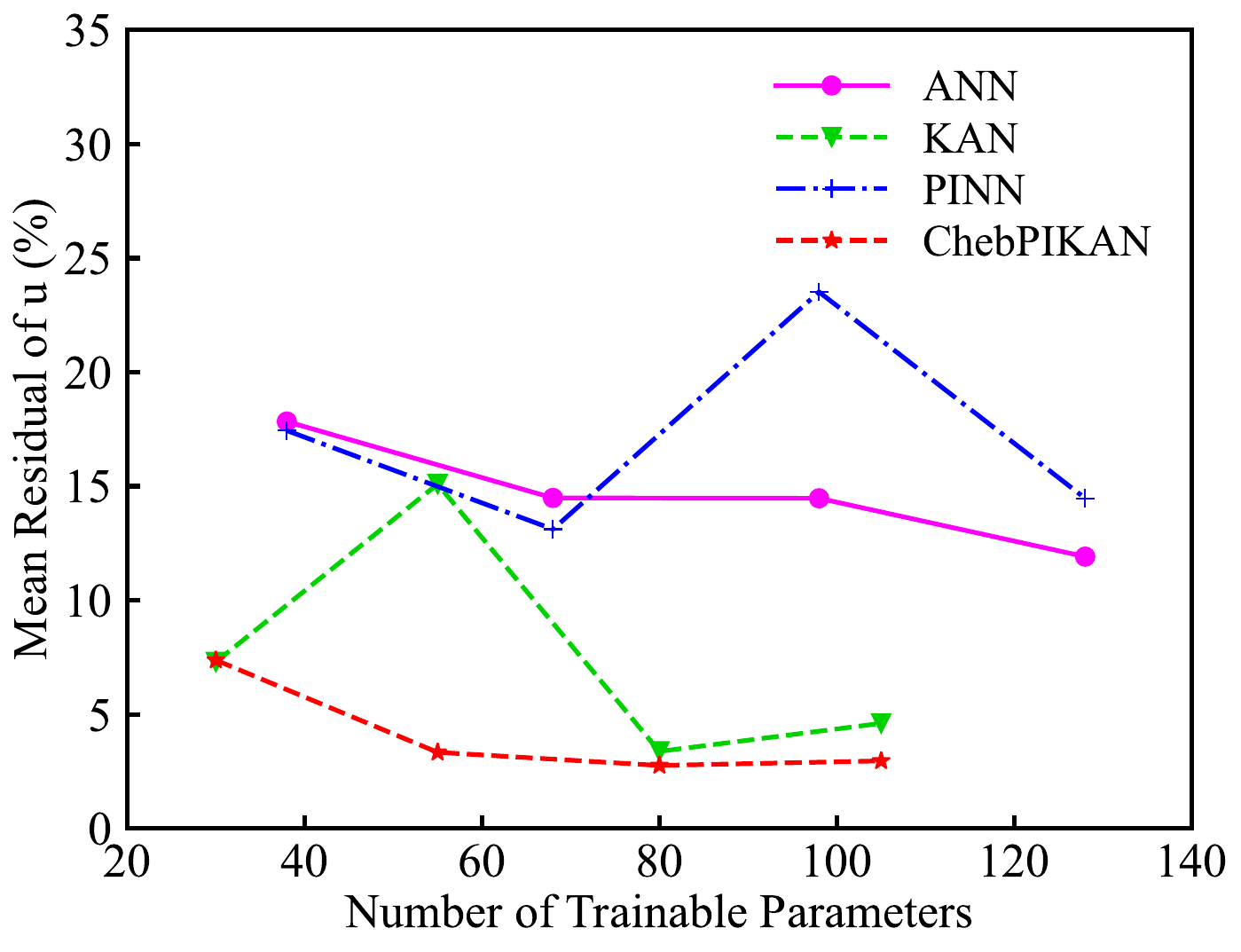}
	\caption{Mean residuals in the velocity $u$ of different neural networks with varying trainable parameters for the unsteady Navier--Stokes equation.}
	\label{ChebPIKAN_PINN_parameters}
\end{figure}

Figure \ref {ChebPIKAN_PINN_parameters} demonstrates that, under identical conditions of ground truth and physical loss function used for training, ChebPIKANs exhibit superior predictive performance compared to PINNs. This advantage arises from the unique functional relationships between nodes in ChebPIKANs, which differ from those in fully connected networks. Eq. (\ref{eq:Chebyshevn}) allows each parameter in ChebPIKANs to capture more complex information, enabling the model to achieve better flow prediction results with fewer parameters. This significantly contributes to reducing the cost of network parameter transfer during future large-scale model migration learning. Notably, when the number of trainable parameters is limited to $\leq$ 140, both ANNs and PINNs underperform compared to KANs. This indicates that under these conditions, the intrinsic fitting capability of the neural network architecture becomes the dominant factor in prediction performance, to such an extent that the networks cannot adequately represent randomly sampled real data. Consequently, some physics-informed PINNs even demonstrate inferior performance to basic ANNs in this scenario. To examine the effects of trainable parameters and physics-informed architectures on neural network performance, we use cavity flow as a benchmark case for steady Navier--Stokes equations.
\begin{figure}[!htbp] 
	\centering
	\includegraphics[height=5cm]{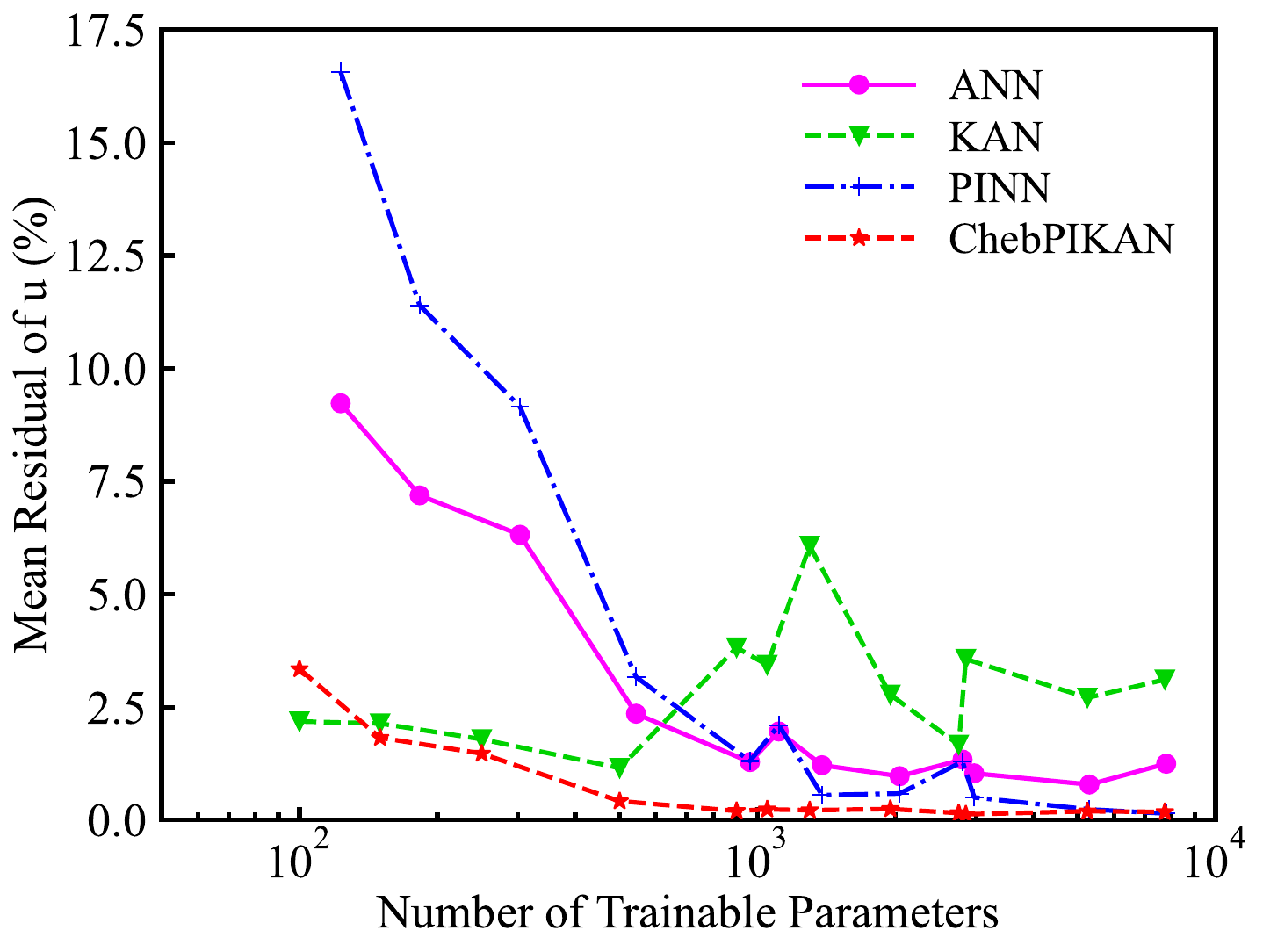}
	\caption{Mean residuals in the velocity $u$ of different neural networks with varying trainable parameters for the steady Navier--Stokes equation.}
	\label{Cavity_parameters}
\end{figure}

As illustrated in Figure \ref {Cavity_parameters}, when the number of trainable parameters is limited, the approximation capability of KANs integrated with Chebyshev polynomials dominates the predictive performance. Consequently, both KANs and ChebPIKANs significantly outperform conventional ANNs and PINNs. As the number of trainable parameters increases, all four neural networks demonstrate improved approximation performance, and the advantages brought by physics-informed guidance become evident. Within the investigated parameter range, ChebPIKANs and PINNs exhibit superior performance compared to the other two neural network types when the maximum number of trainable parameters is employed. Notably, KANs, which showed better performance with the fewest trainable parameters, become the least performant configuration in this scenario. This observation further substantiates that when available ground truth data are scarce, relying solely on strong approximation capability cannot guarantee optimal neural network performance in practical applications, since incorporation of known physical principles remains essential to assist neural networks in prediction tasks.

The Root Mean Square (RMS) is a crucial statistical measure for quantifying the magnitude of variability or dispersion within a dataset. Unlike the arithmetic mean, RMS takes into account both positive and negative values by squaring them, thus offering a more comprehensive measure of central tendency. RMS is especially valuable in scenarios where the direction of deviation (positive or negative) is less important than the overall magnitude of deviation. RMS is defined as
\begin{equation}
	RMS = \sqrt {\frac{{\sum\limits_{i = 1}^n {{{\left( {{u_{i}^{\textit {pred}}} - {u_{i}^{\textit {true}}}} \right)}^2}} }}{n}},
	\label{eq:RMS}
\end{equation}

The RMS values of $u$, $v$ and $p$ for ChebPIAKNs and PINNs under different numbers of trainable parameters were calculated using the Eq. (\ref{eq:RMS}). These are abbreviated as $RMS_{u}$, $RMS_{v}$ and $RMS_{p}$ respectively.
\begin{table}[!ht]
	\centering
	\caption{Root-Mean-Square (RMS) values of ChebPIKANs versus the number of trainable parameters under the unsteady Navier--Stokes equations.}
	\label{tab:RMS_ChebPIKANs}
	\setlength{\tabcolsep}{12pt}
	\begin{tabular}{cccc}
		\hline
		Trainable Parameters & $\mathrm{RMS}(u)$ & $\mathrm{RMS}(v)$ & $\mathrm{RMS}(p)$ \\ \hline
		30 & 9.13$\times 10^{-2}$ & 6.21$\times 10^{-2}$ & 4.04$\times 10^{-2}$  \\ 
		55 & 4.02$\times 10^{-2}$ & 3.56$\times 10^{-2}$ & 1.97$\times 10^{-2}$  \\ 
		80 & 3.36$\times 10^{-2}$ & 3.51$\times 10^{-2}$ & 2.56$\times 10^{-2}$  \\ 
		105 & 3.52$\times 10^{-2}$ & 2.85$\times 10^{-2}$ & 1.93$\times 10^{-2}$  \\ \hline
	\end{tabular}
\end{table}

The data in Table \ref{tab:RMS_ChebPIKANs} demonstrate that the RMS values of the predictions by ChebPIAKNs for the three quantities ($u$, $v$ and $p$) are within the same order of magnitude. Particularly, when the number of trainable parameters is 80, the deviations are relatively close, unlike in Table \ref{tab:Average residuals_NS}, where ChebPIAKNs exhibit superior performance in predicting $u$. This suggests that during the training process of the network, more emphasis is placed on the absolute values of the deviations rather than their percentages. This observation provides a potential direction for future research on ChebPIAKNs, adjusting weights based on the absolute magnitudes of the predicted physical quantities to enhance overall predictive performance.
\begin{table}[!ht]
	\centering
	\caption{Root-Mean-Square (RMS) values of PINNs versus the number of trainable parameters under the unsteady Navier--Stokes equations.}
	\label{tab:RMS_PINNs}
	\setlength{\tabcolsep}{12pt}
	\begin{tabular}{ccccc}
		\hline
		Trainable Parameters & $\mathrm{RMS}(u)$ & $\mathrm{RMS}(v)$ & $\mathrm{RMS}(p)$ \\ \hline
		38 & 2.06$\times 10^{-1}$ & 2.43$\times 10^{-1}$ & 7.27$\times 10^{-2}$  \\ 
		68 & 1.58$\times 10^{-1}$ & 1.22$\times 10^{-1}$ & 7.77$\times 10^{-2}$  \\ 
		98 & 2.85$\times 10^{-1}$ & 2.73$\times 10^{-1}$ & 1.02$\times 10^{-1}$  \\ 
		128 & 1.60$\times 10^{-1}$ & 2.77$\times 10^{-1}$ & 7.46$\times 10^{-2}$  \\ \hline
	\end{tabular}
\end{table}
\begin{figure}[!htbp] 
	\centering
	\includegraphics[height=5cm]{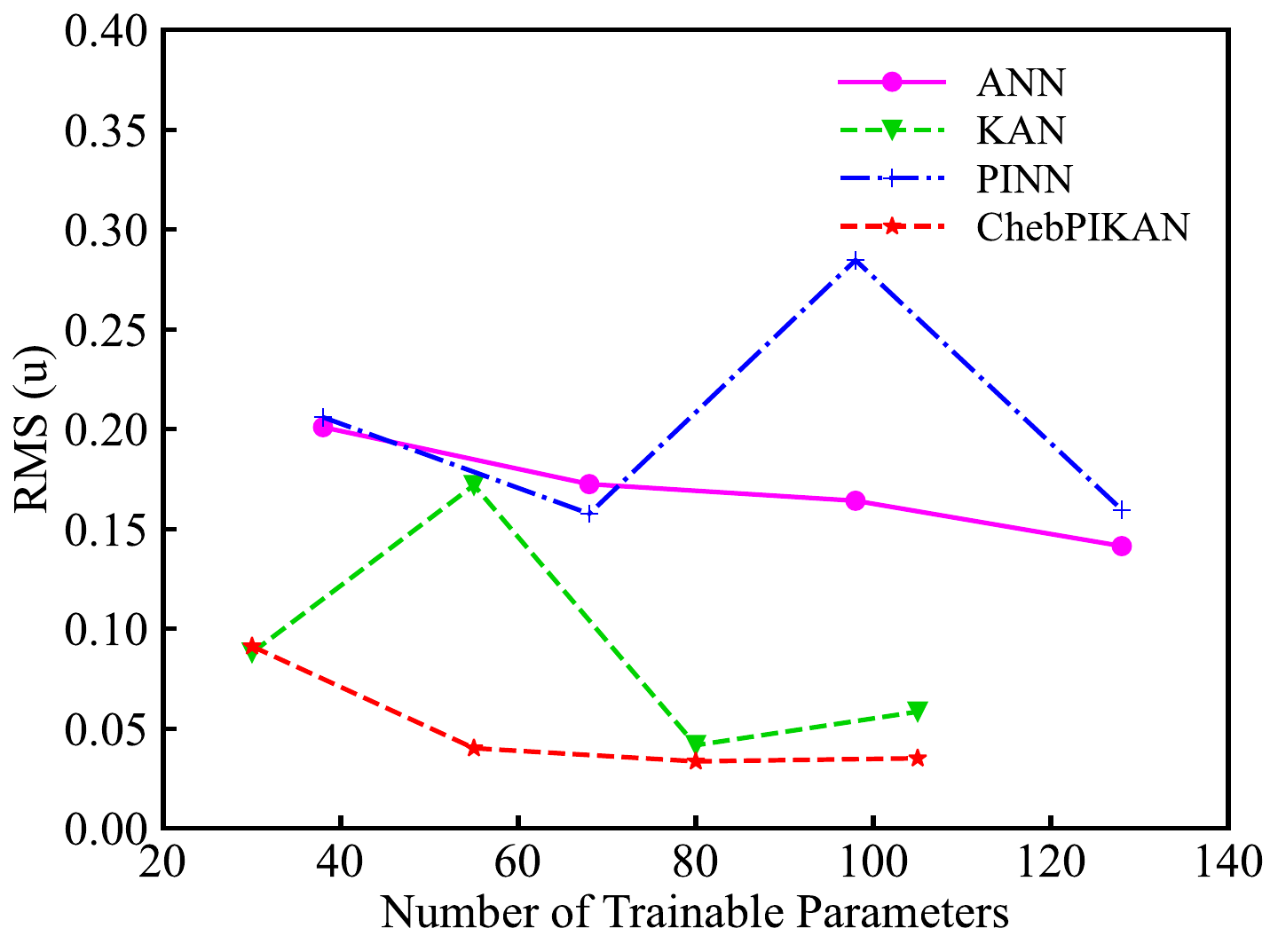}
	\caption{RMS in the velocity $u$ of different neural networks with varying trainable parameters for the unsteady Navier--Stokes equation.}
	\label{ChebPIKAN_KAN_RMS_parameters}
\end{figure}

From the Figure \ref {ChebPIKAN_KAN_RMS_parameters}, it can be observed that when ChebPIKANs and PINNs utilize a comparable number of trainable parameters, the performance of the former significantly surpasses that of the latter. This further underscores the performance enhancement achieved by ChebPIKANs. Moreover, under the condition of being physics-informed, ChebPIKANs demonstrate a superior capability in learning the characteristics of physical equations. Although the RMS results presented for the four neural network types are similar to those of the mean residual, this approach avoids the issue where small absolute errors lead to large percentage errors due to the true value being a relatively small quantity.
\begin{figure}[!htbp] 
	\centering
	\includegraphics[height=5cm]{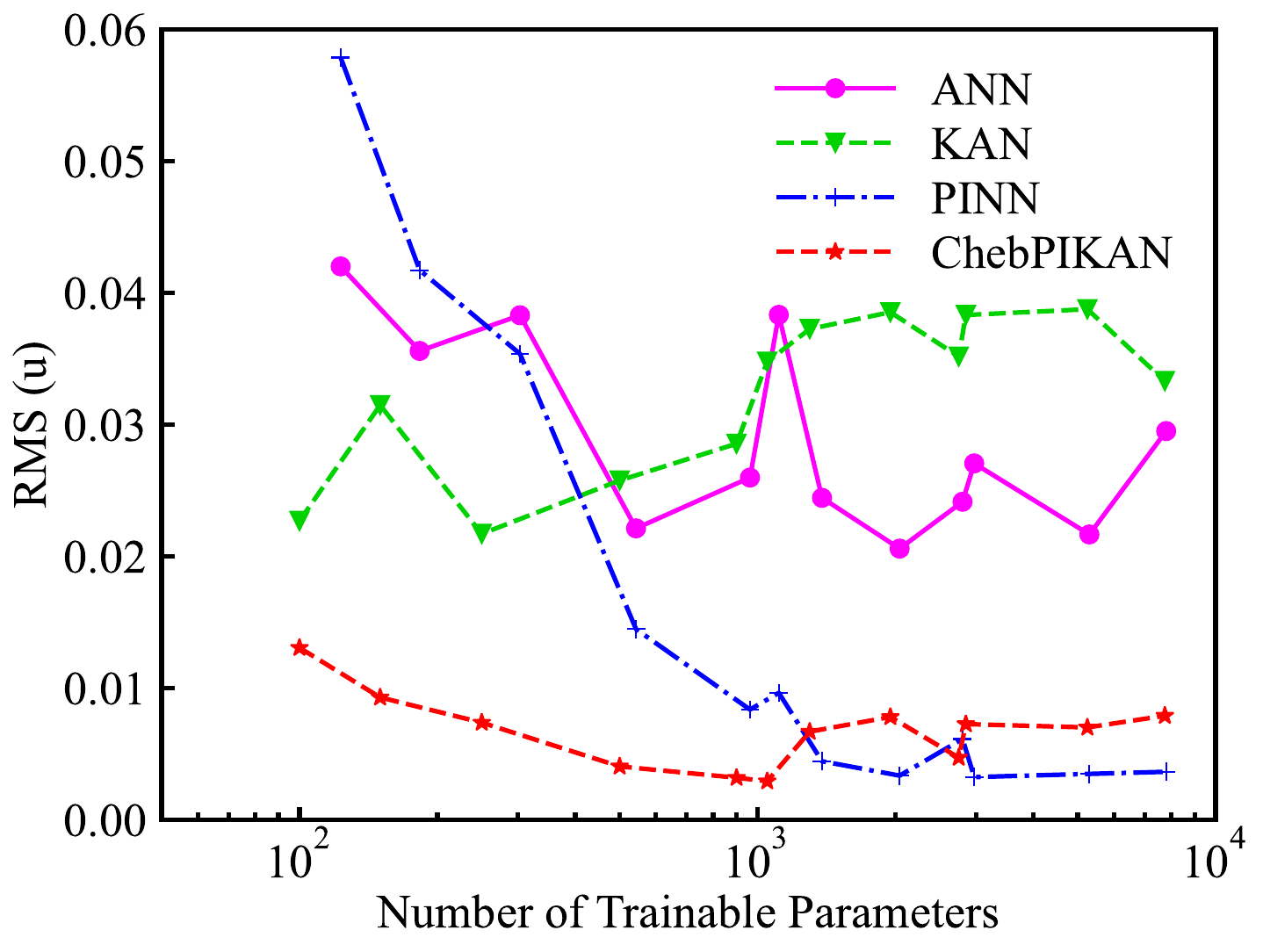}
	\caption{RMS in the velocity $u$ of different neural networks with varying trainable parameters for the steady Navier--Stokes equation.}
	\label{ChebPIKAN_cavity_RMS_parameters}
\end{figure}

For the cavity flow based on the steady Navier--Stokes equations, the RMS results in Figure \ref {ChebPIKAN_cavity_RMS_parameters} exhibit a trend similar to the mean residuals in Figure \ref {Cavity_parameters}. However, it is noteworthy that although ChebPIKANs with 2,850 trainable parameters achieve the best performance across all investigated cases, they slightly underperform PINNs under the RMS metric in certain scenarios with larger trainable parameters. This discrepancy arises from our use of identical training epochs for fair comparison. While the incorporation of Chebyshev polynomials enhances the model’s approximation capacity, their relatively complex inter-node operations inherently increase training difficulty. Consequently, as the number of trainable parameters grows, ChebPIKANs require higher computational costs than PINNs. Addressing this challenge through algorithmic optimization to improve training efficiency will be a critical focus for future research on ChebPIKANs.

\subsection{Flow Field Prediction Performance under Varied Reynolds Number Conditions}

The generalization capability across different Reynolds numbers is crucial for validating the performance of machine learning methods in fluid prediction problems. We systematically evaluate the ChebPIKAN architecture comprising four hidden layers of 30 neurons each, analyzing its flow field prediction capability across four Reynolds number regimes as quantified in Table \ref{tab:RePerformance}.
\begin{table}[!htbp]
	\centering
	\caption{Mean residuals of ChebPIKAN for steady Navier--Stokes equation across different Reynolds numbers.}
	\label{tab:RePerformance}
	\setlength{\tabcolsep}{15pt}
	\begin{tabular}{ccccc}
		\hline
		Re & Hidden Layers & Er(u) & Er(v) & Er(p)  \\ \hline
		400 & 4x30 & 0.09\% & 0.19\% & 0.25\%  \\
		1000 & 4x30 & 0.20\% & 0.24\% & 0.17\%  \\
		3200 & 4x30 & 0.13\% & 0.14\% & 0.10\%  \\
		5000 & 4x30 & 0.29\% & 0.21\% & 0.16\%  \\ \hline
	\end{tabular}
\end{table}

Table \ref{tab:RePerformance} demonstrates that when ChebPIKAN contains sufficient trainable parameters to represent physical flow characteristics, it maintains prediction errors below 0.3\% across sub-5000 Reynolds number cases despite minor performance fluctuations. This confirms the method's generalizability within this Reynolds number range.

The current implementation of ChebPIKAN shares a common limitation with most contemporary intelligent fluid mechanics methodologies, requiring retraining when applied to different Reynolds number regimes. To achieve the ultimate goal of replacing conventional CFD approaches, future research should focus on developing either rapid transfer learning techniques or alternative architectures that directly incorporate the Reynolds number as an input variable. These advancements will be critical for establishing truly generalizable flow prediction systems.

\subsection{The Performance of Flow Field Prediction in Temporal Extrapolation}

Data-driven neural networks have consistently faced the issue of performance degradation or even failure when applied beyond the scope of their training datasets. The physics-informed approach aims to address this by seeking a methodology that not only incorporates data obtained from real experiments but also integrates human-derived physical equations. Physical equations can correct discrepancies in experimental data caused by measurement errors, ensuring adherence to physical principles, while experimental data often contain subtle, higher-order information that may not be fully captured by existing physical models. By leveraging artificial intelligence, researchers can gain a deeper understanding of the true physical world, bridging the gap between empirical data and theoretical knowledge.

Taking the Navier--Stokes equations as an example, 2000 real data points are randomly selected from the first 7 seconds of flow as the training dataset. The prediction errors in the velocity $u$ of ChebPIKANs with 4 hidden layers and KANs with 4 hidden layers are compared over a 12-second period to analyze their temporal evolution.
\begin{figure}[!htbp] 
	\centering
	\includegraphics[height=5cm]{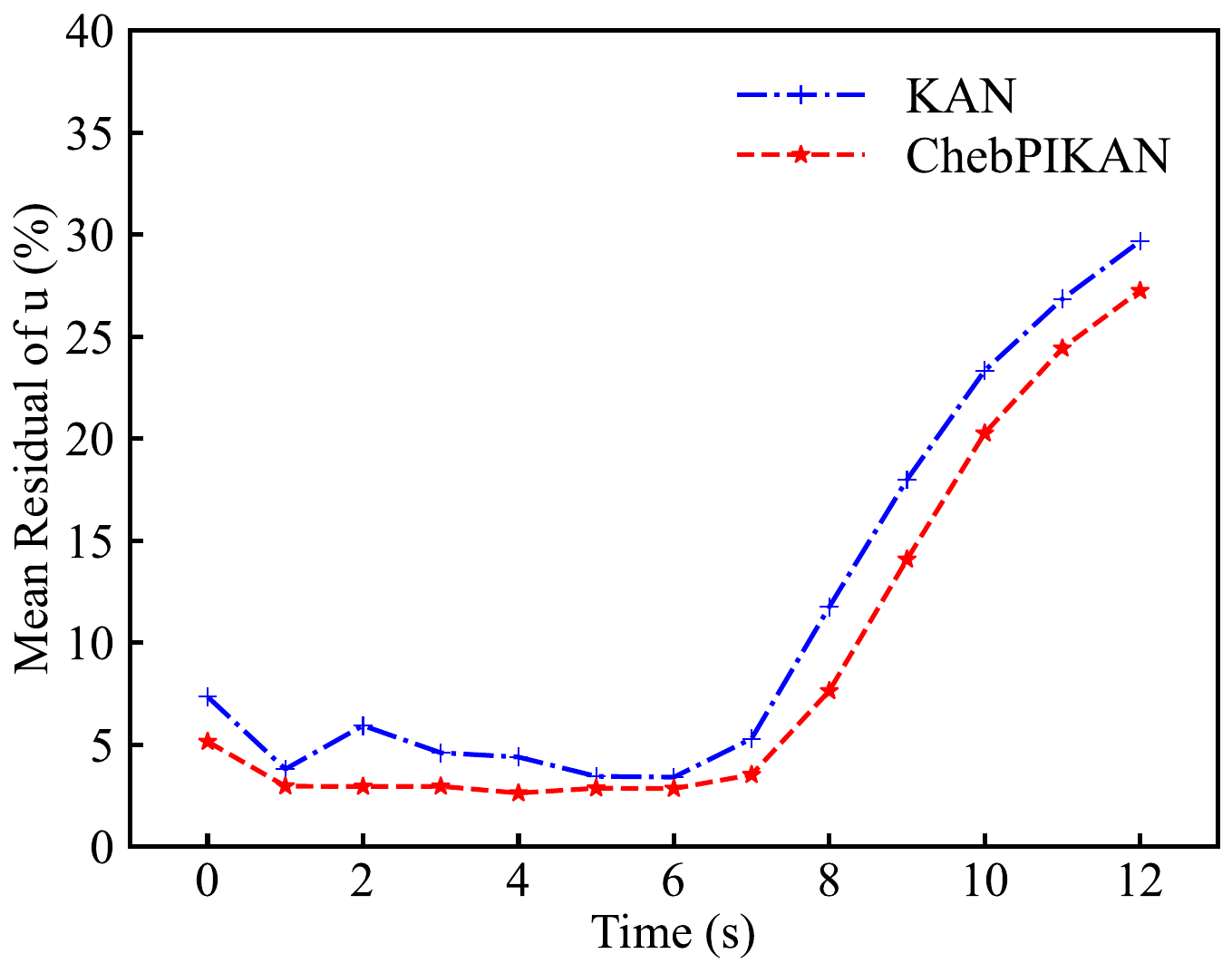}
	\caption{Variation of average residuals in the velocity $u$ of ChebPIKANs and KANs with time for the unsteady Navier--Stokes equation.}
	\label{ChebPIKAN_KAN_time}
\end{figure}

Figure \ref {ChebPIKAN_KAN_time} demonstrates that ChebPIKANs consistently exhibit superior predictive performance compared to KANs within the time range of the training dataset, indicating that the incorporation of physical information effectively enhances forecasting accuracy. However, beyond the time range of the training dataset, the errors of both neural networks increase, with the errors accumulating progressively over time. Nevertheless, overall, ChebPIKANs still outperform KANs, which lack the integration of physical information.
\begin{figure}[!htbp] 
	\centering
	\includegraphics[height=5cm]{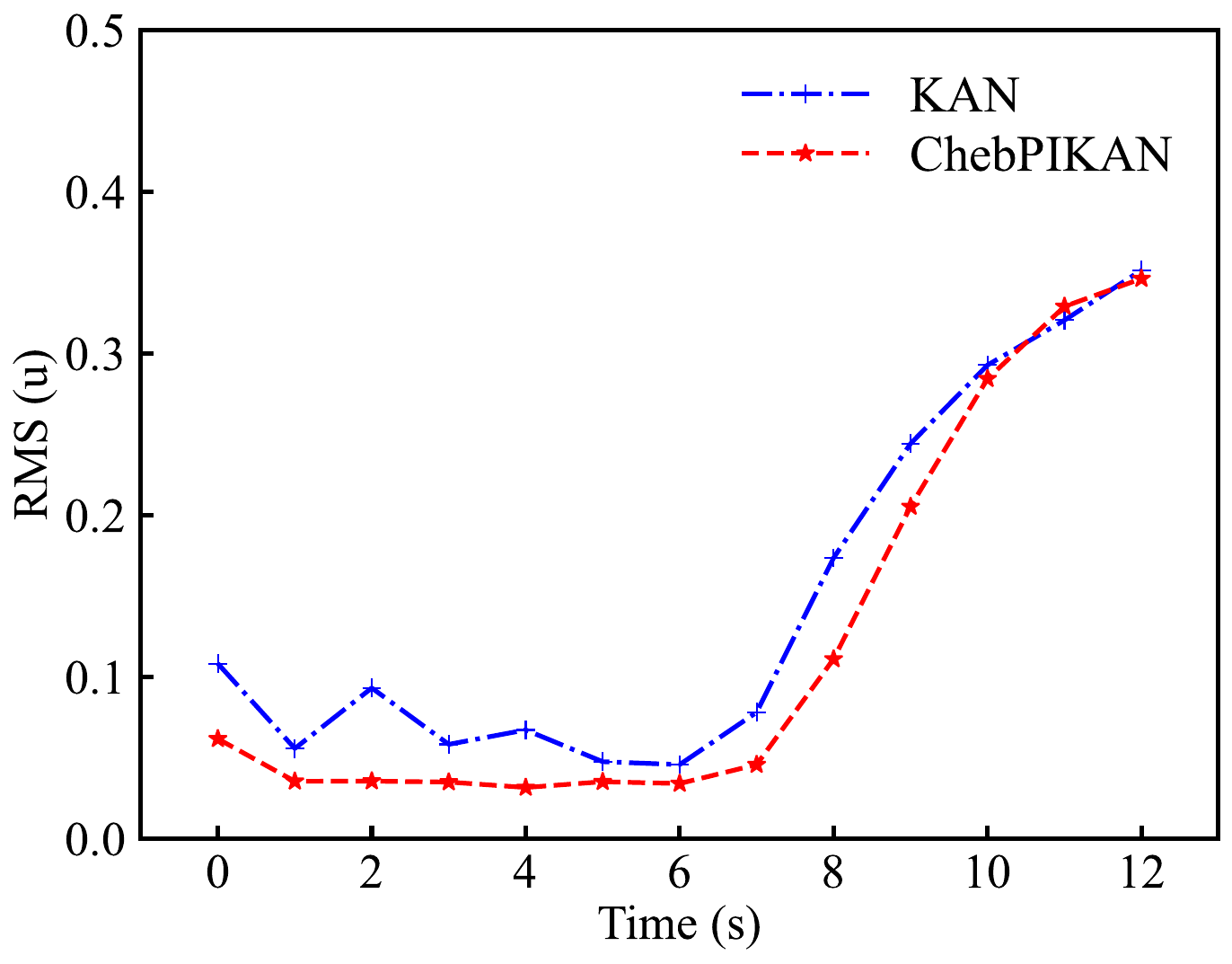}
	\caption{RMS in the velocity $u$ of ChebPIKANs and PINNs with time for the unsteady Navier--Stokes equation.}
	\label{ChebPIKAN_KAN_RMS_time}
\end{figure}

From the Figure \ref {ChebPIKAN_KAN_RMS_time}, it can be observed that within the time range where true data is available, ChebPIKANs outperform KANs, and KANs exhibit relatively unstable predictive performance in the absence of physics-informed guidance. As for the predictive capability of physical quantities at subsequent time steps, both models progressively accumulate errors, indicating that even ChebPIKANs, which incorporate physical information, fail to fully capture the temporal evolution patterns of physical changes across different time steps. Instead, they primarily enhance predictive performance within the considered time range by leveraging physical information. This limitation may stem from the fact that the training data points remain discrete, preventing the models from efficiently learning the relationships between physical information at successive time steps. In future research, it may be beneficial to treat information from adjacent time steps as a unified data entity. This approach could improve the neural network's ability to learn the temporal evolution of physical information and enhance its capability to infer physical quantities at unknown time steps.

\section{Conclusions and Discussion}

Chebyshev physics-informed Kolmogorov–Arnold network (ChebPIKAN) which incorporates physical constraints into KAN through PDE loss terms while replacing the Kolmogorov-Arnold theorem with Chebyshev polynomials, achieves a transformation comparable to the evolution from conventional ANNs to PINNs. In this study, the performance improvement of ChebPIKAN with added physical information relative to KAN was verified by solving partial differential equations, and ChebPIKAN has obvious advantages. It is reflected in the higher stability and generalization of the network compared to KAN without adding physical information.

All flow conditions employ limited experimental data as training sets, where the available datasets cannot fully resolve fine-scale flow field features. This data scarcity realistically mirrors practical experimental constraints in fluid dynamics measurements, thereby enhancing the neural networks' real-world applicability. When neural network architectures contain trainable parameters of the same order of magnitude, KANs incorporating Chebyshev polynomials exhibit superior fitting accuracy on available data. This indicates their ability to encode more information per trainable parameter, which holds significant implications for future large-model transfer learning and substantially reduces hardware requirements for deployed network models. However, as the number of trainable parameters increases, the enhanced fitting capability introduces overfitting challenges. In such cases, the value of physics-informed guidance becomes increasingly apparent. At larger parameter scales, both ChebPIKANs and PINNs outperform the other two conventional neural networks. This finding underscores the essential role of incorporating known physical principles in training neural networks for physics-based AI applications, particularly for experimental scenarios where detailed field data are difficult to obtain. The combination of powerful fitting capacity and effective physical constraints enables ChebPIKANs to deliver satisfactory performance, demonstrating promising potential for practical applications. These results validate that physics-informed approaches can become indispensable components in applying artificial intelligence to physical problems. The inherent architectural advantage of KANs over fully-connected networks results in either reduced computational resource usage during application or achieves higher performance at similar resource costs.

Since the local fitting capability of KAN, ChebPIKAN has better generalization performance. This advantage is evident in the continuous learning performance of the network and its resilience during debugging and training across various tasks, which effectively mitigate the risk of catastrophic forgetting. Furthermore, when future tasks involve overlapping or similar samples, the KAN structure is poised to offer significant benefits.

Despite these strengths, there are several areas for improvement in ChebPIKAN that warrant further exploration:

Mathematical Considerations: Although we excel in adopting more flexible network shapes for different scenarios, the mathematical visualization of the network architecture has not been fully developed due to algorithmic optimizations. As a result, its mathematical interpretability remains questionable. However, compared with traditional multilayer perceptrons (MLPs), KAN inherently has stronger physical significance, which the addition of physical information through PIKAN further enhances. Thus, ChebPIKAN has notably better physical meaning and interpretability than conventional neural networks.

Algorithm Enhancements: We integrate a PDE loss into the KAN framework and establish the basis of ChebPIKAN. This addition, which incorporates physical information, results in multiple loss terms. Properly adjusting the weights of these losses is crucial, since they directly affect the learning trajectory of the neural network. In particular, fine-tuning the correction coefficient for the PDE loss is essential, and inappropriate values can lead to imbalanced learning, where single loss dominates and others become ineffective. Thus, selecting suitable weight coefficients for specific problems is vital for enhancing the predictive performance of ChebPIKAN.

Practical Application: The automatic pruning function of ChebPIKAN has not been fully developed. Currently, the automatic pruning model of PIKAN cannot be directly applied since the hidden parameters lack a one-to-one correspondence. Consequently, we manually defined network architectures for performance testing. Nonetheless, the shape of the automatically trimmed model can serve as a valuable reference to guide the manual network configurations. Although manually setting network parameters is manageable, the ongoing refinement of the automatic pruning function remains an important area for future enhancement.

Overall, ChebPIKAN has demonstrated promising capabilities, but continuous improvements in mathematical interpretability, algorithmic precision, and practical application will be essential for its evolution.

\begin{acknowledgments}
This work was supported by the National Natural Science Foundation of China (NSFC Grant No. 52425111), the Shandong Province Youth Fund Project (Grant No. ZR20240A050), the International Science and Technology Cooperation Project under the Fundamental Research Funds for the Central Universities of Harbin Engineering University (Grant No. 3072024GH2602), the Natural Science Foundation of Hainan Province (Grant No. 425QN376), and the Young Scientist Cultivation Fund of Qingdao Innovation and Development Base of Harbin Engineering University.
\end{acknowledgments}

\section*{Data Availability Statement}

The data that support the findings of this study are available from the corresponding authors upon reasonable request.

\nocite{*}
\bibliography{ChebPIKANs}

\begin{thebibliography}{47}%
\makeatletter
\providecommand \@ifxundefined [1]{%
 \@ifx{#1\undefined}
}%
\providecommand \@ifnum [1]{%
 \ifnum #1\expandafter \@firstoftwo
 \else \expandafter \@secondoftwo
 \fi
}%
\providecommand \@ifx [1]{%
 \ifx #1\expandafter \@firstoftwo
 \else \expandafter \@secondoftwo
 \fi
}%
\providecommand \natexlab [1]{#1}%
\providecommand \enquote  [1]{``#1''}%
\providecommand \bibnamefont  [1]{#1}%
\providecommand \bibfnamefont [1]{#1}%
\providecommand \citenamefont [1]{#1}%
\providecommand \href@noop [0]{\@secondoftwo}%
\providecommand \href [0]{\begingroup \@sanitize@url \@href}%
\providecommand \@href[1]{\@@startlink{#1}\@@href}%
\providecommand \@@href[1]{\endgroup#1\@@endlink}%
\providecommand \@sanitize@url [0]{\catcode `\\12\catcode `\$12\catcode
  `\&12\catcode `\#12\catcode `\^12\catcode `\_12\catcode `\%12\relax}%
\providecommand \@@startlink[1]{}%
\providecommand \@@endlink[0]{}%
\providecommand \url  [0]{\begingroup\@sanitize@url \@url }%
\providecommand \@url [1]{\endgroup\@href {#1}{\urlprefix }}%
\providecommand \urlprefix  [0]{URL }%
\providecommand \Eprint [0]{\href }%
\providecommand \doibase [0]{http://dx.doi.org/}%
\providecommand \selectlanguage [0]{\@gobble}%
\providecommand \bibinfo  [0]{\@secondoftwo}%
\providecommand \bibfield  [0]{\@secondoftwo}%
\providecommand \translation [1]{[#1]}%
\providecommand \BibitemOpen [0]{}%
\providecommand \bibitemStop [0]{}%
\providecommand \bibitemNoStop [0]{.\EOS\space}%
\providecommand \EOS [0]{\spacefactor3000\relax}%
\providecommand \BibitemShut  [1]{\csname bibitem#1\endcsname}%
\let\auto@bib@innerbib\@empty
\bibitem [{\citenamefont {Moin}\ and\ \citenamefont
  {Mahesh}(1998)}]{moin1998direct}%
  \BibitemOpen
  \bibfield  {author} {\bibinfo {author} {\bibfnamefont {P.}~\bibnamefont
  {Moin}}\ and\ \bibinfo {author} {\bibfnamefont {K.}~\bibnamefont {Mahesh}},\
  }\bibfield  {title} {\enquote {\bibinfo {title} {Direct numerical simulation:
  a tool in turbulence research},}\ }\href@noop {} {\bibfield  {journal}
  {\bibinfo  {journal} {Annual review of fluid mechanics}\ }\textbf {\bibinfo
  {volume} {30}},\ \bibinfo {pages} {539--578} (\bibinfo {year}
  {1998})}\BibitemShut {NoStop}%
\bibitem [{\citenamefont {Pirozzoli}\ and\ \citenamefont
  {Orlandi}(2021)}]{PIROZZOLI2021110408}%
  \BibitemOpen
  \bibfield  {author} {\bibinfo {author} {\bibfnamefont {S.}~\bibnamefont
  {Pirozzoli}}\ and\ \bibinfo {author} {\bibfnamefont {P.}~\bibnamefont
  {Orlandi}},\ }\bibfield  {title} {\enquote {\bibinfo {title} {Natural grid
  stretching for {DNS} of wall-bounded flows},}\ }\href {\doibase
  https://doi.org/10.1016/j.jcp.2021.110408} {\bibfield  {journal} {\bibinfo
  {journal} {Journal of Computational Physics}\ }\textbf {\bibinfo {volume}
  {439}},\ \bibinfo {pages} {110408} (\bibinfo {year} {2021})}\BibitemShut
  {NoStop}%
\bibitem [{\citenamefont {Komen}\ \emph {et~al.}(2017)\citenamefont {Komen},
  \citenamefont {Camilo}, \citenamefont {Shams}, \citenamefont {Geurts},\ and\
  \citenamefont {Koren}}]{KOMEN2017565}%
  \BibitemOpen
  \bibfield  {author} {\bibinfo {author} {\bibfnamefont {E.}~\bibnamefont
  {Komen}}, \bibinfo {author} {\bibfnamefont {L.}~\bibnamefont {Camilo}},
  \bibinfo {author} {\bibfnamefont {A.}~\bibnamefont {Shams}}, \bibinfo
  {author} {\bibfnamefont {B.}~\bibnamefont {Geurts}}, \ and\ \bibinfo {author}
  {\bibfnamefont {B.}~\bibnamefont {Koren}},\ }\bibfield  {title} {\enquote
  {\bibinfo {title} {A quantification method for numerical dissipation in
  {quasi-DNS} and under-resolved {DNS}, and effects of numerical dissipation in
  {quasi-DNS} and under-resolved {DNS} of turbulent channel flows},}\ }\href
  {\doibase https://doi.org/10.1016/j.jcp.2017.05.030} {\bibfield  {journal}
  {\bibinfo  {journal} {Journal of Computational Physics}\ }\textbf {\bibinfo
  {volume} {345}},\ \bibinfo {pages} {565--595} (\bibinfo {year}
  {2017})}\BibitemShut {NoStop}%
\bibitem [{\citenamefont {Shoeybi}\ \emph {et~al.}(2010)\citenamefont
  {Shoeybi}, \citenamefont {Svärd}, \citenamefont {Ham},\ and\ \citenamefont
  {Moin}}]{SHOEYBI20105944}%
  \BibitemOpen
  \bibfield  {author} {\bibinfo {author} {\bibfnamefont {M.}~\bibnamefont
  {Shoeybi}}, \bibinfo {author} {\bibfnamefont {M.}~\bibnamefont {Svärd}},
  \bibinfo {author} {\bibfnamefont {F.~E.}\ \bibnamefont {Ham}}, \ and\
  \bibinfo {author} {\bibfnamefont {P.}~\bibnamefont {Moin}},\ }\bibfield
  {title} {\enquote {\bibinfo {title} {An adaptive implicit–explicit scheme
  for the {DNS} and {LES} of compressible flows on unstructured grids},}\
  }\href {\doibase https://doi.org/10.1016/j.jcp.2010.04.027} {\bibfield
  {journal} {\bibinfo  {journal} {Journal of Computational Physics}\ }\textbf
  {\bibinfo {volume} {229}},\ \bibinfo {pages} {5944--5965} (\bibinfo {year}
  {2010})}\BibitemShut {NoStop}%
\bibitem [{\citenamefont {Bhaganagar}, \citenamefont {Rempfer},\ and\
  \citenamefont {Lumley}(2002)}]{BHAGANAGAR2002200}%
  \BibitemOpen
  \bibfield  {author} {\bibinfo {author} {\bibfnamefont {K.}~\bibnamefont
  {Bhaganagar}}, \bibinfo {author} {\bibfnamefont {D.}~\bibnamefont {Rempfer}},
  \ and\ \bibinfo {author} {\bibfnamefont {J.}~\bibnamefont {Lumley}},\
  }\bibfield  {title} {\enquote {\bibinfo {title} {Direct numerical simulation
  of spatial transition to turbulence using fourth-order vertical velocity
  second-order vertical vorticity formulation},}\ }\href {\doibase
  https://doi.org/10.1006/jcph.2002.7088} {\bibfield  {journal} {\bibinfo
  {journal} {Journal of Computational Physics}\ }\textbf {\bibinfo {volume}
  {180}},\ \bibinfo {pages} {200--228} (\bibinfo {year} {2002})}\BibitemShut
  {NoStop}%
\bibitem [{\citenamefont {Zhang}\ \emph {et~al.}(2021)\citenamefont {Zhang},
  \citenamefont {Dwight}, \citenamefont {Schmelzer}, \citenamefont {Gómez},
  \citenamefont {hua Han},\ and\ \citenamefont {Hickel}}]{ZHANG2021110153}%
  \BibitemOpen
  \bibfield  {author} {\bibinfo {author} {\bibfnamefont {Y.}~\bibnamefont
  {Zhang}}, \bibinfo {author} {\bibfnamefont {R.~P.}\ \bibnamefont {Dwight}},
  \bibinfo {author} {\bibfnamefont {M.}~\bibnamefont {Schmelzer}}, \bibinfo
  {author} {\bibfnamefont {J.~F.}\ \bibnamefont {Gómez}}, \bibinfo {author}
  {\bibfnamefont {Z.}~\bibnamefont {hua Han}}, \ and\ \bibinfo {author}
  {\bibfnamefont {S.}~\bibnamefont {Hickel}},\ }\bibfield  {title} {\enquote
  {\bibinfo {title} {Customized data-driven rans closures for bi-fidelity
  les–rans optimization},}\ }\href {\doibase
  https://doi.org/10.1016/j.jcp.2021.110153} {\bibfield  {journal} {\bibinfo
  {journal} {Journal of Computational Physics}\ }\textbf {\bibinfo {volume}
  {432}},\ \bibinfo {pages} {110153} (\bibinfo {year} {2021})}\BibitemShut
  {NoStop}%
\bibitem [{\citenamefont {Clerici}, \citenamefont {Spalart},\ and\
  \citenamefont {Alauzet}(2024)}]{CLERICI2024113191}%
  \BibitemOpen
  \bibfield  {author} {\bibinfo {author} {\bibfnamefont {F.}~\bibnamefont
  {Clerici}}, \bibinfo {author} {\bibfnamefont {P.}~\bibnamefont {Spalart}}, \
  and\ \bibinfo {author} {\bibfnamefont {F.}~\bibnamefont {Alauzet}},\
  }\bibfield  {title} {\enquote {\bibinfo {title} {Turbulence driven
  goal-oriented anisotropic mesh adaptation for {RANS} simulations in
  aerodynamics},}\ }\href {\doibase https://doi.org/10.1016/j.jcp.2024.113191}
  {\bibfield  {journal} {\bibinfo  {journal} {Journal of Computational
  Physics}\ }\textbf {\bibinfo {volume} {514}},\ \bibinfo {pages} {113191}
  (\bibinfo {year} {2024})}\BibitemShut {NoStop}%
\bibitem [{\citenamefont {Xiao}\ and\ \citenamefont
  {Jenny}(2012)}]{XIAO20121848}%
  \BibitemOpen
  \bibfield  {author} {\bibinfo {author} {\bibfnamefont {H.}~\bibnamefont
  {Xiao}}\ and\ \bibinfo {author} {\bibfnamefont {P.}~\bibnamefont {Jenny}},\
  }\bibfield  {title} {\enquote {\bibinfo {title} {A consistent dual-mesh
  framework for hybrid {LES/RANS} modeling},}\ }\href {\doibase
  https://doi.org/10.1016/j.jcp.2011.11.009} {\bibfield  {journal} {\bibinfo
  {journal} {Journal of Computational Physics}\ }\textbf {\bibinfo {volume}
  {231}},\ \bibinfo {pages} {1848--1865} (\bibinfo {year} {2012})}\BibitemShut
  {NoStop}%
\bibitem [{\citenamefont {Jin}\ \emph {et~al.}(2021)\citenamefont {Jin},
  \citenamefont {Cai}, \citenamefont {Li},\ and\ \citenamefont
  {Karniadakis}}]{jin2021nsfnets}%
  \BibitemOpen
  \bibfield  {author} {\bibinfo {author} {\bibfnamefont {X.}~\bibnamefont
  {Jin}}, \bibinfo {author} {\bibfnamefont {S.}~\bibnamefont {Cai}}, \bibinfo
  {author} {\bibfnamefont {H.}~\bibnamefont {Li}}, \ and\ \bibinfo {author}
  {\bibfnamefont {G.~E.}\ \bibnamefont {Karniadakis}},\ }\bibfield  {title}
  {\enquote {\bibinfo {title} {{NSFnets} (navier-stokes flow nets):
  Physics-informed neural networks for the incompressible {Navier-Stokes}
  equations},}\ }\href@noop {} {\bibfield  {journal} {\bibinfo  {journal}
  {Journal of Computational Physics}\ }\textbf {\bibinfo {volume} {426}},\
  \bibinfo {pages} {109951} (\bibinfo {year} {2021})}\BibitemShut {NoStop}%
\bibitem [{\citenamefont {Srinivasan}\ \emph {et~al.}(2019)\citenamefont
  {Srinivasan}, \citenamefont {Guastoni}, \citenamefont {Azizpour},
  \citenamefont {Schlatter},\ and\ \citenamefont
  {Vinuesa}}]{srinivasan2019predictions}%
  \BibitemOpen
  \bibfield  {author} {\bibinfo {author} {\bibfnamefont {P.~A.}\ \bibnamefont
  {Srinivasan}}, \bibinfo {author} {\bibfnamefont {L.}~\bibnamefont
  {Guastoni}}, \bibinfo {author} {\bibfnamefont {H.}~\bibnamefont {Azizpour}},
  \bibinfo {author} {\bibfnamefont {P.}~\bibnamefont {Schlatter}}, \ and\
  \bibinfo {author} {\bibfnamefont {R.}~\bibnamefont {Vinuesa}},\ }\bibfield
  {title} {\enquote {\bibinfo {title} {Predictions of turbulent shear flows
  using deep neural networks},}\ }\href@noop {} {\bibfield  {journal} {\bibinfo
   {journal} {Physical Review Fluids}\ }\textbf {\bibinfo {volume} {4}},\
  \bibinfo {pages} {054603} (\bibinfo {year} {2019})}\BibitemShut {NoStop}%
\bibitem [{\citenamefont {Bhatnagar}\ \emph {et~al.}(2019)\citenamefont
  {Bhatnagar}, \citenamefont {Afshar}, \citenamefont {Pan}, \citenamefont
  {Duraisamy},\ and\ \citenamefont {Kaushik}}]{bhatnagar2019prediction}%
  \BibitemOpen
  \bibfield  {author} {\bibinfo {author} {\bibfnamefont {S.}~\bibnamefont
  {Bhatnagar}}, \bibinfo {author} {\bibfnamefont {Y.}~\bibnamefont {Afshar}},
  \bibinfo {author} {\bibfnamefont {S.}~\bibnamefont {Pan}}, \bibinfo {author}
  {\bibfnamefont {K.}~\bibnamefont {Duraisamy}}, \ and\ \bibinfo {author}
  {\bibfnamefont {S.}~\bibnamefont {Kaushik}},\ }\bibfield  {title} {\enquote
  {\bibinfo {title} {Prediction of aerodynamic flow fields using convolutional
  neural networks},}\ }\href@noop {} {\bibfield  {journal} {\bibinfo  {journal}
  {Computational Mechanics}\ }\textbf {\bibinfo {volume} {64}},\ \bibinfo
  {pages} {525--545} (\bibinfo {year} {2019})}\BibitemShut {NoStop}%
\bibitem [{\citenamefont {Zhang}\ \emph {et~al.}(2024)\citenamefont {Zhang},
  \citenamefont {Sun}, \citenamefont {Zhang}, \citenamefont {Wang},
  \citenamefont {Zhang}, \citenamefont {Deng}, \citenamefont {Lin},\ and\
  \citenamefont {Chen}}]{10.1063/5.0195824}%
  \BibitemOpen
  \bibfield  {author} {\bibinfo {author} {\bibfnamefont {X.}~\bibnamefont
  {Zhang}}, \bibinfo {author} {\bibfnamefont {G.}~\bibnamefont {Sun}}, \bibinfo
  {author} {\bibfnamefont {P.}~\bibnamefont {Zhang}}, \bibinfo {author}
  {\bibfnamefont {Y.}~\bibnamefont {Wang}}, \bibinfo {author} {\bibfnamefont
  {J.}~\bibnamefont {Zhang}}, \bibinfo {author} {\bibfnamefont
  {L.}~\bibnamefont {Deng}}, \bibinfo {author} {\bibfnamefont {J.}~\bibnamefont
  {Lin}}, \ and\ \bibinfo {author} {\bibfnamefont {J.}~\bibnamefont {Chen}},\
  }\bibfield  {title} {\enquote {\bibinfo {title} {A residual graph
  convolutional network for setting initial flow field in computational fluid
  dynamics simulations},}\ }\href {\doibase 10.1063/5.0195824} {\bibfield
  {journal} {\bibinfo  {journal} {Physics of Fluids}\ }\textbf {\bibinfo
  {volume} {36}},\ \bibinfo {pages} {037150} (\bibinfo {year}
  {2024})}\BibitemShut {NoStop}%
\bibitem [{\citenamefont {Santos}\ \emph {et~al.}(2020)\citenamefont {Santos},
  \citenamefont {Xu}, \citenamefont {Jo}, \citenamefont {Landry}, \citenamefont
  {Prodanovi{\'c}},\ and\ \citenamefont {Pyrcz}}]{santos2020poreflow}%
  \BibitemOpen
  \bibfield  {author} {\bibinfo {author} {\bibfnamefont {J.~E.}\ \bibnamefont
  {Santos}}, \bibinfo {author} {\bibfnamefont {D.}~\bibnamefont {Xu}}, \bibinfo
  {author} {\bibfnamefont {H.}~\bibnamefont {Jo}}, \bibinfo {author}
  {\bibfnamefont {C.~J.}\ \bibnamefont {Landry}}, \bibinfo {author}
  {\bibfnamefont {M.}~\bibnamefont {Prodanovi{\'c}}}, \ and\ \bibinfo {author}
  {\bibfnamefont {M.~J.}\ \bibnamefont {Pyrcz}},\ }\bibfield  {title} {\enquote
  {\bibinfo {title} {Poreflow-net: A {3D} convolutional neural network to
  predict fluid flow through porous media},}\ }\href@noop {} {\bibfield
  {journal} {\bibinfo  {journal} {Advances in Water Resources}\ }\textbf
  {\bibinfo {volume} {138}},\ \bibinfo {pages} {103539} (\bibinfo {year}
  {2020})}\BibitemShut {NoStop}%
\bibitem [{\citenamefont {Kim}\ and\ \citenamefont
  {Lee}(2020)}]{kim2020prediction}%
  \BibitemOpen
  \bibfield  {author} {\bibinfo {author} {\bibfnamefont {J.}~\bibnamefont
  {Kim}}\ and\ \bibinfo {author} {\bibfnamefont {C.}~\bibnamefont {Lee}},\
  }\bibfield  {title} {\enquote {\bibinfo {title} {Prediction of turbulent heat
  transfer using convolutional neural networks},}\ }\href@noop {} {\bibfield
  {journal} {\bibinfo  {journal} {Journal of Fluid Mechanics}\ }\textbf
  {\bibinfo {volume} {882}},\ \bibinfo {pages} {A18} (\bibinfo {year}
  {2020})}\BibitemShut {NoStop}%
\bibitem [{\citenamefont {Chen}\ \emph {et~al.}(2021)\citenamefont {Chen},
  \citenamefont {Wang}, \citenamefont {Hesthaven},\ and\ \citenamefont
  {Zhang}}]{CHEN2021110666}%
  \BibitemOpen
  \bibfield  {author} {\bibinfo {author} {\bibfnamefont {W.}~\bibnamefont
  {Chen}}, \bibinfo {author} {\bibfnamefont {Q.}~\bibnamefont {Wang}}, \bibinfo
  {author} {\bibfnamefont {J.~S.}\ \bibnamefont {Hesthaven}}, \ and\ \bibinfo
  {author} {\bibfnamefont {C.}~\bibnamefont {Zhang}},\ }\bibfield  {title}
  {\enquote {\bibinfo {title} {Physics-informed machine learning for
  reduced-order modeling of nonlinear problems},}\ }\href {\doibase
  https://doi.org/10.1016/j.jcp.2021.110666} {\bibfield  {journal} {\bibinfo
  {journal} {Journal of Computational Physics}\ }\textbf {\bibinfo {volume}
  {446}},\ \bibinfo {pages} {110666} (\bibinfo {year} {2021})}\BibitemShut
  {NoStop}%
\bibitem [{\citenamefont {Erichson}\ \emph {et~al.}(2020)\citenamefont
  {Erichson}, \citenamefont {Mathelin}, \citenamefont {Yao}, \citenamefont
  {Brunton}, \citenamefont {Mahoney},\ and\ \citenamefont
  {Kutz}}]{erichson2020shallow}%
  \BibitemOpen
  \bibfield  {author} {\bibinfo {author} {\bibfnamefont {N.~B.}\ \bibnamefont
  {Erichson}}, \bibinfo {author} {\bibfnamefont {L.}~\bibnamefont {Mathelin}},
  \bibinfo {author} {\bibfnamefont {Z.}~\bibnamefont {Yao}}, \bibinfo {author}
  {\bibfnamefont {S.~L.}\ \bibnamefont {Brunton}}, \bibinfo {author}
  {\bibfnamefont {M.~W.}\ \bibnamefont {Mahoney}}, \ and\ \bibinfo {author}
  {\bibfnamefont {J.~N.}\ \bibnamefont {Kutz}},\ }\bibfield  {title} {\enquote
  {\bibinfo {title} {Shallow neural networks for fluid flow reconstruction with
  limited sensors},}\ }\href@noop {} {\bibfield  {journal} {\bibinfo  {journal}
  {Proceedings of the Royal Society A}\ }\textbf {\bibinfo {volume} {476}},\
  \bibinfo {pages} {20200097} (\bibinfo {year} {2020})}\BibitemShut {NoStop}%
\bibitem [{\citenamefont {Kashefi}, \citenamefont {Rempe},\ and\ \citenamefont
  {Guibas}(2021)}]{kashefi2021point}%
  \BibitemOpen
  \bibfield  {author} {\bibinfo {author} {\bibfnamefont {A.}~\bibnamefont
  {Kashefi}}, \bibinfo {author} {\bibfnamefont {D.}~\bibnamefont {Rempe}}, \
  and\ \bibinfo {author} {\bibfnamefont {L.~J.}\ \bibnamefont {Guibas}},\
  }\bibfield  {title} {\enquote {\bibinfo {title} {A point-cloud deep learning
  framework for prediction of fluid flow fields on irregular geometries},}\
  }\href@noop {} {\bibfield  {journal} {\bibinfo  {journal} {Physics of
  Fluids}\ }\textbf {\bibinfo {volume} {33}} (\bibinfo {year}
  {2021})}\BibitemShut {NoStop}%
\bibitem [{\citenamefont {Raissi}, \citenamefont {Perdikaris},\ and\
  \citenamefont {Karniadakis}(2019)}]{raissi2019physics}%
  \BibitemOpen
  \bibfield  {author} {\bibinfo {author} {\bibfnamefont {M.}~\bibnamefont
  {Raissi}}, \bibinfo {author} {\bibfnamefont {P.}~\bibnamefont {Perdikaris}},
  \ and\ \bibinfo {author} {\bibfnamefont {G.~E.}\ \bibnamefont
  {Karniadakis}},\ }\bibfield  {title} {\enquote {\bibinfo {title}
  {Physics-informed neural networks: A deep learning framework for solving
  forward and inverse problems involving nonlinear partial differential
  equations},}\ }\href@noop {} {\bibfield  {journal} {\bibinfo  {journal}
  {Journal of Computational physics}\ }\textbf {\bibinfo {volume} {378}},\
  \bibinfo {pages} {686--707} (\bibinfo {year} {2019})}\BibitemShut {NoStop}%
\bibitem [{\citenamefont {Alzubaidi}\ \emph {et~al.}(2023)\citenamefont
  {Alzubaidi}, \citenamefont {Bai}, \citenamefont {Al-Sabaawi}, \citenamefont
  {Santamar{'i}a}, \citenamefont {Albahri}, \citenamefont {Al-dabbagh},
  \citenamefont {Fadhel}, \citenamefont {Manoufali}, \citenamefont {Zhang},
  \citenamefont {Al-Timemy} \emph {et~al.}}]{alzubaidi2023survey}%
  \BibitemOpen
  \bibfield  {author} {\bibinfo {author} {\bibfnamefont {L.}~\bibnamefont
  {Alzubaidi}}, \bibinfo {author} {\bibfnamefont {J.}~\bibnamefont {Bai}},
  \bibinfo {author} {\bibfnamefont {A.}~\bibnamefont {Al-Sabaawi}}, \bibinfo
  {author} {\bibfnamefont {J.}~\bibnamefont {Santamar{'i}a}}, \bibinfo {author}
  {\bibfnamefont {A.~S.}\ \bibnamefont {Albahri}}, \bibinfo {author}
  {\bibfnamefont {B.~S.~N.}\ \bibnamefont {Al-dabbagh}}, \bibinfo {author}
  {\bibfnamefont {M.~A.}\ \bibnamefont {Fadhel}}, \bibinfo {author}
  {\bibfnamefont {M.}~\bibnamefont {Manoufali}}, \bibinfo {author}
  {\bibfnamefont {J.}~\bibnamefont {Zhang}}, \bibinfo {author} {\bibfnamefont
  {A.~H.}\ \bibnamefont {Al-Timemy}},  \emph {et~al.},\ }\bibfield  {title}
  {\enquote {\bibinfo {title} {A survey on deep learning tools dealing with
  data scarcity: definitions, challenges, solutions, tips, and applications},}\
  }\href@noop {} {\bibfield  {journal} {\bibinfo  {journal} {Journal of Big
  Data}\ }\textbf {\bibinfo {volume} {10}},\ \bibinfo {pages} {46} (\bibinfo
  {year} {2023})}\BibitemShut {NoStop}%
\bibitem [{\citenamefont {Arzani}, \citenamefont {Wang},\ and\ \citenamefont
  {D'Souza}(2021)}]{arzani2021uncovering}%
  \BibitemOpen
  \bibfield  {author} {\bibinfo {author} {\bibfnamefont {A.}~\bibnamefont
  {Arzani}}, \bibinfo {author} {\bibfnamefont {J.-X.}\ \bibnamefont {Wang}}, \
  and\ \bibinfo {author} {\bibfnamefont {R.~M.}\ \bibnamefont {D'Souza}},\
  }\bibfield  {title} {\enquote {\bibinfo {title} {Uncovering near-wall blood
  flow from sparse data with physics-informed neural networks},}\ }\href@noop
  {} {\bibfield  {journal} {\bibinfo  {journal} {Physics of Fluids}\ }\textbf
  {\bibinfo {volume} {33}} (\bibinfo {year} {2021})}\BibitemShut {NoStop}%
\bibitem [{\citenamefont {Chen}, \citenamefont {Liu},\ and\ \citenamefont
  {Sun}(2021)}]{chen2021physics}%
  \BibitemOpen
  \bibfield  {author} {\bibinfo {author} {\bibfnamefont {Z.}~\bibnamefont
  {Chen}}, \bibinfo {author} {\bibfnamefont {Y.}~\bibnamefont {Liu}}, \ and\
  \bibinfo {author} {\bibfnamefont {H.}~\bibnamefont {Sun}},\ }\bibfield
  {title} {\enquote {\bibinfo {title} {Physics-informed learning of governing
  equations from scarce data},}\ }\href@noop {} {\bibfield  {journal} {\bibinfo
   {journal} {Nature communications}\ }\textbf {\bibinfo {volume} {12}},\
  \bibinfo {pages} {6136} (\bibinfo {year} {2021})}\BibitemShut {NoStop}%
\bibitem [{\citenamefont {Ramabathiran}\ and\ \citenamefont
  {Ramachandran}(2021)}]{ramabathiran2021spinn}%
  \BibitemOpen
  \bibfield  {author} {\bibinfo {author} {\bibfnamefont {A.~A.}\ \bibnamefont
  {Ramabathiran}}\ and\ \bibinfo {author} {\bibfnamefont {P.}~\bibnamefont
  {Ramachandran}},\ }\bibfield  {title} {\enquote {\bibinfo {title} {{SPINN}:
  sparse, physics-based, and partially interpretable neural networks for
  {PDEs}},}\ }\href@noop {} {\bibfield  {journal} {\bibinfo  {journal} {Journal
  of Computational Physics}\ }\textbf {\bibinfo {volume} {445}},\ \bibinfo
  {pages} {110600} (\bibinfo {year} {2021})}\BibitemShut {NoStop}%
\bibitem [{\citenamefont {Hanrahan}, \citenamefont {Kozul},\ and\ \citenamefont
  {Sandberg}(2023)}]{hanrahan2023studying}%
  \BibitemOpen
  \bibfield  {author} {\bibinfo {author} {\bibfnamefont {S.}~\bibnamefont
  {Hanrahan}}, \bibinfo {author} {\bibfnamefont {M.}~\bibnamefont {Kozul}}, \
  and\ \bibinfo {author} {\bibfnamefont {R.}~\bibnamefont {Sandberg}},\
  }\bibfield  {title} {\enquote {\bibinfo {title} {Studying turbulent flows
  with physics-informed neural networks and sparse data},}\ }\href@noop {}
  {\bibfield  {journal} {\bibinfo  {journal} {International Journal of Heat and
  Fluid Flow}\ }\textbf {\bibinfo {volume} {104}},\ \bibinfo {pages} {109232}
  (\bibinfo {year} {2023})}\BibitemShut {NoStop}%
\bibitem [{\citenamefont {Xu}\ \emph {et~al.}(2023)\citenamefont {Xu},
  \citenamefont {Sun}, \citenamefont {Huang}, \citenamefont {Guo},
  \citenamefont {Yang},\ and\ \citenamefont {Ju}}]{xu2023practical}%
  \BibitemOpen
  \bibfield  {author} {\bibinfo {author} {\bibfnamefont {S.}~\bibnamefont
  {Xu}}, \bibinfo {author} {\bibfnamefont {Z.}~\bibnamefont {Sun}}, \bibinfo
  {author} {\bibfnamefont {R.}~\bibnamefont {Huang}}, \bibinfo {author}
  {\bibfnamefont {D.}~\bibnamefont {Guo}}, \bibinfo {author} {\bibfnamefont
  {G.}~\bibnamefont {Yang}}, \ and\ \bibinfo {author} {\bibfnamefont
  {S.}~\bibnamefont {Ju}},\ }\bibfield  {title} {\enquote {\bibinfo {title} {A
  practical approach to flow field reconstruction with sparse or incomplete
  data through physics informed neural network},}\ }\href@noop {} {\bibfield
  {journal} {\bibinfo  {journal} {Acta Mechanica Sinica}\ }\textbf {\bibinfo
  {volume} {39}},\ \bibinfo {pages} {322302} (\bibinfo {year}
  {2023})}\BibitemShut {NoStop}%
\bibitem [{\citenamefont {Rasht-Behesht}\ \emph {et~al.}(2022)\citenamefont
  {Rasht-Behesht}, \citenamefont {Huber}, \citenamefont {Shukla},\ and\
  \citenamefont {Karniadakis}}]{rasht2022physics}%
  \BibitemOpen
  \bibfield  {author} {\bibinfo {author} {\bibfnamefont {M.}~\bibnamefont
  {Rasht-Behesht}}, \bibinfo {author} {\bibfnamefont {C.}~\bibnamefont
  {Huber}}, \bibinfo {author} {\bibfnamefont {K.}~\bibnamefont {Shukla}}, \
  and\ \bibinfo {author} {\bibfnamefont {G.~E.}\ \bibnamefont {Karniadakis}},\
  }\bibfield  {title} {\enquote {\bibinfo {title} {Physics-informed neural
  networks ({PINNs}) for wave propagation and full waveform inversions},}\
  }\href@noop {} {\bibfield  {journal} {\bibinfo  {journal} {Journal of
  Geophysical Research: Solid Earth}\ }\textbf {\bibinfo {volume} {127}},\
  \bibinfo {pages} {e2021JB023120} (\bibinfo {year} {2022})}\BibitemShut
  {NoStop}%
\bibitem [{\citenamefont {Eivazi}\ \emph {et~al.}(2022)\citenamefont {Eivazi},
  \citenamefont {Tahani}, \citenamefont {Schlatter},\ and\ \citenamefont
  {Vinuesa}}]{eivazi2022physics}%
  \BibitemOpen
  \bibfield  {author} {\bibinfo {author} {\bibfnamefont {H.}~\bibnamefont
  {Eivazi}}, \bibinfo {author} {\bibfnamefont {M.}~\bibnamefont {Tahani}},
  \bibinfo {author} {\bibfnamefont {P.}~\bibnamefont {Schlatter}}, \ and\
  \bibinfo {author} {\bibfnamefont {R.}~\bibnamefont {Vinuesa}},\ }\bibfield
  {title} {\enquote {\bibinfo {title} {Physics-informed neural networks for
  solving {Reynolds-averaged} {Navier--Stokes} equations},}\ }\href@noop {}
  {\bibfield  {journal} {\bibinfo  {journal} {Physics of Fluids}\ }\textbf
  {\bibinfo {volume} {34}} (\bibinfo {year} {2022})}\BibitemShut {NoStop}%
\bibitem [{\citenamefont {Wang}, \citenamefont {Liu},\ and\ \citenamefont
  {Wang}(2022)}]{wang2022dense}%
  \BibitemOpen
  \bibfield  {author} {\bibinfo {author} {\bibfnamefont {H.}~\bibnamefont
  {Wang}}, \bibinfo {author} {\bibfnamefont {Y.}~\bibnamefont {Liu}}, \ and\
  \bibinfo {author} {\bibfnamefont {S.}~\bibnamefont {Wang}},\ }\bibfield
  {title} {\enquote {\bibinfo {title} {Dense velocity reconstruction from
  particle image velocimetry/particle tracking velocimetry using a
  physics-informed neural network},}\ }\href@noop {} {\bibfield  {journal}
  {\bibinfo  {journal} {Physics of fluids}\ }\textbf {\bibinfo {volume} {34}}
  (\bibinfo {year} {2022})}\BibitemShut {NoStop}%
\bibitem [{\citenamefont {McClenny}\ and\ \citenamefont
  {Braga-Neto}(2023)}]{MCCLENNY2023111722}%
  \BibitemOpen
  \bibfield  {author} {\bibinfo {author} {\bibfnamefont {L.~D.}\ \bibnamefont
  {McClenny}}\ and\ \bibinfo {author} {\bibfnamefont {U.~M.}\ \bibnamefont
  {Braga-Neto}},\ }\bibfield  {title} {\enquote {\bibinfo {title}
  {Self-adaptive physics-informed neural networks},}\ }\href {\doibase
  https://doi.org/10.1016/j.jcp.2022.111722} {\bibfield  {journal} {\bibinfo
  {journal} {Journal of Computational Physics}\ }\textbf {\bibinfo {volume}
  {474}},\ \bibinfo {pages} {111722} (\bibinfo {year} {2023})}\BibitemShut
  {NoStop}%
\bibitem [{\citenamefont {Lin}\ and\ \citenamefont
  {Chen}(2022)}]{LIN2022111053}%
  \BibitemOpen
  \bibfield  {author} {\bibinfo {author} {\bibfnamefont {S.}~\bibnamefont
  {Lin}}\ and\ \bibinfo {author} {\bibfnamefont {Y.}~\bibnamefont {Chen}},\
  }\bibfield  {title} {\enquote {\bibinfo {title} {A two-stage physics-informed
  neural network method based on conserved quantities and applications in
  localized wave solutions},}\ }\href {\doibase
  https://doi.org/10.1016/j.jcp.2022.111053} {\bibfield  {journal} {\bibinfo
  {journal} {Journal of Computational Physics}\ }\textbf {\bibinfo {volume}
  {457}},\ \bibinfo {pages} {111053} (\bibinfo {year} {2022})}\BibitemShut
  {NoStop}%
\bibitem [{\citenamefont {Yuan}\ \emph {et~al.}(2022)\citenamefont {Yuan},
  \citenamefont {Ni}, \citenamefont {Deng},\ and\ \citenamefont
  {Hao}}]{yuan2022pinn}%
  \BibitemOpen
  \bibfield  {author} {\bibinfo {author} {\bibfnamefont {L.}~\bibnamefont
  {Yuan}}, \bibinfo {author} {\bibfnamefont {Y.-Q.}\ \bibnamefont {Ni}},
  \bibinfo {author} {\bibfnamefont {X.-Y.}\ \bibnamefont {Deng}}, \ and\
  \bibinfo {author} {\bibfnamefont {S.}~\bibnamefont {Hao}},\ }\bibfield
  {title} {\enquote {\bibinfo {title} {{A-PINN}: Auxiliary physics informed
  neural networks for forward and inverse problems of nonlinear
  integro-differential equations},}\ }\href@noop {} {\bibfield  {journal}
  {\bibinfo  {journal} {Journal of Computational Physics}\ }\textbf {\bibinfo
  {volume} {462}},\ \bibinfo {pages} {111260} (\bibinfo {year}
  {2022})}\BibitemShut {NoStop}%
\bibitem [{\citenamefont {Rezaei}\ \emph {et~al.}(2022)\citenamefont {Rezaei},
  \citenamefont {Harandi}, \citenamefont {Moeineddin}, \citenamefont {Xu},\
  and\ \citenamefont {Reese}}]{rezaei2022mixed}%
  \BibitemOpen
  \bibfield  {author} {\bibinfo {author} {\bibfnamefont {S.}~\bibnamefont
  {Rezaei}}, \bibinfo {author} {\bibfnamefont {A.}~\bibnamefont {Harandi}},
  \bibinfo {author} {\bibfnamefont {A.}~\bibnamefont {Moeineddin}}, \bibinfo
  {author} {\bibfnamefont {B.-X.}\ \bibnamefont {Xu}}, \ and\ \bibinfo {author}
  {\bibfnamefont {S.}~\bibnamefont {Reese}},\ }\bibfield  {title} {\enquote
  {\bibinfo {title} {A mixed formulation for physics-informed neural networks
  as a potential solver for engineering problems in heterogeneous domains:
  Comparison with finite element method},}\ }\href@noop {} {\bibfield
  {journal} {\bibinfo  {journal} {Computer Methods in Applied Mechanics and
  Engineering}\ }\textbf {\bibinfo {volume} {401}},\ \bibinfo {pages} {115616}
  (\bibinfo {year} {2022})}\BibitemShut {NoStop}%
\bibitem [{\citenamefont {Zhao}\ \emph {et~al.}(2025)\citenamefont {Zhao},
  \citenamefont {Li}, \citenamefont {Fan}, \citenamefont {Wang}, \citenamefont
  {Yang},\ and\ \citenamefont {Wang}}]{ZHAO2025114125}%
  \BibitemOpen
  \bibfield  {author} {\bibinfo {author} {\bibfnamefont {S.}~\bibnamefont
  {Zhao}}, \bibinfo {author} {\bibfnamefont {Z.}~\bibnamefont {Li}}, \bibinfo
  {author} {\bibfnamefont {B.}~\bibnamefont {Fan}}, \bibinfo {author}
  {\bibfnamefont {Y.}~\bibnamefont {Wang}}, \bibinfo {author} {\bibfnamefont
  {H.}~\bibnamefont {Yang}}, \ and\ \bibinfo {author} {\bibfnamefont
  {J.}~\bibnamefont {Wang}},\ }\bibfield  {title} {\enquote {\bibinfo {title}
  {Lesnets (large-eddy simulation nets): Physics-informed neural operator for
  large-eddy simulation of turbulence},}\ }\href {\doibase
  https://doi.org/10.1016/j.jcp.2025.114125} {\bibfield  {journal} {\bibinfo
  {journal} {Journal of Computational Physics}\ }\textbf {\bibinfo {volume}
  {537}},\ \bibinfo {pages} {114125} (\bibinfo {year} {2025})}\BibitemShut
  {NoStop}%
\bibitem [{\citenamefont {Jeong}\ \emph {et~al.}(2023)\citenamefont {Jeong},
  \citenamefont {Bai}, \citenamefont {Batuwatta-Gamage}, \citenamefont
  {Rathnayaka}, \citenamefont {Zhou},\ and\ \citenamefont
  {Gu}}]{jeong2023physics}%
  \BibitemOpen
  \bibfield  {author} {\bibinfo {author} {\bibfnamefont {H.}~\bibnamefont
  {Jeong}}, \bibinfo {author} {\bibfnamefont {J.}~\bibnamefont {Bai}}, \bibinfo
  {author} {\bibfnamefont {C.~P.}\ \bibnamefont {Batuwatta-Gamage}}, \bibinfo
  {author} {\bibfnamefont {C.}~\bibnamefont {Rathnayaka}}, \bibinfo {author}
  {\bibfnamefont {Y.}~\bibnamefont {Zhou}}, \ and\ \bibinfo {author}
  {\bibfnamefont {Y.}~\bibnamefont {Gu}},\ }\bibfield  {title} {\enquote
  {\bibinfo {title} {A physics-informed neural network-based topology
  optimization ({PINNTO}) framework for structural optimization},}\ }\href@noop
  {} {\bibfield  {journal} {\bibinfo  {journal} {Engineering Structures}\
  }\textbf {\bibinfo {volume} {278}},\ \bibinfo {pages} {115484} (\bibinfo
  {year} {2023})}\BibitemShut {NoStop}%
\bibitem [{\citenamefont {Liu}\ \emph {et~al.}(2024{\natexlab{a}})\citenamefont
  {Liu}, \citenamefont {Chen}, \citenamefont {Ding},\ and\ \citenamefont
  {Chen}}]{liu2024adaptive}%
  \BibitemOpen
  \bibfield  {author} {\bibinfo {author} {\bibfnamefont {Y.}~\bibnamefont
  {Liu}}, \bibinfo {author} {\bibfnamefont {L.}~\bibnamefont {Chen}}, \bibinfo
  {author} {\bibfnamefont {J.}~\bibnamefont {Ding}}, \ and\ \bibinfo {author}
  {\bibfnamefont {Y.}~\bibnamefont {Chen}},\ }\bibfield  {title} {\enquote
  {\bibinfo {title} {An adaptive sampling method based on expected improvement
  function and residual gradient in {PINNs}},}\ }\href@noop {} {\bibfield
  {journal} {\bibinfo  {journal} {IEEE Access}\ } (\bibinfo {year}
  {2024}{\natexlab{a}})}\BibitemShut {NoStop}%
\bibitem [{\citenamefont {Zhou}, \citenamefont {Miwa},\ and\ \citenamefont
  {Okamoto}(2024)}]{10.1063/5.0180770}%
  \BibitemOpen
  \bibfield  {author} {\bibinfo {author} {\bibfnamefont {W.}~\bibnamefont
  {Zhou}}, \bibinfo {author} {\bibfnamefont {S.}~\bibnamefont {Miwa}}, \ and\
  \bibinfo {author} {\bibfnamefont {K.}~\bibnamefont {Okamoto}},\ }\bibfield
  {title} {\enquote {\bibinfo {title} {Advancing fluid dynamics simulations: A
  comprehensive approach to optimizing physics-informed neural networks},}\
  }\href {\doibase 10.1063/5.0180770} {\bibfield  {journal} {\bibinfo
  {journal} {Physics of Fluids}\ }\textbf {\bibinfo {volume} {36}},\ \bibinfo
  {pages} {013615} (\bibinfo {year} {2024})}\BibitemShut {NoStop}%
\bibitem [{\citenamefont {Bai}\ \emph {et~al.}(2023)\citenamefont {Bai},
  \citenamefont {Rabczuk}, \citenamefont {Gupta}, \citenamefont {Alzubaidi},\
  and\ \citenamefont {Gu}}]{bai2023physics}%
  \BibitemOpen
  \bibfield  {author} {\bibinfo {author} {\bibfnamefont {J.}~\bibnamefont
  {Bai}}, \bibinfo {author} {\bibfnamefont {T.}~\bibnamefont {Rabczuk}},
  \bibinfo {author} {\bibfnamefont {A.}~\bibnamefont {Gupta}}, \bibinfo
  {author} {\bibfnamefont {L.}~\bibnamefont {Alzubaidi}}, \ and\ \bibinfo
  {author} {\bibfnamefont {Y.}~\bibnamefont {Gu}},\ }\bibfield  {title}
  {\enquote {\bibinfo {title} {A physics-informed neural network technique
  based on a modified loss function for computational {2D} and {3D} solid
  mechanics},}\ }\href@noop {} {\bibfield  {journal} {\bibinfo  {journal}
  {Computational Mechanics}\ }\textbf {\bibinfo {volume} {71}},\ \bibinfo
  {pages} {543--562} (\bibinfo {year} {2023})}\BibitemShut {NoStop}%
\bibitem [{\citenamefont {Gao}, \citenamefont {Sun},\ and\ \citenamefont
  {Wang}(2021)}]{GAO2021110079}%
  \BibitemOpen
  \bibfield  {author} {\bibinfo {author} {\bibfnamefont {H.}~\bibnamefont
  {Gao}}, \bibinfo {author} {\bibfnamefont {L.}~\bibnamefont {Sun}}, \ and\
  \bibinfo {author} {\bibfnamefont {J.-X.}\ \bibnamefont {Wang}},\ }\bibfield
  {title} {\enquote {\bibinfo {title} {{PhyGeoNet}: Physics-informed
  geometry-adaptive convolutional neural networks for solving parameterized
  steady-state {PDEs} on irregular domain},}\ }\href {\doibase
  https://doi.org/10.1016/j.jcp.2020.110079} {\bibfield  {journal} {\bibinfo
  {journal} {Journal of Computational Physics}\ }\textbf {\bibinfo {volume}
  {428}},\ \bibinfo {pages} {110079} (\bibinfo {year} {2021})}\BibitemShut
  {NoStop}%
\bibitem [{\citenamefont {Wang}\ \emph
  {et~al.}(2025{\natexlab{a}})\citenamefont {Wang}, \citenamefont {Sankaran},
  \citenamefont {Stinis},\ and\ \citenamefont
  {Perdikaris}}]{wang2025simulatingthreedimensionalturbulencephysicsinformed}%
  \BibitemOpen
  \bibfield  {author} {\bibinfo {author} {\bibfnamefont {S.}~\bibnamefont
  {Wang}}, \bibinfo {author} {\bibfnamefont {S.}~\bibnamefont {Sankaran}},
  \bibinfo {author} {\bibfnamefont {P.}~\bibnamefont {Stinis}}, \ and\ \bibinfo
  {author} {\bibfnamefont {P.}~\bibnamefont {Perdikaris}},\ }\bibfield  {title}
  {\enquote {\bibinfo {title} {Simulating three-dimensional turbulence with
  physics-informed neural networks},}\ }\href@noop {} {\bibfield  {journal}
  {\bibinfo  {journal} {arXiv preprint arXiv:2507.08972}\ } (\bibinfo {year}
  {2025}{\natexlab{a}})}\BibitemShut {NoStop}%
\bibitem [{\citenamefont {Liu}\ \emph {et~al.}(2023)\citenamefont {Liu},
  \citenamefont {Liu}, \citenamefont {Yan}, \citenamefont {Guo},\ and\
  \citenamefont {an~Zhang}}]{LIU2023112291}%
  \BibitemOpen
  \bibfield  {author} {\bibinfo {author} {\bibfnamefont {Y.}~\bibnamefont
  {Liu}}, \bibinfo {author} {\bibfnamefont {W.}~\bibnamefont {Liu}}, \bibinfo
  {author} {\bibfnamefont {X.}~\bibnamefont {Yan}}, \bibinfo {author}
  {\bibfnamefont {S.}~\bibnamefont {Guo}}, \ and\ \bibinfo {author}
  {\bibfnamefont {C.}~\bibnamefont {an~Zhang}},\ }\bibfield  {title} {\enquote
  {\bibinfo {title} {Adaptive transfer learning for {PINN}},}\ }\href {\doibase
  https://doi.org/10.1016/j.jcp.2023.112291} {\bibfield  {journal} {\bibinfo
  {journal} {Journal of Computational Physics}\ }\textbf {\bibinfo {volume}
  {490}},\ \bibinfo {pages} {112291} (\bibinfo {year} {2023})}\BibitemShut
  {NoStop}%
\bibitem [{\citenamefont {Faroughi}\ \emph {et~al.}(2024)\citenamefont
  {Faroughi}, \citenamefont {Pawar}, \citenamefont {Fernandes}, \citenamefont
  {Raissi}, \citenamefont {Das}, \citenamefont {Kalantari},\ and\ \citenamefont
  {Kourosh~Mahjour}}]{faroughi2024physics}%
  \BibitemOpen
  \bibfield  {author} {\bibinfo {author} {\bibfnamefont {S.~A.}\ \bibnamefont
  {Faroughi}}, \bibinfo {author} {\bibfnamefont {N.~M.}\ \bibnamefont {Pawar}},
  \bibinfo {author} {\bibfnamefont {C.}~\bibnamefont {Fernandes}}, \bibinfo
  {author} {\bibfnamefont {M.}~\bibnamefont {Raissi}}, \bibinfo {author}
  {\bibfnamefont {S.}~\bibnamefont {Das}}, \bibinfo {author} {\bibfnamefont
  {N.~K.}\ \bibnamefont {Kalantari}}, \ and\ \bibinfo {author} {\bibfnamefont
  {S.}~\bibnamefont {Kourosh~Mahjour}},\ }\bibfield  {title} {\enquote
  {\bibinfo {title} {Physics-guided, physics-informed, and physics-encoded
  neural networks and operators in scientific computing: Fluid and solid
  mechanics},}\ }\href@noop {} {\bibfield  {journal} {\bibinfo  {journal}
  {Journal of Computing and Information Science in Engineering}\ }\textbf
  {\bibinfo {volume} {24}},\ \bibinfo {pages} {040802} (\bibinfo {year}
  {2024})}\BibitemShut {NoStop}%
\bibitem [{\citenamefont {Liu}\ \emph {et~al.}(2024{\natexlab{b}})\citenamefont
  {Liu}, \citenamefont {Wang}, \citenamefont {Vaidya}, \citenamefont {Ruehle},
  \citenamefont {Halverson}, \citenamefont {Solja{v{c}}i{'c}}, \citenamefont
  {Hou},\ and\ \citenamefont {Tegmark}}]{liu2024kan}%
  \BibitemOpen
  \bibfield  {author} {\bibinfo {author} {\bibfnamefont {Z.}~\bibnamefont
  {Liu}}, \bibinfo {author} {\bibfnamefont {Y.}~\bibnamefont {Wang}}, \bibinfo
  {author} {\bibfnamefont {S.}~\bibnamefont {Vaidya}}, \bibinfo {author}
  {\bibfnamefont {F.}~\bibnamefont {Ruehle}}, \bibinfo {author} {\bibfnamefont
  {J.}~\bibnamefont {Halverson}}, \bibinfo {author} {\bibfnamefont
  {M.}~\bibnamefont {Solja{v{c}}i{'c}}}, \bibinfo {author} {\bibfnamefont
  {T.~Y.}\ \bibnamefont {Hou}}, \ and\ \bibinfo {author} {\bibfnamefont
  {M.}~\bibnamefont {Tegmark}},\ }\bibfield  {title} {\enquote {\bibinfo
  {title} {Kan: Kolmogorov-arnold networks},}\ }\href@noop {} {\bibfield
  {journal} {\bibinfo  {journal} {arXiv preprint arXiv:2404.19756}\ } (\bibinfo
  {year} {2024}{\natexlab{b}})}\BibitemShut {NoStop}%
\bibitem [{\citenamefont {de~Fran{c{c}}a}(2018)}]{de2018greedy}%
  \BibitemOpen
  \bibfield  {author} {\bibinfo {author} {\bibfnamefont {F.~O.}\ \bibnamefont
  {de~Fran{c{c}}a}},\ }\bibfield  {title} {\enquote {\bibinfo {title} {A greedy
  search tree heuristic for symbolic regression},}\ }\href@noop {} {\bibfield
  {journal} {\bibinfo  {journal} {Information Sciences}\ }\textbf {\bibinfo
  {volume} {442}},\ \bibinfo {pages} {18--32} (\bibinfo {year}
  {2018})}\BibitemShut {NoStop}%
\bibitem [{\citenamefont {Willard}\ \emph {et~al.}(2022)\citenamefont
  {Willard}, \citenamefont {Jia}, \citenamefont {Xu}, \citenamefont
  {Steinbach},\ and\ \citenamefont {Kumar}}]{willard2022integrating}%
  \BibitemOpen
  \bibfield  {author} {\bibinfo {author} {\bibfnamefont {J.}~\bibnamefont
  {Willard}}, \bibinfo {author} {\bibfnamefont {X.}~\bibnamefont {Jia}},
  \bibinfo {author} {\bibfnamefont {S.}~\bibnamefont {Xu}}, \bibinfo {author}
  {\bibfnamefont {M.}~\bibnamefont {Steinbach}}, \ and\ \bibinfo {author}
  {\bibfnamefont {V.}~\bibnamefont {Kumar}},\ }\bibfield  {title} {\enquote
  {\bibinfo {title} {Integrating scientific knowledge with machine learning for
  engineering and environmental systems},}\ }\href@noop {} {\bibfield
  {journal} {\bibinfo  {journal} {ACM Computing Surveys}\ }\textbf {\bibinfo
  {volume} {55}},\ \bibinfo {pages} {1--37} (\bibinfo {year}
  {2022})}\BibitemShut {NoStop}%
\bibitem [{\citenamefont {Li}, \citenamefont {Du},\ and\ \citenamefont
  {Martins}(2022)}]{li2022machine}%
  \BibitemOpen
  \bibfield  {author} {\bibinfo {author} {\bibfnamefont {J.}~\bibnamefont
  {Li}}, \bibinfo {author} {\bibfnamefont {X.}~\bibnamefont {Du}}, \ and\
  \bibinfo {author} {\bibfnamefont {J.~R.}\ \bibnamefont {Martins}},\
  }\bibfield  {title} {\enquote {\bibinfo {title} {Machine learning in
  aerodynamic shape optimization},}\ }\href@noop {} {\bibfield  {journal}
  {\bibinfo  {journal} {Progress in Aerospace Sciences}\ }\textbf {\bibinfo
  {volume} {134}},\ \bibinfo {pages} {100849} (\bibinfo {year}
  {2022})}\BibitemShut {NoStop}%
\bibitem [{\citenamefont {Yang}, \citenamefont {Meng},\ and\ \citenamefont
  {Karniadakis}(2021)}]{YANG2021109913}%
  \BibitemOpen
  \bibfield  {author} {\bibinfo {author} {\bibfnamefont {L.}~\bibnamefont
  {Yang}}, \bibinfo {author} {\bibfnamefont {X.}~\bibnamefont {Meng}}, \ and\
  \bibinfo {author} {\bibfnamefont {G.~E.}\ \bibnamefont {Karniadakis}},\
  }\bibfield  {title} {\enquote {\bibinfo {title} {{B-PINNs}: Bayesian
  physics-informed neural networks for forward and inverse {PDE} problems with
  noisy data},}\ }\href {\doibase https://doi.org/10.1016/j.jcp.2020.109913}
  {\bibfield  {journal} {\bibinfo  {journal} {Journal of Computational
  Physics}\ }\textbf {\bibinfo {volume} {425}},\ \bibinfo {pages} {109913}
  (\bibinfo {year} {2021})}\BibitemShut {NoStop}%
\bibitem [{\citenamefont {Koenig}, \citenamefont {Kim},\ and\ \citenamefont
  {Deng}(2024)}]{KOENIG2024117397}%
  \BibitemOpen
  \bibfield  {author} {\bibinfo {author} {\bibfnamefont {B.~C.}\ \bibnamefont
  {Koenig}}, \bibinfo {author} {\bibfnamefont {S.}~\bibnamefont {Kim}}, \ and\
  \bibinfo {author} {\bibfnamefont {S.}~\bibnamefont {Deng}},\ }\bibfield
  {title} {\enquote {\bibinfo {title} {{KAN-ODEs}: {Kolmogorov–Arnold}
  network ordinary differential equations for learning dynamical systems and
  hidden physics},}\ }\href {\doibase
  https://doi.org/10.1016/j.cma.2024.117397} {\bibfield  {journal} {\bibinfo
  {journal} {Computer Methods in Applied Mechanics and Engineering}\ }\textbf
  {\bibinfo {volume} {432}},\ \bibinfo {pages} {117397} (\bibinfo {year}
  {2024})}\BibitemShut {NoStop}%
\bibitem [{\citenamefont {Wang}\ \emph
  {et~al.}(2025{\natexlab{b}})\citenamefont {Wang}, \citenamefont {Sun},
  \citenamefont {Bai}, \citenamefont {Anitescu}, \citenamefont {Eshaghi},
  \citenamefont {Zhuang}, \citenamefont {Rabczuk},\ and\ \citenamefont
  {Liu}}]{WANG2025117518}%
  \BibitemOpen
  \bibfield  {author} {\bibinfo {author} {\bibfnamefont {Y.}~\bibnamefont
  {Wang}}, \bibinfo {author} {\bibfnamefont {J.}~\bibnamefont {Sun}}, \bibinfo
  {author} {\bibfnamefont {J.}~\bibnamefont {Bai}}, \bibinfo {author}
  {\bibfnamefont {C.}~\bibnamefont {Anitescu}}, \bibinfo {author}
  {\bibfnamefont {M.~S.}\ \bibnamefont {Eshaghi}}, \bibinfo {author}
  {\bibfnamefont {X.}~\bibnamefont {Zhuang}}, \bibinfo {author} {\bibfnamefont
  {T.}~\bibnamefont {Rabczuk}}, \ and\ \bibinfo {author} {\bibfnamefont
  {Y.}~\bibnamefont {Liu}},\ }\bibfield  {title} {\enquote {\bibinfo {title}
  {Kolmogorov–arnold-informed neural network: A physics-informed deep
  learning framework for solving forward and inverse problems based on
  kolmogorov–arnold networks},}\ }\href {\doibase
  https://doi.org/10.1016/j.cma.2024.117518} {\bibfield  {journal} {\bibinfo
  {journal} {Computer Methods in Applied Mechanics and Engineering}\ }\textbf
  {\bibinfo {volume} {433}},\ \bibinfo {pages} {117518} (\bibinfo {year}
  {2025}{\natexlab{b}})}\BibitemShut {NoStop}%
\end{thebibliography}%

\appendix
\renewcommand{\thefigure}{A\arabic{figure}}
\setcounter{figure}{0}
\section{Performance Observations of the Original KAN}

\begin{figure}[!htbp]
	\centering
	\includegraphics[height=5cm]{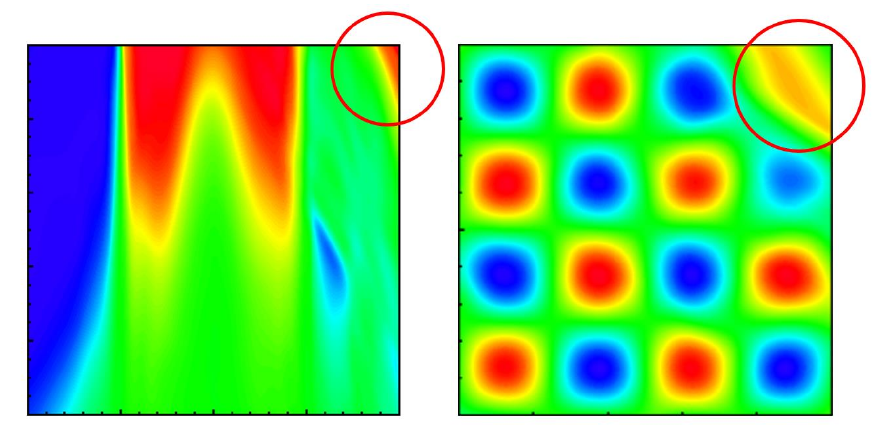}
	\caption{Phenomena observed during flow field prediction using the original KAN.}
	\label{KANerror}
\end{figure}

Following its introduction, the Kolmogorov-Arnold Network (KAN) has garnered significant attention in the scientific community, prompting our research team to promptly engage in its investigation. However, during our application of the original KAN framework to flow field prediction tasks, we consistently observed prediction inaccuracies occurring at domain boundaries, as illustrated in Figure \ref{KANerror}.

Figure \ref{KANerror} clearly shows significant errors occurring at the boundaries of the computational domain when using the original KAN for flow prediction, which led us to investigate the causes of this phenomenon. From the mathematical perspective, B-$splines$ exhibit potential limitations when applied to the complex problems investigated in this study. Primarily, their non-orthogonal basis functions exhibit linear dependence, which may lead to solution instability when strong inter-basis correlations occur. Second, while spline curves are commonly used in research to prevent Runge's phenomenon, they still exhibit severe oscillations when processing highly irregular node distributions. In contrast, Chebyshev polynomials inherently overcome these limitations through their dual advantages of mutual orthogonality and endpoint-clustering characteristics, prompting us to incorporate them into the original KAN framework.

\section{Prediction Performance under Different Weighting Coefficients $\lambda$}
\renewcommand{\thetable}{B\arabic{table}}
\setcounter{table}{0}
\setcounter{figure}{0}
\renewcommand{\thefigure}{B\arabic{figure}}

For neural networks with physical information loss terms, the weighting coefficient $\lambda$ determines the relative contribution of $Los{s_{data}}$ and $Los{s_{phy}}$ in guiding the optimization direction during training, and remains a critical hyperparameter affecting model performance. In the main study, we first conduct pretraining with $\lambda$ = 1, then determine the final weighting coefficient $\lambda$ = 0.1 based on the ratio between $Los{s_{data}}$ and $Los{s_{phy}}$.

In this Appendix B, we investigate the impact of different $\lambda$ values on neural network performance. We evaluate both ChebPIKAN and PINN architectures with 2×30 hidden layers using $\lambda$ values ranging from $10^{-4}$ to $10^{2}$, with their predictive performance summarized in the Table \ref{tab:lamda}.
\begin{table}[!htbp]
	\centering
	\caption{Mean residuals of ChebPIKANs and PINNs for Navier--Stokes equations under different weighting coefficients $\lambda$.}
	\label{tab:lamda}
	\setlength{\tabcolsep}{12pt}
	\begin{tabular}{cccccc}
		\hline
		Name & Hidden Layers & $\lambda$ & Er(u) & Er(v) & Er(p)  \\ \hline
		ChebPIKANs & 2x30 & $10^{-4}$ & 3.05\% & 0.75\% & 3.70\%  \\ 
		~ & 2x30 & $10^{-3}$ & 1.78\% & 1.05\% & 2.01\%  \\
		~ & 2x30 & $10^{-2}$ & 0.44\% & 0.38\% & 0.49\%  \\
		~ & 2x30 & $10^{-1}$ & 0.24\% & 0.35\% & 0.33\%  \\ 
		~ & 2x30 & $10^{0}$ & 1.09\% & 1.39\% & 1.16\%  \\
		~ & 2x30 & $10^{1}$ & 8.86\% & 12.40\% & 16.80\%  \\ 
		~ & 2x30 & $10^{2}$ & 29.15\% & 44.96\% & 54.03\%  \\
		PINNs & 2x30 & $10^{-4}$ & 2.41\% & 2.73\% & 4.31\%  \\ 
		~ & 2x30 & $10^{-3}$ & 1.93\% & 2.82\% & 3.89\%  \\
		~ & 2x30 & $10^{-2}$ & 2.32\% & 2.72\% & 3.93\%  \\ 
		~ & 2x30 & $10^{-1}$ & 2.10\% & 2.52\% & 2.34\%  \\ 
		~ & 2x30 & $10^{0}$ & 7.65\% & 8.60\% & 5.53\%  \\ 
		~ & 2x30 & $10^{-1}$ & 21.83\% & 32.97\% & 40.04\%  \\ 
		~ & 2x30 & $10^{-2}$ & 37.07\% & 54.99\% & 57.91\%  \\ \hline
	\end{tabular}
\end{table}
\begin{figure}[!htbp]
	\centering
	\includegraphics[height=5cm]{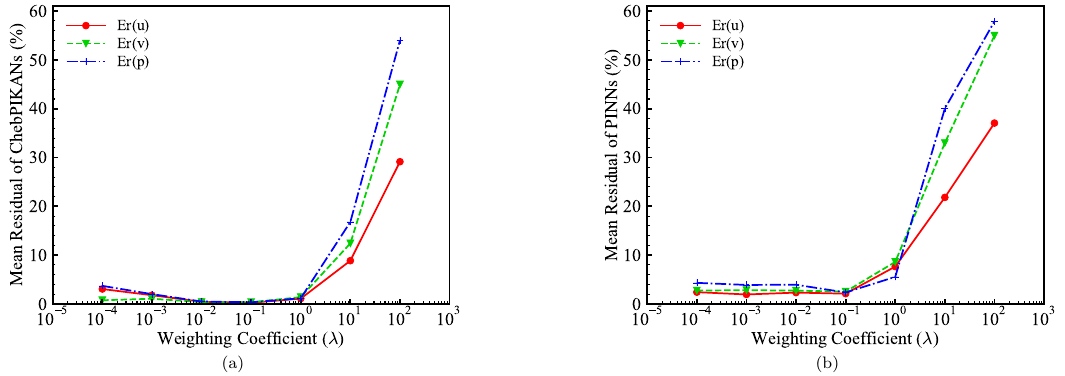}
	\caption{Mean residuals of $u$, $v$, and $p$ of ChebPIKANs and PINNs with different weighting coefficients $\lambda$ for Navier--Stokes equations. (a) ChebPIKANs, (b) PINNs.}
	\label{B1}
\end{figure}

As shown in Table \ref{tab:lamda} and Figure \ref{B1}, ChebPIKANs consistently demonstrate superior performance to PINNs under various $\lambda$ conditions. Both architectures achieve their best performance at $\lambda$ = 0.1 among the discrete values we examined, suggesting that the optimal $\lambda$ probably lies around this value. Therefore, we select $\lambda$ = 0.1 in our research. The results demonstrate that network performance degrades significantly when the weighting coefficient $\lambda$ becomes either too small or too large, revealing an inherent trade-off in balancing the two loss terms. At lower $\lambda$ values, the $Los{s_{data}}$ dominates the optimization process, causing the network to behave similarly to conventional KANs and ANNs without physical constraints. Conversely, higher $\lambda$ values may drive the training process away from the training data distribution as the model prioritizes minimizing the $Los{s_{phy}}$, potentially leading to constant-field solutions that satisfy the physical constraints while degrading the prediction accuracy.

This analysis highlights how the pretraining process with limited $\lambda$ variations may fail to identify optimal weighting coefficients, since the initial choice of $\lambda$ itself influences the final parameter selection. The challenge of determining appropriate hyperparameter values remains a persistent issue in physics-informed neural network development. Consequently, the creation of adaptive algorithms that can automatically balance these competing loss terms during training represents a valuable direction for future research in this field. We will continue to develop our algorithm specifically to address this issue.

\end{document}